\documentclass[a4paper,10pt]{article}

\usepackage[margin=1.1in]{geometry}

\usepackage{graphicx}
\usepackage{epstopdf}
\usepackage{amsmath,amssymb,pdflscape,color}
\usepackage{hyperref}
\usepackage[normalem]{ulem}
\usepackage{xr}

\usepackage{cleveref}
\crefformat{section}{\S#2#1#3} 
\crefformat{subsection}{\S#2#1#3}
\crefformat{subsubsection}{\S#2#1#3}

\usepackage{caption}
\usepackage[symbol]{footmisc}

\usepackage[numbers,sort&compress]{natbib}

\begin{document}

\title{Implementing measurement error models with mechanistic mathematical models in a likelihood-based framework for estimation, identifiability analysis, and prediction in the life sciences\date{}}
	
\author{Ryan J. Murphy$^{1,*}$, Oliver J. Maclaren$^{2}$, Matthew J. Simpson$^{1}$}

\maketitle

\vspace{-0.25in}
\begin{center} 
	\textit{$^{1}$ Mathematical Sciences, Queensland University of Technology, Brisbane, Australia}\\
		\textit{$^{2}$ The Department of Engineering Science and Biomedical Engineering, University of Auckland, Auckland, New Zealand}
\end{center}

\begin{abstract}
\noindent Throughout the life sciences we routinely seek to interpret measurements and observations using parameterised mechanistic mathematical models. A fundamental and often overlooked choice in this approach involves relating the solution of a mathematical model with noisy and incomplete measurement data. This is often achieved by assuming that the data are noisy measurements of the solution of a deterministic mathematical model, and that measurement errors are additive and normally distributed. While this assumption of additive Gaussian noise is extremely common and simple to implement and interpret, it is often unjustified and can lead to poor parameter estimates and non-physical predictions. One way to overcome this challenge is to implement a different measurement error model. In this review, we demonstrate how to implement a range of measurement error models in a likelihood-based framework for estimation, identifiability analysis, and prediction, called Profile-Wise Analysis. This frequentist approach to uncertainty quantification for mechanistic models leverages the profile likelihood for targeting parameters and understanding their influence on predictions. Case studies, motivated by simple caricature models routinely used in systems biology and mathematical biology literature, illustrate how the same ideas apply to different types of mathematical models. Open-source Julia code to reproduce results is available on \href{https://github.com/ryanmurphy42/Murphy2023ErrorModels}{GitHub}.

\end{abstract}


\footnotetext[1]{Corresponding author: r23.murphy@qut.edu.au}

\newpage
\section{Introduction}

Mechanistic mathematical modelling and statistical uncertainty quantification are powerful tools for interpreting noisy incomplete data and facilitate decision making across a wide range of applications in the life sciences. Interpreting such data using mathematical models involves many different types of modelling choices, each of which can impact results and their interpretation. One of the simplest examples of connecting a mathematical model to data involves the use of a straight line model. A common approach to estimate a best-fit straight line involves linear regression and the method of ordinary least squares \cite{Gelman2006,Montgomery2012,Seber2003,Weisberg2005}. In this example, the mathematical model is chosen to be a straight line, $y = mx + c$, and the noisy data are assumed to be normally distributed with zero mean and constant positive variance about the true straight line. This assumption of additive Gaussian noise is a modelling choice that we refer to as an additive Gaussian measurement error model. Measurement error models are primarily used to describe uncertainties in the measurement process, and to a lesser extent random intrinsic variation \cite{Simpson2022b}. Other similar terminologies include noise model, error model, and observation error model, but here we will refer to this as a measurement error model. Here and throughout, we assume that measurement errors are uncorrelated, independent and identically distributed. Ordinary least squares best-fit model parameters, $\hat{m}$ and $\hat{c}$, are estimated by minimising the sum of the squared residuals, $E(m,c) = \sum_{i=1}^{I}(y_i^{\mathrm{o}} - y_i)^2$, where the $i^{\mathrm{th}}$ residual, for $i=1,2,\ldots, I$, is the distance in the $y$-direction between the $i^{\mathrm{th}}$ data point, $y_{i}^{\mathrm{o}}$, and the corresponding point on the best-fit straight line, $y_i$. Hence the name method of least squares. The best-fit straight line is then the mathematical model evaluated at the best-fit model parameters, i.e. $y = \hat{m}x + \hat{c}$, where $\hat{m}$ and $\hat{c}$ are the values of the slope and intercept that minimises $E(m,c)$. Uncertainty in this example can be captured through the use of confidence intervals for model parameters, a confidence interval for the straight line based on the uncertainty in the model parameters, and a prediction interval for future observations \cite{Gelman2006,Montgomery2012,Seber2003,Weisberg2005}. 
 
In this review we present a general framework extending these concepts to mechanistic mathematical models, in the form of systems of ordinary differential equations (ODEs) and systems of partial differential equations (PDEs), that are often considered in the systems biology literature and the mathematical biology literature, respectively. In particular, our primary focus is on the fundamental question of how to connect the output of a mathematical model to data using a variety of measurement error models.

 The additive Gaussian measurement error model is ubiquitous and simple to interpret for mechanistic mathematical models, and often relates to estimating a best-fit model solution using nonlinear regression and a least-squares estimation problem \cite{Motulsky1987,Seber2003a}. Nonlinear regression extends the concept of linear regression to models where there is a nonlinear dependence between model parameters and model outputs that is typical for many deterministic ODEs and PDEs. Use of an additive Gaussian error model is often justified via the central limit theorem. However, the assumption of additive Gaussian noise is often unjustified in practice and, as we demonstrate, this can have important consequences because this assumption can lead to poor parameter estimates and non-physical predictions. Furthermore, even when the additive Gaussian error model is a reasonable choice it may not always be the most appropriate. In general there are many ways in which noise could impact a system. For example, multiplicative noise models are often thought to be more relevant to problems in some parts of the systems biology literature \cite{Agamennoni2012,Firth1987,Furusawa2005,Kreutz2007,Lacey1997,Limpert2001,Raue2013a}. One approach to tackle such challenges is to implement a different measurement error model. Here, we present a practical guide to implement a variety of measurement error models. Then, using illustrative case studies, we explain how to interpret results. Our approach in this review is not to claim that one noise model is superior to another, but to illustrate how relatively straightforward it can be to implement different noise models with different types of mathematical models.

All modelling choices, including the choice of a relevant mechanistic mathematical model and the choice of how to connect the mathematical model to data, should be considered on a case-by-case basis. As our focus is on the implementation of different error measurement models for ease of exposition we choose to explore simple caricature mathematical models from the systems biology literature and the mathematical biology literature  rather than focusing on very specific models that might be relevant to a smaller audience. The kinds of mathematical models we explore include systems biology-type systems of ODEs \cite{Alon2019,Kreutz2013,Schmidt2006}, mathematical biology-type systems of PDEs \cite{Britton2005,EdelsteinKeshet2005,Murray2002a,Murray2002b,Kot2001}, and difference equations \cite{Murray2002a,Murray2002b,AugerMethe2021,Hefley2013a,Ricker1954,Valpine2002}.  Mathematical models of greater complexity are straightforward to explore using the methods presented in this study and our open source software can be adapted to deal with more biologically complicated models as required. Measurement error models can take many forms, for example discrete, continuous, additive, and multiplicative, and the framework is well-suited to explore these different options. We do not preference any particular measurement error model, however we do illustrate that the framework can be used to help distinguish between the suitability of different choices of error model, such as choosing an error model that ensures non-negative predictions for quantities like concentrations or population densities.

\begin{figure}[p]
	\centering
	\includegraphics[width=\textwidth]{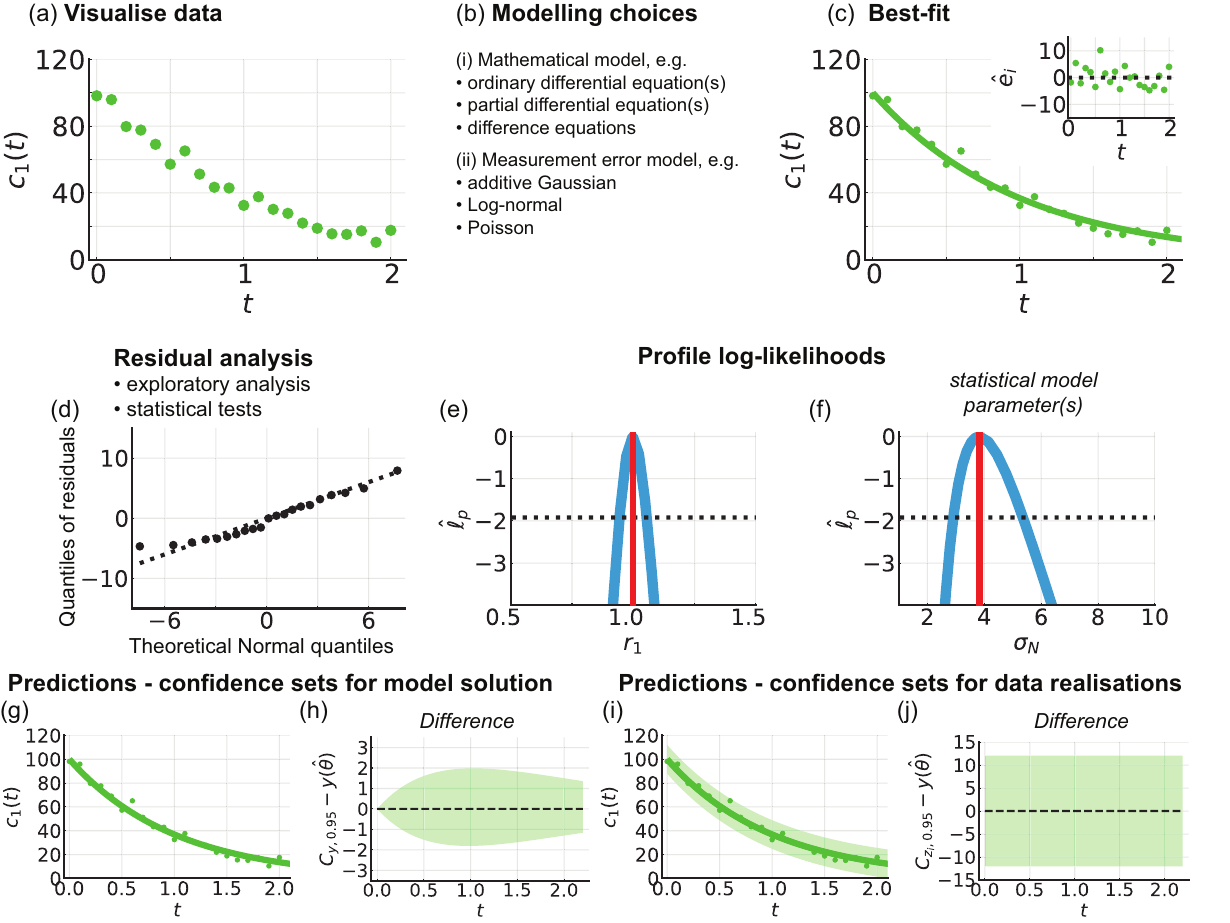}
	\caption{Implementing a variety of measurement error models in a profile likelihood-based framework for parameter estimation, identifiability analysis, and prediction.  (a) Synthetic data (circles). (b) The framework is applicable to a range of  mathematical models and measurement error models. Schematics show results for a simple exponential decay ordinary differential model, $\mathrm{d}c_{1}(t)/\mathrm{d}t = -r_{1}c_{1}(t)$, the additive Gaussian measurement error model, known model parameters $\theta=(r_{1}, \sigma_{\mathrm{N}}) = (1.0, 5.0)$, fixed initial condition $c_{1}(0)=100.0$, and observed data at twenty one equally--spaced time points from $t=0.0$ to $t=2.0$. (c) Mathematical model simulated with the MLE $\hat{\theta}=(r_{1}, \sigma_{\mathrm{N}}) = (0.99, 3.83)$ (solid line).  Inset of (c) residuals $\hat{e}_{i} = y_{i}^{\mathrm{o}} - y_{i}(\hat{\theta})$ with time, $t$. (d) Residual analysis can take many forms, results show a normal quantile-quantile plot of residuals. (e-g) Profile log-likelihoods (blue) shown for (e) $r_{1}$, and (f) $\sigma_{\mathrm{N}}$ with MLE (vertical red) and an approximate $95\%$ confidence interval threshold (horizontal black-dashed). Predictions in the form of (g) union of profile-wise confidence sets for the model solution and (i) the union of profile-wise Bonferroni correction-based confidence sets for data realisations. (g,i) show the mathematical model simulated with the MLE (solid), synthetic data (circles), and confidence sets (shaded regions). (h,j) To examine the confidence sets in detail we show the difference between the respective confidence sets and the mathematical model simulated with the MLE.}
	\label{fig:fig1}
\end{figure}

We now outline the Profile-Wise Analysis (PWA) \cite{Simpson2023} approach to estimation, identifiability analysis, and prediction for a set of data that takes the form of a time series of chemical concentrations, as is often the case in applications in systems biology. Crucial first steps are to visualise the data (Fig \ref{fig:fig1}a) and to implement certain modelling choices such as choosing between a continuous ODE or discrete difference model (Fig \ref{fig:fig1}b). As always, the choice of mathematical model should be considered with respect to structural identifiability of its parameters \cite{Audoly2001,Cheung2013,Chis2011,Raue2009,Hengl2007,Wieland2021}. Structural parameter non-identifiability means that there is a non-unique choice of model parameters that lead to the same model solution, and this can severely impede our ability to interpret results mechanistically since our ability to understand and interpret data mechanistically is often related to parameter estimation. For example, suppose one seeks to estimate two parameters $\lambda$ and $D$ but only the product $\lambda D$ is identifiable in the model \cite{Murray2002b,Maini2004a,Maini2004b}. In such a situation, we will be unable to estimate the value of the individual parameters irrespective of the number of measurements. Tools to assess structural identifiability of ODEs are reviewed in \cite{Barreiro2022}, including DAISY \cite{DAISY}, GENSSI2 \cite{GenSSI}, and the \texttt{StructuralIdentifiability} Julia package  \cite{Dong2022}.

\newpage
\textit{Parameter estimation for the mathematical model and measurement error model.} Given a mathematical model and a measurement error model we generate a best-fit model solution, analogous to a best-fit curve (Fig \ref{fig:fig1}c). To estimate the best-fit model solution we work within a likelihood-based framework. The likelihood function, $L(\theta \mid D)$, is related to the probability of observing data $D$ as a function of the parameters $\theta$ \cite{Pawitan2001}. In this setting the best-fit model solution corresponds to the output of the mathematical model simulated at the model parameters which are found to be `best' in the sense of those parameters that maximise $L(\theta \mid D)$. Parameters can be used to describe the mathematical model, such as $m$ and $c$ in the straight line example, and as well as describing the noise, such as the variance $\sigma_{N}^{2}$ in the additive Gaussian measurement error model. In this work we estimate both mathematical model parameters and statistical noise parameters simultaneously. Comparing the best-fit model solution with the data, and analysing residuals helps us to understand whether modelling choices are appropriate (Fig \ref{fig:fig1}d). Techniques to analyse standard additive residuals are reviewed in \cite{Motulsky1987,Seber2003a}.

\textit{Practical parameter identifiability.} While point estimates of best-fit model parameters are insightful, we often seek to understand how well parameters can be identified given a finite set of noisy incomplete data \cite{Maclaren2020,Kreutz2013,Raue2009,Simpson2023}. This question of practical parameter identifiability, and the subsequent components of the framework, can be explored using frequentist \cite{Kreutz2013,Raue2009,Pawitan2001,Simpson2023} or Bayesian methods \cite{Gelman2013,Hines2014,Linden2022,Raue2013,Raue2014,Siekmann2012}. While both approaches are generally interested in uncertainty quantification, we choose to work with a frequentist profile likelihood-based method that employs numerical optimisation procedures \cite{Kreutz2013,Raue2009,Pawitan2001,Simpson2023,Campbell2013,Eisenberg2014,Frohlich2014,Simpson2022a}. The optimisation procedures tend to be more computationally efficient than sampling-based methods for problems considered in this study \cite{Simpson2023,Simpson2020,Simpson2022c}. We also choose to work with a frequentist framework since there are many estimation, identifiability, and prediction workflows in Bayesian frameworks, but corresponding frequentist workflows that include prediction have received much less attention. Similarities and differences between our frequentist PWA workflow and Bayesian workflows are explored in \cite{Simpson2023}. While working with a full likelihood-based approach is relatively straightforward for models with a small number of parameters, this approach becomes computationally challenging for more complicated models with many parameters. By using a profile likelihood-based method we can target individual parameters of interest, explore their practical identifiability, and form approximate confidence intervals (Fig \ref{fig:fig1}e-f) \cite{Pawitan2001}. 

\textit{Prediction.} Given a set of estimated model parameters, together with an estimate of the uncertainty in our estimates, it is natural to seek to understand how uncertainty in model parameters impacts predictions of model solutions (mathematical model trajectories) and data realisations (unobserved measurements). This is important because practitioners are most likely to be interested in understanding the variability in predictions rather than variability in parameter estimates. In this framework we show that using parameter estimates to generate predictions is a powerful tool to assess the appropriateness of modelling choices and to interpret results. Predictions in the form of profile-wise confidence sets for model solutions are introduced in \cite{Simpson2023,Simpson2022c,Murphy2022c} and allow for predictions at a finer resolution than the data (Fig \ref{fig:fig1}g-h). These methods are simpler to implement and interpret in comparison to previous prediction methods that can involve additional constrained optimisation problems or integration based techniques \cite{Kreutz2013,Bjornstad1990,Hass2016,Kreutz2012,Villaverde2022,Wu2022}. An approach to form likelihood-based confidence sets for model realisations, where the model is composed of a mechanistic mathematical model and a measurement error model, was introduced in \cite{Simpson2023} and here we present concrete examples (Fig \ref{fig:fig1}i-j). We also demonstrate how to assess statistical coverage properties that are often of interest, including curvewise and pointwise coverage properties for predictions, and make comparisons to a gold-standard full likelihood-based approach \cite{Simpson2023}.

This review is structured as follows. In \cref{sec:methods}, we detail how to implement different measurement error models for parameter estimation, identifiability analysis, and prediction using profile likelihood-based techniques. In \cref{sec:resultsdiscussion}, we demonstrate the generality of the framework by exploring a variety of measurement error models using illustrative case studies motivated by systems biology-type models and mathematical biology-type models. In \cref{sec:coverage} we present an explicit example of how to evaluate statistical coverage properties. Supplementary material presents additional results including a comparison to a full likelihood-based approach \cite{Simpson2023}. To aid with understanding and reproducibility, all open source Julia code used to generate results is freely available on \href{https://github.com/ryanmurphy42/Murphy2023ErrorModels}{GitHub}.

\newpage
\section{Parameter estimation, identifiability analysis, and prediction}\label{sec:methods}

Here we detail the PWA profile likelihood-based framework for parameter estimation, identifiability analysis, and prediction. Throughout, we assume that experimental measurements are noisy observations of a deterministic mechanistic mathematical model. This framework is very general as it applies to cases where measurement error models may be additive, multiplicative, discrete, or continuous. As illustrative examples, we explicitly discuss and implement additive Gaussian noise, multiplicative log-normal and Poisson noise models. Mechanistic mathematical models may take many forms, for example systems of ODEs, systems of PDEs, and systems of difference equations. We choose to work with simple models to focus on the implementation of the framework and to make this work of interest to the broadest possible audience, as opposed to focusing on the details of specific mathematical models that are likely to be of interest to a smaller community.  Our hope is that by focusing on fundamental mathematical models and providing open source code that readers can adapt these ideas to suit specific models for their particular area of interest.

\subsection{Data}\label{sec:methods_data}

We consider temporal data that are often reported in the systems biology literature and are often interpreted in terms of models of chemical reaction networks and gene regulatory networks, and spatio-temporal data that are often reported in mathematical biology literature and interpreted using reaction-diffusion models. Temporal data are recorded at specified times. Spatio-temporal data are recorded at specified times and spatial positions. We let $y_{i}^{\mathrm{o}}$ denote the $i^{\mathrm{th}}$ experimental measurement at time $t_{i}$ and spatial position $x_{i}$. The superscript `o' is used to distinguish the observed data from mechanistic mathematical model predictions. The spatial position, $x_{i}$, may be a scalar or vector, and is omitted for temporal data. We represent multiple measurements at the same time and spatial position using distinct subscript indices. Assuming $I$ experimental measurements, we collect the individual noisy measurements into a vector $y_{1:I}^{\mathrm{o}}$, collect the observation times into a vector $t_{1:I}$, and, for spatio-temporal data, collect the spatial positions into a vector $x_{1:I}$.

 \subsection{Mechanistic mathematical model}\label{sec:methods_mathematicalmodel}

We consider a variety of temporal and spatio-temporal mechanistic mathematical models. Temporal models in systems biology often take the form of systems of ODEs \cite{Alon2019,Kreutz2013,Schmidt2006},
\begin{equation}\label{eqn:models_odes}
	\frac{\mathrm{d}y(t)}{\mathrm{d}t} = f(y(t); \theta_{\mathrm{M}}),
\end{equation}
where $y(t)=\left(y^{(1)}(t), y^{(2)}(t), \ldots, y^{(n)}(t)\right)$ represents an $n$-dimensional vector of model solutions at time $t$, and $\theta_{\mathrm{M}}$ represents a vector of mathematical model parameters. Noise free mathematical model solutions are evaluated at each $t_{i}$, denoted $y_{i}(\theta_{\mathrm{M}})=y(t_{i};\theta_{\mathrm{M}})$, and collected into a vector $y_{1:I}(\theta_{\mathrm{M}})$.

\newpage 
Spatio-temporal models often take the form of systems of PDEs. In mathematical biology we often consider systems of advection-diffusion-reaction equations \cite{Britton2005,EdelsteinKeshet2005,Murray2002a,Murray2002b,Kot2001},
\begin{equation}\label{eqn:models_pdes}
	\frac{\mathrm{\partial}y(t,x)}{\mathrm{\partial}t} = f\left(y(t,x),\frac{\partial y(t,x)}{\partial x}, \frac{\partial^{2}y(t,x)}{\partial x^{2}}; \theta_{\mathrm{M}}\right),
\end{equation}
where $y(t,x)=\left(y^{(1)}(t,x), y^{(2)}(t,x), \ldots, y^{(n)}(t,x) \right)$ represents an $n$-dimensional vector of model solutions at time $t$ and position $x$, and $\theta_{\mathrm{M}}$ represents a vector of mathematical model parameters. Noise free mathematical model solutions, evaluated at $t_{i}$ and $x_{i}$ are denoted $y_{i}(\theta_{\mathrm{M}})=y(t_{i},x_{i};\theta_{\mathrm{M}})$, and collected into a vector $y_{1:I}(\theta_{\mathrm{M}})$. The framework is well-suited to consider natural extensions of Eq (\ref{eqn:models_pdes}), for example additional mechanisms such as nonlinear diffusion or non-local diffusion  or PDE models in higher dimensions or in different coordinate systems \cite{Murray2002a,Murray2002b}. The framework is also well-suited to consider many more mechanistic mathematical models, for example difference equations (Supplementary S4). In all such examples the noise free output of the mathematical model can be collected into a vector $y_{1:I}(\theta_{\mathrm{M}})$.

\subsection{Measurement error models}\label{sec:methods_measurementerror}

  Measurement error models are a powerful tool to describe and interpret the relationship between experimental measurements, $y_{i}^{\mathrm{o}}$, and noise free mathematical model solutions, $y_{i}(\theta_{\mathrm{M}})$.  We take the common approach and assume that experimental measurements are noisy observations of a deterministic mechanistic mathematical model. This often corresponds to uncorrelated, independent, and identically distributed additive errors or multiplicative errors, in which case measurement errors are of the form $e_{i} = y_{i}^{\mathrm{o}}-y_{i}(\theta_{\mathrm{M}})$ or $e_{i} = y_{i}^{\mathrm{o}}/y_{i}(\theta_{\mathrm{M}})$, respectively. Good agreement between the data and the solution of a mathematical model corresponds to $e_{i}=0$ for additive errors and $e_{i}=1$ for multiplicative noise. In practice, the true model solution $y(\theta_{\mathrm{M}})$ is unknown and we use a prediction of the best-fit model solution $y(\hat{\theta})$. Therefore, for additive errors we analyse standard additive residuals taking the form $\hat{e}_{i} = y_{i}^{\mathrm{o}}-y_{i}(\hat{\theta})$. While it is common to analyse multiplicative noise via additive residuals in log-transformed variables, i.e. $\log(y_{i}^{\mathrm{o}})-\log(y_{i}(\theta))= \hat{e}_{i}$ \cite{Kreutz2007}, here we take a more direct approach and analyse the ratio $\hat{e}_{i} = y_{i}^{\mathrm{o}}/y_{i}(\hat{\theta})$. Error models can take many forms, including discrete or continuous models, and are typically characterised by a vector of parameters $\theta_{\mathrm{E}}$. The full model, comprising the mathematical model and  measurement error model, is then characterised by $\theta = (\theta_{\mathrm{M}},\theta_{\mathrm{E}})$. We will demonstrate that it is straightforward to implement a range of measurement error models using three illustrative examples.

\subsubsection{Additive Gaussian model}
The additive Gaussian model is ubiquitous, simple to interpret, and captures random errors and measurement uncertainties in a wide range of applications. Measurement errors are assumed to be additive, independent, and normally distributed with zero mean and constant variance, $\sigma_{N}^{2} > 0$. Therefore, experimental measurements, $y_{i}^{\mathrm{o}}$, are assumed to be independent and normally distributed about the noise free model solution, $y_{i}(\theta_{\mathrm{M}})$,
 \begin{equation}\label{eqn:noise_normal}
 	y_i^{\textrm{o}} \mid \theta \sim \mathcal{N}(y_i(\theta_{\mathrm{M}}), \sigma_{\mathrm{N}}^2).
 \end{equation}
Under this noise model the mean, median, and mode of the distribution of possible values of $y_i^{\textrm{o}} \mid \theta$ are identical and equal to $y_i(\theta)$. The variance is $\sigma_{\mathrm{N}}^2$ and $\theta_{\mathrm{E}}=\sigma_{N}$. Using this error model to obtain a best-fit solution of the mathematical model to the data, in the form of a maximum likelihood estimate, reduces to a nonlinear least squares problem. However, this error model is not always appropriate. Data in systems and mathematical biology are often non-negative, for example chemical concentrations or population densities. Implementing the additive Gaussian error model for data close to zero can be problematic and lead to non-negative physically unrealistic predictions as we will explore later in several case studies.

 \subsubsection{Log-normal model}

The log-normal model is employed to ensure non-negative and right-skewed errors in a range of biological applications \cite{Furusawa2005,Kreutz2007,Lacey1997,Limpert2001,Raue2013a}. This error model is multiplicative and we write
\begin{equation}\label{eqn:noise_lognormal1}
 	y_i^{\textrm{o}} \mid \theta = y_i(\theta)\eta_{i} \quad \mathrm{where} \quad  \eta_{i} \sim \mathrm{LogNormal}(0,\sigma_{\mathrm{L}}^2).
\end{equation}
 Here, $\theta_{\mathrm{E}}=\sigma_{L}$ and $\eta_{i}$ are assumed to be independent. Eq (\ref{eqn:noise_lognormal1}) can also be written as $y_i^{\textrm{o}} \mid \theta \sim \mathrm{LogNormal}\left( \log\left(y_{i}(\theta) \right), \sigma_{L}^{2} \right)$. Key statistics for the distribution of possible values of  $y_i^{\textrm{o}}\mid \theta$ include the mean $y_i(\theta)\exp(\sigma_{\mathrm{L}}^2 / 2)$, median $y_{i}(\theta)$, mode $y_{i}(\theta)\exp(-\sigma_{\mathrm{L}}^{2})$, and variance $(y_{i}(\theta))^{2}\exp(\sigma_{\mathrm{L}}^{2})\left[ \exp(\sigma_{\mathrm{L}}^{2}) - 1 \right]$. In contrast to the additive Gaussian model which has constant variability over time, with the log-normal model variability increases as $y_i(\theta)$ increases and variability vanishes as $y_i(\theta) \to 0^{+}$.  The log-normal error model can also be written as $y_i^{\textrm{o}}\mid \theta = y_i(\theta)\exp(\varepsilon_{i})$  where $\varepsilon_{i} \sim \mathcal{N}(0,\sigma_{\mathrm{L}}^2)$ and is equivalent to implementing an additive Gaussian error model for log-transformed experimental measurements and log-transformed noise free model solutions, i.e. $\log(y_i^{\textrm{o}}) \mid \theta \sim \mathcal{N}(\log(y_i(\theta_{\mathrm{M}})), \sigma_{\mathrm{L}}^2)$. 
 
 \subsubsection{Poisson model}
 
 The Poisson model is commonly employed to analyse non-negative count data \cite{Simpson2023,Auger2021,Hilbe2014}. Unlike the previous two measurement error model models, we do not introduce additional parameters to describe this error model, so $\theta =\theta_{\mathrm{M}}$, and we write
 \begin{equation}\label{eqn:noise_poisson}
 	y_{i}^{\textrm{o}} \mid \theta \sim  \mathrm{Pois}(y_{i}(\theta)).
 \end{equation}
The Poisson distribution in Eq (\ref{eqn:noise_poisson}) is a discrete probability density function that is neither additive or multiplicative. The model is only appropriate when observed data, $y_{i}^{\textrm{o}}$, are non-negative integers. However, there are no such technical restrictions for the output of the mathematical model and $y_{i}(\theta)$ may take any non-negative value. When $y_{i}(\theta)=0$ we consider the limit of Poisson distribution such that the only possible outcome is $y_{i}^{\mathrm{o}}=0$ \cite{Said1958}. Under the Poisson model key statistics for the distribution of possible values of $y_i^{\textrm{o}} \mid \theta$ include the mean $y_i(\theta)$; the median lies between $\lfloor y_i(\theta)-1 \rfloor$ and $\lfloor y_i(\theta)+1 \rfloor$; the modes are $y_i(\theta)$ and $y_i(\theta)-1$ when $y_i(\theta)$ is a positive integer and $\lfloor y_i(\theta) \rfloor$ when $y_i(\theta)$ is a positive non-integer; and the variance is $y_i(\theta)$ \cite{Johnson2005}. In contrast to the additive Gaussian model which has approximately constant variability over time, with the Poisson model variability increases as $y_i(\theta)$ increases and variability vanishes as $y_i(\theta) \to 0^{+}$.

\subsection{Parameter estimation}

We perform parameter estimation for the full model that comprises two components: (i) a mechanistic mathematical model; and, (ii) a measurement error model. We take a general approach and simultaneously estimate the full model parameters $\theta$. This means that we estimate the mathematical model parameters, $\theta_{\mathrm{M}}$, and measurement error model parameters, $\theta_{E}$, simultaneously. It is straightforward to consider special cases of this approach where a subset of the full model parameters $\theta$ may be pre-specified or assumed known, for example in cases where the measurement error model parameters $\theta_{E}$ can be pre-specified\cite{Hines2014,Simpson2020}.

Taking a likelihood-based approach to parameter estimation, we use the log-likelihood,
 \begin{equation}\label{eqn:methods_parameterestimation_loglikelihood}
	\ell(\theta \mid y_{1:I}^{\mathrm{o}}) = \sum_{i=1}^{I} \log\left[ \phi\left(y_{i}^{\mathrm{o}}; y_{i}(\theta), \theta \right) \right],
\end{equation}
where $\phi\left(y_{i}^{\mathrm{o}}; y_{i}(\theta), \theta \right)$ represents the probability density function related to the measurement error model. For the additive Gaussian error model $\phi\left(y_{i}^{\mathrm{o}}; y_{i}(\theta), \theta \right) = \hat{\phi}\left(y_{i}^{\mathrm{o}}; y_{i}(\theta), \sigma_{N}^{2}(\theta) \right)$, where $ \hat{\phi}(x; \mu, \sigma^{2})$ represents the Gaussian probability density function with mean $\mu$ and variance $\sigma^{2}$. For the log-normal error model $\phi\left(y_{i}^{\mathrm{o}}; y_{i}(\theta), \theta \right) = \hat{\phi}\left(y_{i}^{\mathrm{o}}; \log{\left(y_{i}(\theta)\right)},\sigma_{L}^{2}(\theta) \right)$, where $\hat{\phi}\left(x; \mu, \sigma \right)$ represents the probability density function of the $\mathrm{Lognormal}(\mu,\sigma^{2})$ distribution. For the Poisson error model, $\phi\left(y_{i}^{\mathrm{o}}; y_{i}(\theta), \theta \right) = \hat{\phi}\left(y_{i}^{\mathrm{o}}; y_{i}(\theta)\right)$, where $\hat{\phi}(x;\lambda)$ represents the probability density function for the Poisson distribution with rate parameter $\lambda$. 

To obtain a point-estimate of $\theta$ that gives the best match to the data, in the sense of the highest likelihood, we seek the maximum likelihood estimate (MLE),
 \begin{equation}\label{eqn:methods_parameterestimation_MLE}
	\hat{\theta} =\underset{\theta}{\mathrm{argmax}} \ \ell(\theta \mid y_{1:I}^{\mathrm{o}}). 
\end{equation}
We estimate $\hat{\theta}$, subject to bound constraints, using numerical optimisation.

\subsection{Identifiability analysis using the profile likelihood}

We are often interested in the range of parameters that give a similar match to the data as the MLE. This is analogous to asking whether parameters can be uniquely identified given the data. There are two approaches to address this question of parameter identifiability: structural identifiability and practical identifiability. Structural identifiability explores whether parameters are uniquely identifiable given continuous noise free observations of model solutions. Many software tools, utilising symbolic calculations, have been developed to analyse structural identifiability for systems of ODEs as reviewed in \cite{Barreiro2022}. Tools to assess structural identifiability of systems of PDEs have not been widely developed \cite{Renardy2022}, and structural identifiability analysis of PDE models is an active area of research.

Practical identifiability assesses how well model parameters can be identified given a finite set of noisy incomplete data. To explore practical identifiability we use a profile likelihood-based approach and work with the normalised log-likelihood, 
\begin{equation}\label{eqn:methods_parameterestimation_normalisedloglikelihood}
\hat{\ell}(\theta \mid y_{1:I}^{\mathrm{o}}) = \ell(\theta \mid y_{1:I}^{\mathrm{o}}) -\ell(\hat{\theta} \mid y_{1:I}^{\mathrm{o}}).
\end{equation}
Normalising the log-likelihood means that $\hat{l}( \theta \mid y_{1:I}^{\mathrm{o}}) \leq 0$ and $\hat{l}( \hat{\theta} \mid y_{1:I}^{\mathrm{o}})=0$.

To assess practically identifiability of parameters within the full parameter vector, $\theta$, we partition $\theta$ as $\theta = (\psi, \lambda)$ where $\psi$ can represent any combination of parameters and $\lambda$ represents the complement \cite{Pawitan2001,Casella2001,Pace1997,Simpson2023}. In this section, we assess whether each parameter within the full parameter vector is practically identifiable in turn. We consider $\psi$ to represent a scalar parameter of interest and $\lambda$ to represent a vector of the remaining nuisance parameters. This allows us to focus on univariate profile likelihoods. We now work with the profile log-likelihood for the scalar interest parameter $\psi$ \cite{Pawitan2001,Cox2006},
\begin{equation}\label{eqn:methods_parameterestimation_profileloglikelihood}
	\hat{\ell}_{p}(\psi \mid y_{1:I}^{\mathrm{o}}) = \underset{\lambda \mid \psi}{\mathrm{sup}}  \ \hat{\ell}(\psi, \lambda \mid y_{1:I}^{\mathrm{o}}),
\end{equation}
where the subscript $p$ is introduced to denote the profile log-likelihood. Therefore, the profile log-likelihood maximises the normalised log-likelihood for each value of the scalar $\psi$. This process implicitly defines a function $\lambda^{*}(\psi)$ of optimal values of $\lambda$ for each $\psi$, and defines a curve with points $(\psi,\lambda^{*}(\psi) )$ in parameter space that includes the MLE, $\hat{\theta} = (\hat{\psi}, \hat{\lambda})$. To estimate $\hat{\ell}_{p}(\psi \mid y_{1:I}^{\mathrm{o}})$ we define a mesh of $2N$ points for $\psi$ comprising $N$ equally--spaced points from a pre-specified lower bound, $\psi_{\mathrm{L}}$, to $\hat{\psi}$ and $N$ equally--spaced points from $\hat{\psi}$ to a pre-specified upper bound, $\psi_{\mathrm{U}}$. We choose the lower and upper bounds to capture approximate confidence intervals. We choose the number of mesh points so that there are many points within the approximate confidence interval, typically we choose $N=20$. Further details on how the choice of $N$ impacts coverage properties are presented in Section \ref{sec:methods_confidencesetsparameters}. For each value of $\psi$ in the mesh we estimate $\hat{\ell}_{p}(\psi \mid y_{1:I}^{\mathrm{o}})$, subject to the bound constraints for $\lambda$, using numerical maximisation.

 Univariate profile log-likelihoods for scalar interest parameters, referred to as profiles for brevity, provide a visual and quantitative tool to assess practical identifiability. A narrow univariate profile that is well-formed about a single peak corresponds to a parameter of interest that is practically identifiable, while a wide flat profile indicates that the parameter of interest is not practically identifiable. We assess narrow and wide relative to log-likelihood-based approximate confidence intervals. We define the log-likelihood-based approximate confidence interval for the scalar $\psi$ from the profile log-likelihood, 
 \begin{equation}\label{eqn:methods_parameterconfidenceinterval}
 	C_{\psi, 1-\alpha}(y_{1:I}^{\mathrm{o}}) = \left\{ \psi \mid \hat{\ell}_{p}(\psi \mid y_{1:I}^{\mathrm{o}}) \geq \ell_{c} \right\},
 \end{equation}
where the threshold parameter $\ell_{c}$ is chosen such that the confidence interval has an approximate asymptotic coverage probability of $1-\alpha$. Many studies report  $90\%$, $95\%$, $99\%$ or $99.9\%$ confidence intervals for univariate profiles \cite{Pawitan2001,Royston2007}. These thresholds are calibrated using the $\chi^{2}$ distribution, which is reasonable for sufficiently regular problems \cite{Pawitan2001,Royston2007}. In particular, $\ell_{c}=-\Delta_{\nu,1-\alpha}/2$, where $\Delta_{\nu,1-\alpha}$ refers to the $(1-\alpha)$ quantile of a $\chi^{2}$ distribution with $\nu$ degrees of freedom set equal to the dimension of the interest parameter, e.g $\nu=1$ for univariate profiles. It is straightforward to extend this approach to consider a vector valued interest parameters, for example to generate bivariate profiles \cite{Simpson2023}.

\subsection{Predictions}

We generate predictions for model solutions, $y=y(t; \theta)$, and data realisations, $z_{i}$, using a profile log-likelihood-based approach. These predictions propagate forward uncertainties in interest parameters and allow us to understand and interpret the contribution of each model parameter, or unions of parameters, to uncertainties in predictions. This step is very important when using mathematical models to interpret data and to communicate with collaborators from other disciplines simply because predictions and variability in predictions are likely to be of greater interest than estimates of parameter values in a mathematical model.

\subsubsection{Confidence sets for deterministic model solutions}\label{sec:methods_confidencesetsparameters}

We now propagate forward uncertainty in a scalar interest parameter, $\psi$, to understand and interpret the uncertainty in predictions of the model solution, $y=y(t; \theta)$. The approximate profile-wise log-likelihood for the model solution, $y$, is obtained by taking the maximum profile log-likelihood value over all values of $\psi$ consistent with $y(t; (\psi,\lambda^{*}(\psi) ))=y$, i.e.,
\begin{equation}\label{eqn:methods_parameterestimation_profileloglikelihoodmodelsolution}
	\hat{\ell}_{p}\Bigl( y \left(t; \left(\psi,\lambda^{*}\left(\psi\right) \right) \right) = y \ \Big| \ y_{1:I}^{\mathrm{o}}\Bigr) = \underset{\psi \mid y(t; (\psi,\lambda^{*}(\psi) )) = y}{\mathrm{sup}}  \ \hat{\ell}_{p}(\psi \mid y_{1:I}^{\mathrm{o}}).
\end{equation}
Here, $y(t; (\psi,\lambda^{*}(\psi) ))$ corresponds to the output or solution of the mechanistic mathematical model solved with parameter values $\theta = (\psi,\lambda^{*}(\psi) )$.  The confidence set for the model solution, $y$, propagated from the scalar interest parameter $\psi$ is 
\begin{equation}
	C_{y, 1-\alpha}^{\psi}(y_{1:I}^{\mathrm{o}}) = \left\{ y \ \bigg|\  \hat{\ell}_{p}\Bigl( y\left(t; \left(\psi,\lambda^{*}\left(\psi\right) \right) \right) = y \ \Big| \ y_{1:I}^{\mathrm{o}}\Bigr)  \geq \ell_{c} \right\}.
\end{equation}
In practice, we form an approximate $(1-\alpha)$\% confidence interval, $C_{y, 1-\alpha }^{\psi}(y_{1:I}^{\mathrm{o}})$, by simulating $y\left(t; \left(\psi,\lambda^{*}\left(\psi\right) \right) \right)$ for each $\psi \in C_{\psi, 1-\alpha}(y_{1:I}^{\mathrm{o}})$. This confidence set can be used to reveal the influence of uncertainty in $\psi$ on predictions of the model solution. From an implementation perspective, this is where the number of mesh points used to compute profiles can be important and should be considered on a case-by-case basis. If there are not enough mesh points in the confidence interval then the confidence sets will not have good coverage properties. For example, in the extreme case of only one mesh point in the confidence interval the confidence set would only be the mathematical model simulated at the MLE and would not provide any insight into uncertainty. 

Each parameter in $\theta$ can be treated in turn as an interest parameter. Therefore, for each parameter in $\theta$ we can construct an approximate confidence interval $C_{y, 1-\alpha }^{\psi}(y_{1:I}^{\mathrm{o}})$. Comparing approximate confidence intervals constructed for different parameters in $\theta$ illustrates which parameters contribute to greater uncertainty in model solutions \cite{Murphy2022c}. This can be important for understanding how to improve predictions and for experimental design. However, optimising out nuisance parameters in this profile log-likelihood-based approach typically leads to lower coverage than other methods that consider all uncertainties simultaneously, especially when the model solution has weak dependence on the interest parameter and non-trivial dependence on the nuisance parameters \cite{Simpson2022c}.  More conservative approximate confidence sets, relative to the individual profile-wise confidence sets, can be constructed by taking the union of individual profile-wise confidence sets for the model solution,
\begin{equation}\label{eqn:methods_parameterestimation_confidencesetsunion}
	C_{y, 1-\alpha}(y_{1:I}^{\mathrm{o}})  \approx  \bigcup_{\psi} C_{y, 1-\alpha}^{\psi}(y_{1:I}^{\mathrm{o}}).
\end{equation}
Equation (\ref{eqn:methods_parameterestimation_confidencesetsunion}) provides insight into the uncertainty due to all model parameters across the solution of the mathematical model. As we will demonstrate, this approach is a simple, computationally efficient, and an intuitive model diagnostic tool. Furthermore, the method can be repeated with vector-valued interest parameters and increasing the dimension results in closer agreement to full likelihood-based methods \cite{Simpson2023}. As an example, the union of profile-wise confidence sets  for two-dimensional interest parameters can be constructed by considering bivariate profiles for all pairs of parameters \cite{Simpson2023}. This approach can also be generalised beyond that of predictions of the model solution to predictions of data distribution parameters \cite{Simpson2023}. Note that for the additive Gaussian and Poisson measurement error models the model solution is the mean of the data distribution and for the log-normal measurement error model the model solution is the median of data distribution. These methods are simpler to implement and interpret in comparison to previous methods, such as those that involve additional constrained optimisation problems \cite{Bjornstad1990,Hass2016,Kreutz2012,Villaverde2022}.

\subsubsection{Confidence sets for noisy data realisations}\label{sec:methods_confidencesetsrealisations}

In practice we are often interested in using mathematical models to generate predictions of noisy data realisations, since an individual experiment measurement can be thought of as a noisy data realisation. These predictions allow us to explore what we would expect to observe if we were to repeat the experiment or if we were to measure at different times and/or spatial positions. By building our framework on parameterised mechanistic mathematical models we can also predict beyond the data based on a mechanistic understanding. In contrast to confidence sets for deterministic model solutions where it is naturally to consider continuous trajectories, data are naturally defined at discrete time points therefore here we consider confidence sets for noisy single time observations.

To form approximate $(1-\alpha)\%$ confidence sets for model realisations we consider a number of approaches: (i) a simple MLE-based approach that may not reach the desired coverage level; and (ii) Bonferroni correction-based approaches that are likely to exceed the desired coverage level. To explain these approaches consider the problem of forming a $(1-\alpha)\%$ confidence set for a single unknown data realisation $z_{i}$ at time $t_{i}$ for $i=1,2,\ldots,J$, where the variable $z_{i}$ is used to distinguish the unknown data realisation from an observed data realisation $y_{i}^{\mathrm{o}}$ at time $t_{i}$. These predictions can be made at the same time points as observed data and can also be made at time points where observed data is not collected. In this review, to visualise the uncertainty throughout time, we generate predictions at a higher temporal resolution in comparison to the observed data. If the mathematical model, mathematical model parameters, measurement error model, and measurement error model parameters are all known then it is straightforward to form a confidence set for each $z_{i}$. The bounds of the $(1-\alpha)\%$ confidence set are obtained by computing the $\alpha/2$ and $1-\alpha/2$ quantiles of the probability distribution associated with the measurement error model and mathematical model solution at time $t_{i}$. This procedure can be repeated for each unknown data realisation at each time $t_{i}$. For example, consider a scalar valued model solution, $y$, that depends only on time, with an additive Gaussian measurement error model where $\sigma_{N}$ is known. The lower and upper bounds of the prediction set can be estimated at each point in time $t_{i}$ by calculating the $\alpha/2$ and $1-\alpha/2$ quantiles of the normal distribution with mean $y(t_{i})$ and standard deviation $\sigma_{N}$. This computational approach naturally extends to other measurement error models, including the Poisson and log-normal models.  In practice however, we typically face a more challenging scenario where the true model parameters and true mathematical model solution, $y=y(t;\theta)$, are all unknown, and we now outline two approaches for dealing with this situation.

\textit{MLE-based approach.} When the true model parameters and true mathematical model solution are unknown a simple approach is to assume that the model parameters are given by the MLE, $\hat{\theta}$, and the true solution of the mathematical model is given by evaluating the solution of the model at the MLE, $y(t; \hat{\theta})$. With this assumption, it is then straightforward to generate a $(1-\alpha)\%$ confidence set as previously described. In practice, it is unlikely that the MLE, $\hat{\theta}$, will be identical to the true model parameters, $\theta$, so this approach may not reach the desired coverage level. However, when uncertainty due to statistical noise is large relative to the difference between $y(t;\theta)$ and $y(t; \hat{\theta})$ this simple MLE-based approach can work well.

\textit{Bonferroni correction-based approaches.} A more conservative approach for forming confidence sets for model realisations involves propagating forward uncertainty in model parameters. The following approach was introduced in \cite{Simpson2023}, and here we present concrete examples. Consider a scalar interest parameter $\psi$ and a corresponding confidence set for the model solution, $C_{y, 1-\alpha/2}^{\psi}(y_{1:I}^{\mathrm{o}})$. For each $y \in C_{y, 1-\alpha/2}^{\psi}(y_{1:I}^{\mathrm{o}})$ we construct a prediction set $\mathcal{A}_{y,1-\alpha/2}^{\psi}(y_{1:I}^{\mathrm{o}})$ such that the probability of observing a measurement $z_{i} \in \mathcal{A}_{y,1-\alpha/2}^{\psi}(y_{1:I}^{\mathrm{o}})$ is $1-\alpha/2$. Computationally, $\mathcal{A}_{y,1-\alpha/2}^{\psi}(y_{1:I}^{\mathrm{o}})$ can be constructed in a pointwise manner by estimating the $\alpha/4$ and $1-\alpha/4$ quantiles of the probability distribution associated with the measurement error model. Taking the union for each $y \in C^\psi_{y, 1-\alpha/2}(y^{\textrm{o}}_{i:I})$ we obtain a conservative $(1-\alpha)\%$ confidence set for model realisations from the interest parameter $\psi$,
\begin{equation}\label{eqn:methods_parameterestimation_confidencesetsunionArealisation}
	C_{z_{i}, 1-\alpha}^{\psi}(y_{1:I}^{\mathrm{o}})  =  \bigcup_{y \in C_{y, 1-\alpha/2}^{\psi}(y_{1:I}^{\mathrm{o}})} \mathcal{A}_{y, 1-\alpha/2}^{\psi}(y_{1:I}^{\mathrm{o}}).
\end{equation}
This approach employs a Bonferroni correction method \cite{Simpson2023,Miller1981}.

Equation (\ref{eqn:methods_parameterestimation_confidencesetsunionArealisation}) represents a conservative confidence set for the data realisations $z_{i}$ at the level of the individual interest parameter $\psi$. Treating each parameter in $\theta$ in turn as an interest parameter and taking the union results in a confidence set for the overall uncertainty in data realisations,
\begin{equation}\label{eqn:methods_parameterestimation_confidencesetsunionpsirealisation}
	C_{z_{i}, 1-\alpha}(y_{1:I}^{\mathrm{o}})  \approx  \bigcup_{\psi}	C_{z_{i}, 1-\alpha}^{\psi}(y_{1:I}^{\mathrm{o}}).
\end{equation}

\subsection{Coverage properties}\label{sec:methods_coverage}

Coverage properties of confidence intervals and confidence sets are defined formally, but for likelihood-based confidence sets coverage properties are expected to only hold asymptotically in data size. In practice, we can evaluate approximate statistical coverage properties numerically by repeated sampling. In particular, we can generate, and then analyse, many data sets using the same mathematical model, measurement error model, and true model parameters, $\theta$. A detailed illustrative example for temporal data is discussed in \cref{sec:coverage}. The procedure is applicable to a range of models and data.

\clearpage
\newpage
\section{Case studies}\label{sec:resultsdiscussion}

We will now implement the general framework using simple caricature mathematical models routinely used in the systems biology literature and the mathematical biology literature. The full models are formed by (i) a deterministic mathematical model and (ii) a measurement error model. Example mathematical models that we consider include systems of linear and nonlinear temporal ODEs often used in the systems biology literature and systems of spatio-temporal PDEs often used in the mathematical biology literature. Example measurement error models that we consider include additive Gaussian, log-normal, and Poisson.

\subsection{Temporal linear models}\label{sec:results_ode1}

Consider a chemical reaction network with two chemical species $C_{1}$ and $C_{2}$. We assume that $C_{1}$ decays to form $C_{2}$ at a rate $r_{1}$, and that $C_{2}$ decays at a rate $r_{2}$. Within this modelling framework we do not explicitly model the decay products from the second reaction. Applying the law of mass action, the concentrations of $C_{1}$ and $C_{2}$ at time $t$, denoted $c_{1}(t)$ and $c_{2}(t)$, respectively, are governed by the following system of ODEs,
\begin{equation}\label{eqn:linearodes}
	\begin{split}
		\frac{\mathrm{d}c_{1}(t)}{\mathrm{d}t} &= -r_{1}c_{1}(t),\\
		\frac{\mathrm{d}c_{2}(t)}{\mathrm{d}t} &= r_{1}c_{1}(t) - r_{2}c_{2}(t).
	\end{split}
\end{equation}
We refer to the terms on the right-hand side of Eq (\ref{eqn:linearodes}) as the reaction terms, which are linear in this simple case. Equation (\ref{eqn:linearodes}) has an analytical solution, which for $r_{1} \neq r_{2}$ can be written as,
\begin{equation}\label{eqn:linearodes_sol}
	\begin{split}
		c_{1}(t)  &= c_{1}(0)\exp\left(-r_{1}t\right),\\
		c_{2}(t) &= c_{1}(0)\exp(-r_{1}t)\left( \frac{r_{1}}{r_{2}-r_{1}} \right) + \left[c_{2}(0) - \frac{c_{1}(0)r_{1}}{r_{2}-r_{1}} \right] \exp(-r_{2}t).
	\end{split}
\end{equation}
In the special case $r_{1}=r_{2}$ we can write the exact solution in a different format where $c_2(t)$ is proportional to $c_1(t)$. We treat the initial conditions $c_{1}(0)$ and $c_{2}(0)$ as known so that Eqs (\ref{eqn:linearodes})-(\ref{eqn:linearodes_sol}) are characterised by two parameters $r_{1}$ and $r_{2}$ that we will estimate. Here, $r_{1}$ and $r_{2}$ are structurally identifiable. Initial conditions can also easily be treated as unknowns within this framework \cite{Murphy2022c,Warne2017}. For parameter estimation we solve Eq (\ref{eqn:linearodes}) numerically which is convenient because we do not have to consider the cases $r_{1} \neq r_{2}$ and $r_{1} = r_{2}$ separately in our numerical implementation.

\begin{figure}[p]
	\centering
	\includegraphics[width=0.9\textwidth]{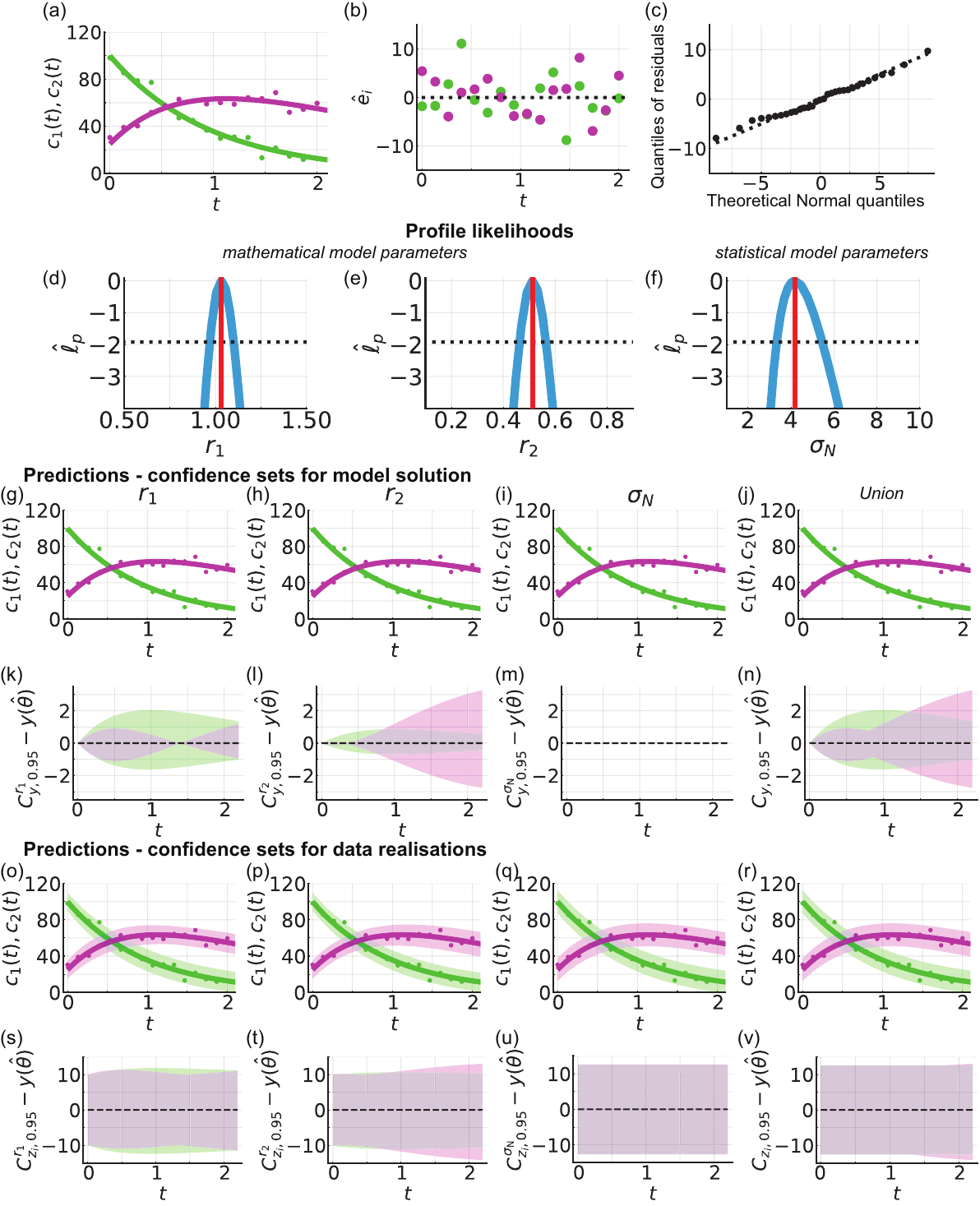}
	\caption{Caricature ODE model with linear reactions (Eq (\ref{eqn:linearodes})). (a) Synthetic data (circles) at sixteen equally--spaced time points from $t=0.0$ to $t=2.0$ are generated by simulating Eq (\ref{eqn:linearodes}), the additive Gaussian measurement error model, known model parameters $\theta=(r_{1}, r_{2}, \sigma_{\mathrm{N}}) = (1.0, 0.5, 5.0)$, fixed initial conditions $(c_{1}(0), c_{2}(0))=(100.0, 25.0)$. The MLE, computed assuming an additive Gaussian measurement error model, is $\hat{\theta}=(r_{1}, r_{2}, \sigma_{\mathrm{L}}) = (1.03, 0.51, 4.18)$. Equation (\ref{eqn:linearodes}) simulated with the MLE (solid). Throughout $c_{1}(t)$ (solid green) and $c_{2}(t)$ (solid magenta). (b) Residuals $\hat{e}_{i}=y_{i}^{\mathrm{o}}-y_{i}(\hat{\theta})$ with time $t$. (c) Normal quantile-quantile plot of residuals. (d)-(f) Profile log-likelihoods (blue) for (d) $r_{1}$, (e) $r_{2}$, and (f) $\sigma_{\mathrm{N}}$ with MLE (red-dashed), an approximate $95\%$ confidence interval threshold (horizontal black-dashed). (g)-(j) Profile-wise confidence sets for the model solution (g) $r_{1}$, (h) $r_{2}$, (i) $\sigma_{\mathrm{N}}$, and (i) their union. (k-n) Difference between confidence set for model solution and the solution of the mathematical model evaluated at the MLE. (o-r) Profile-wise confidence sets for data realisations (shaded) for (o) $r_{1}$, (p) $r_{2}$, (q) $\sigma_{\mathrm{N}}$, and (r) their union.  (s-v) Difference between confidence set for model realisations and the solution of the mathematical model evaluated at the MLE.}
	\label{fig:fig2}
\end{figure}

We now explore a simple example shown in Fig \ref{fig:fig2} and specify $(c_{1}(0), c_{2}(0))=(100.0, 25.0)$. We generate synthetic data using Eq (\ref{eqn:linearodes_sol}), the additive Gaussian error model, and model parameters $\theta=(r_{1}, r_{2}, \sigma_{\mathrm{N}}) = (1.0, 0.5, 5.0)$ (Fig \ref{fig:fig2}a). Then, to demonstrate that the framework accurately recovers these known parameter values and to generate predictions, we use Eq (\ref{eqn:linearodes_sol}) and the additive Gaussian error model. Computing the maximum likelihood estimate (MLE) of the model parameters we obtain $\hat{\theta}=(r_{1}, r_{2}, \sigma_{\mathrm{N}}) = (1.03, 0.51, 4.18)$. Simulating the deterministic mathematical model with MLE we observe excellent agreement with the data (Fig \ref{fig:fig2}a). Inspecting the residuals, $\hat{e}_{i}=y_{i}^{\mathrm{o}}-y_{i}(\hat{\theta})$, suggests that they appear, visually at least, to be independent and normally distributed (Fig \ref{fig:fig2}b). There are many techniques to analyse standard additive residuals in greater detail should a simple visual interpretation lead us to conclude that the residuals are not independent \cite{Motulsky1987,Seber2003a,Simpson2022a}. We take a simple and common graphical approach. We plot the residuals on a normal quantile-quantile plot (Fig \ref{fig:fig2}c). As the residuals appear close to the reference line on the normal quantile-quantile plot, the assumption of normally distributed residuals appears reasonable.

In practice it is often crucial to understand whether model parameters can be approximately identified or whether many combinations of parameter values result in a similar fit to the data. To address this question of practical identifiability we compute univariate profile log-likelihoods for $r_{1}$, $r_{2},$ and $\sigma_{\mathrm{N}}$. Each profile is well-formed around a single central peak (Fig \ref{fig:fig2}d-f). This suggests that each model parameter is well identified by the data. Using the profile log-likelihoods we compute approximate $95\%$ confidence intervals, $r_{1} \in (0.97, 1.10)$, $r_{2} \in (0.45, 0.56)$ and $\sigma_{\mathrm{N}} \in (3.33, 5.46)$. These confidence intervals indicate the range of values for which we are $95\%$ confident that the true values lie within. On this occasion each component of the known parameter $\theta$ is contained within the respective confidence interval.

Thus far we have obtained estimates of best-fit parameters and associated uncertainties. To connect estimates of best-fit parameters and associated uncertainties to data we need to understand how uncertainty in $\theta$ propagates forward to uncertainties in the dependent variables, here $c_{1}(t)$ and $c_{2}(t)$, as this is what is measured in reality. There are many predictions of $c_{1}(t)$ and $c_{2}(t)$ that one could make. We consider two key forms of predictions: confidence sets for deterministic model solutions and Bonferroni correction-based confidence sets for noisy data realisations. For each parameter we generate confidence sets for the model solution and explore the difference between the confidence sets and the mathematical model simulated with the MLE (Fig \ref{fig:fig2}g-n). Results in Fig \ref{fig:fig2}g-i,k-m reveal the influence of individual model parameters on predictions of the model solution. For example, uncertainty in the parameter $r_{2}$ corresponds to increasing uncertainty in the model solution for $c_{2}(t)$ as time increases, i.e. $C_{y,0.95}^{r_{2}}- y(\hat{\theta})$ increases with time for $c_{2}(t)$ (Fig \ref{fig:fig2}h,l). However, uncertainty in the measurement error model parameter, $\sigma_{N}$, does not contribute to uncertainty in predictions of the model solution (Fig \ref{fig:fig2}i,m), since the noise is additive. Furthermore, we can observe that for $t \geq 1$ uncertainty in $r_{2}$ contributes to greater uncertainty in $c_{2}(t)$ than uncertainty in $r_{1}$ (Fig \ref{fig:fig2}g,h,k,l). Predictions in the form of Bonferroni correction-based confidence sets for data realisations take into account the measurement error model (Fig \ref{fig:fig2}o-v). These can be generated for each individual parameter and an understanding of the overall uncertainty can be obtained by taking their union. Overall, results in Fig \ref{fig:fig2} show that the framework recovers known parameter values and generates sensible predictions when the mathematical model and measurement error model are both known.

\begin{figure}[p]
	\centering
	\includegraphics[width=\textwidth]{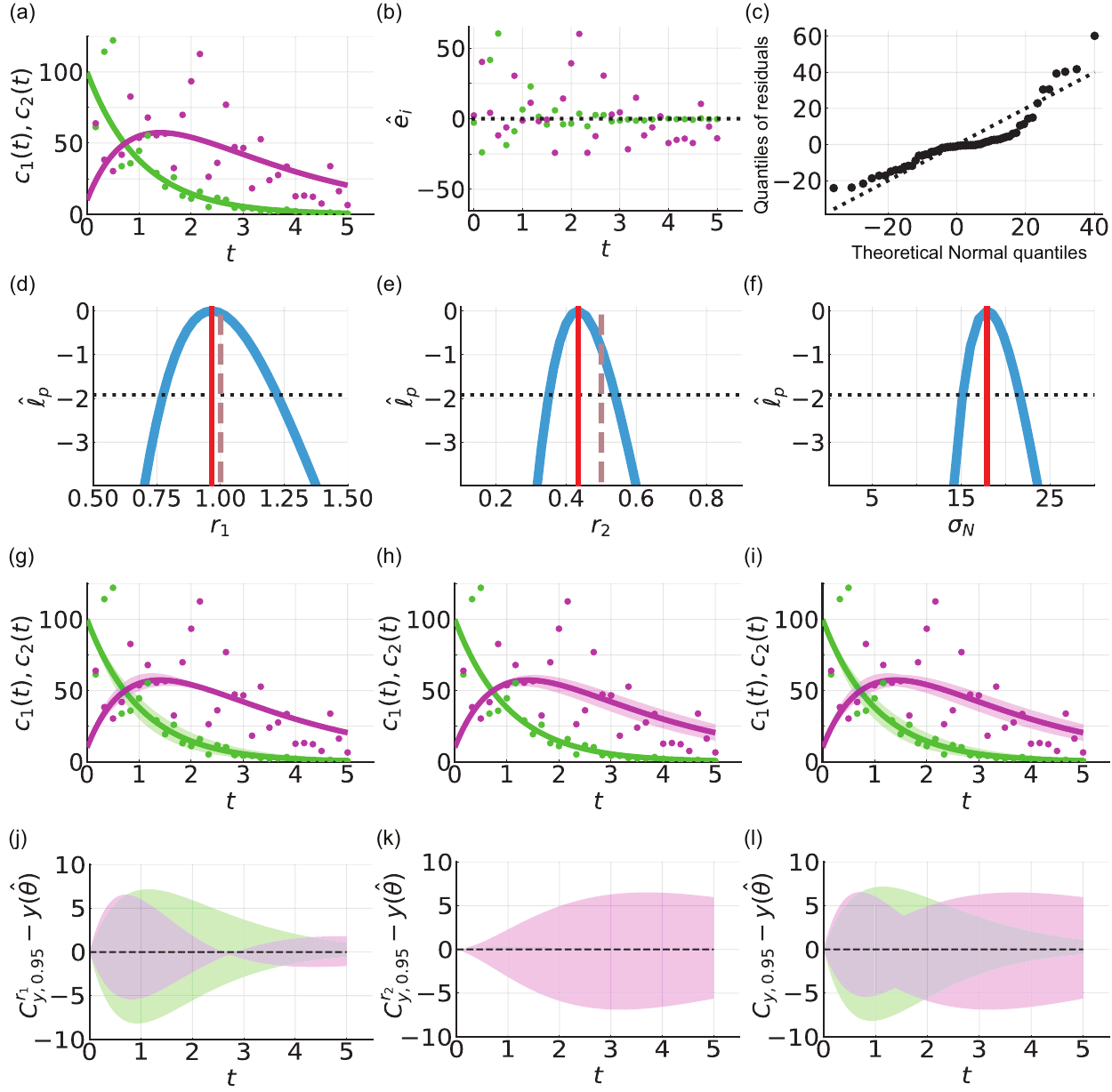}
	\caption{Caricature ODE model with linear reactions (Eq (\ref{eqn:linearodes})) and intentional misspecification of the measurement error model. (a) Synthetic data (circles) at thirty-one equally--spaced time points from $t=0.0$ to $t=5.0$ are generated by simulating Eq (\ref{eqn:linearodes}), the log-normal measurement error model, known model parameters $\theta = (r_{1}, r_{2}, \sigma_{\mathrm{L}}) = (1.0, 0.5, 0.4)$, and initial conditions $(c_{1}(0), c_{2}(0)) = (100.0, 10.0)$. The MLE, computed assuming an additive Gaussian measurement error model, is $\hat{\theta}=(r_{1}, r_{2}, \sigma_{\mathrm{L}}) = (0.97, 0.43, 17.9)$. Equation (\ref{eqn:linearodes}) simulated with the MLE (solid). Throughout $c_{1}(t)$ (solid green) and $c_{2}(t)$ (solid magenta). (b) Residuals $\hat{e}_{i}=y_{i}^{\mathrm{o}}-y_{i}(\hat{\theta})$ with time $t$. (c) Normal quantile-quantile plot of residuals. (d)-(f) Profile log-likelihoods (blue) for (d) $r_{1}$, (e) $r_{2}$, and (f) $\sigma_{\mathrm{N}}$ with MLE (red-dashed), an approximate $95\%$ confidence interval threshold (horizontal black-dashed) and known model parameters (vertical brown dashed). (g)-(i) Profile-wise confidence sets for the model solution (g) $r_{1}$, (h) $r_{2}$, and (i) their union. (k-l) Difference between confidence set and the solution of the mathematical model evaluated at the MLE.}
	\label{fig:fig3}
\end{figure}

\newpage
In practice faced with experimental data, we do not know which measurement model is appropriate. An extremely common approach in this situation is to assume an additive Gaussian measurement error model as we do in Figure \ref{fig:fig2}. This choice is simple to implement and interpret but the suitability of this choice is often unjustified. We now explore an example where assuming additive Gaussian errors is inappropriate and leads to physically-unrealistic predictions. In Fig \ref{fig:fig3}a we present synthetic data generated by simulating Eq (\ref{eqn:linearodes_sol}) and the log-normal error model with known parameter values, $\theta=(r_{1}, r_{2}, \sigma_{\mathrm{L}}) = (1.0, 0.5, 0.4)$, and initial conditions, $(c_{1}(0), c_{2}(0))=(100.0, 10.0)$. To estimate model parameters and generate predictions, we assume that the true mathematical model is known and intentionally misspecify the measurement error model. 

Assuming an additive Gaussian error model, the MLE is $\hat{\theta}=(r_{1}, r_{2}, \sigma_{\mathrm{N}}) = (0.97, 0.43, 17.90)$. Evaluating Eq (\ref{eqn:linearodes}) with the MLE we observe good agreement with the data (Fig \ref{fig:fig3}a). However, plotting the residuals, $\hat{e}_{i}=y_{i}^{\mathrm{o}}-y_{i}(\hat{\theta})$, on a normal quantile-quantile plot shows a visually distinct deviation from the reference line with points representing the tails of the residuals above the reference line and points close the the median of the residuals below the reference line (Fig \ref{fig:fig3}c). This suggests that the additive Gaussian measurement error model may be inappropriate. Nevertheless, we proceed with the additive Gaussian error model to demonstrate further issues that can arise and subsequent opportunities to detect the misspecified measurement error model. Profile log-likelihoods for $r_{1}$, $r_{2}$, and  $\sigma_{\mathrm{N}}$ suggest that these parameters are practically identifiable and approximate $95\%$ confidence intervals, $r_{1} \in (0.77, 1.22)$ and $r_{2} \in (0.35, 0.54)$, capture known parameter values. Due to the error model misspecification, we are unable to compare the approximate confidence interval for $\sigma_{\mathrm{N}}$ to a known value. 

\begin{figure}[h!]
	\centering
	\includegraphics[width=\textwidth]{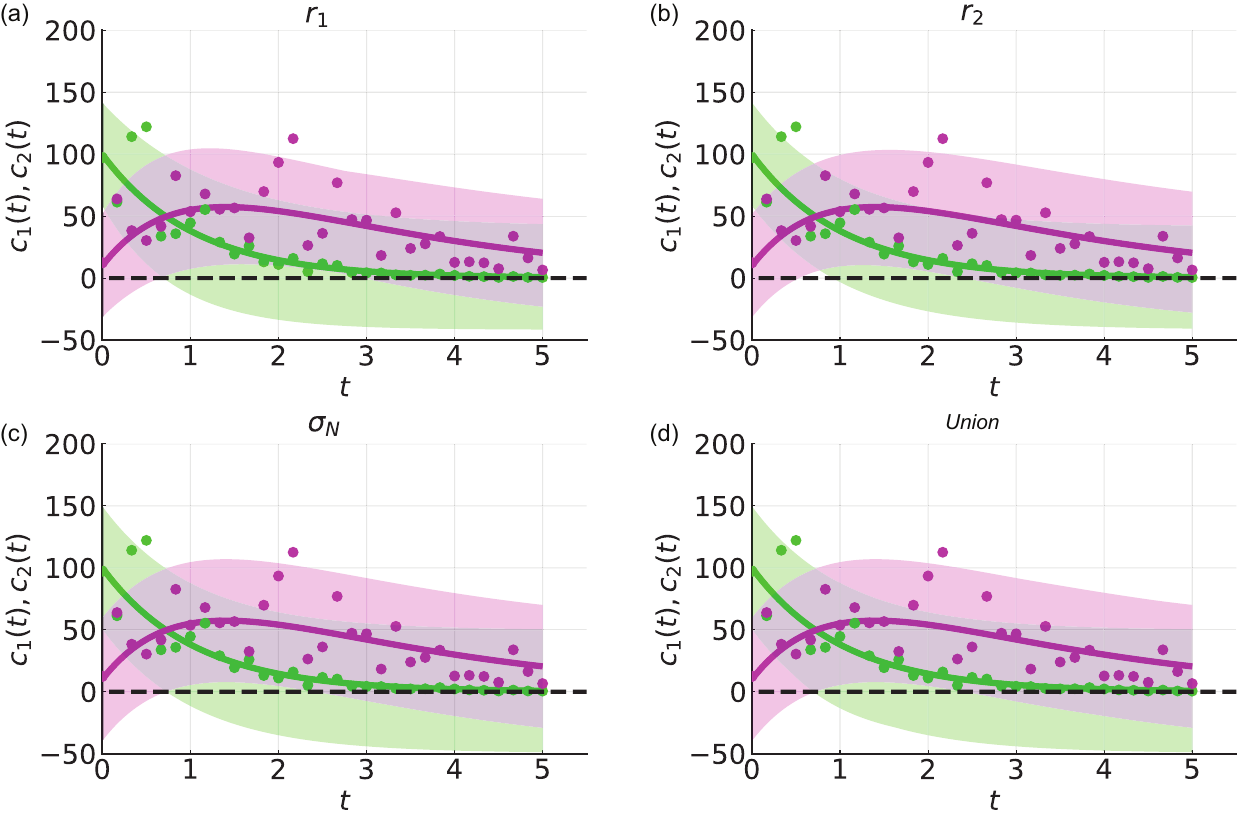}
	\caption{Bonferroni correction-based confidence sets for data realisations for the caricature ODE model with linear reactions (Eq (\ref{eqn:linearodes})) and intentional misspecification of the measurement error model. (a-d) Profile-wise confidence sets for data realisations (shaded) for (a) $r_{1}$, (b) $r_{2}$  (c) $\sigma_{\mathrm{N}}$, and (d) their union. Predictions suggest negative concentrations which are non-physical. The model, parameter values, and colours are identical to Fig \ref{fig:fig3}. The black-dashed line corresponds to zero concentration, predictions below this level are not physically realistic.}
	\label{fig:fig4}
\end{figure}

We now generate a range of predictions. Profile-wise confidence sets for the mean reveal how uncertainty in estimates of mathematical model parameters, $r_{1}$ and $r_{2}$, result in uncertainty in predictions (Fig \ref{fig:fig3}g,h,j,k). For example, Figs \ref{fig:fig3}g,j show that uncertainty in $r_{1}$ results in greater uncertainty in $c_{2}(t)$ close to $t=1$ as opposed to close to $t=0$ and $t=5$. In contrast, Figs \ref{fig:fig3}h,k show that uncertainty in $r_{2}$ results in greater uncertainty in $c_{2}(t)$ for $t\geq 1$ than $0 < t < 1$. In addition, we observe that uncertainty in $r_{1}$ contributes to greater uncertainty in predictions for $c_{1}(t)$ than uncertainty in $r_{2}$ (Fig \ref{fig:fig3}g,h). Taking the union of the profile-wise confidence sets for the model solution we observe the overall uncertainty due to mathematical model parameters (Fig \ref{fig:fig3}i). Thus far these results appear to be physically realistic. However, now we consider Bonferroni correction-based profile-wise confidence sets for data realisations, and their union, that incorporate uncertainty in both the mathematical model parameters and measurement error model parameters (Fig \ref{fig:fig4}). These predictions of data realisations generate results with negative concentrations (Fig \ref{fig:fig3}). Such non-physical predictions are a direct consequence of using the additive Gaussian error model which suggests that this error model may not be appropriate in this situation.

\clearpage

\begin{figure}[h!]
	\centering
	\includegraphics[width=0.9\textwidth]{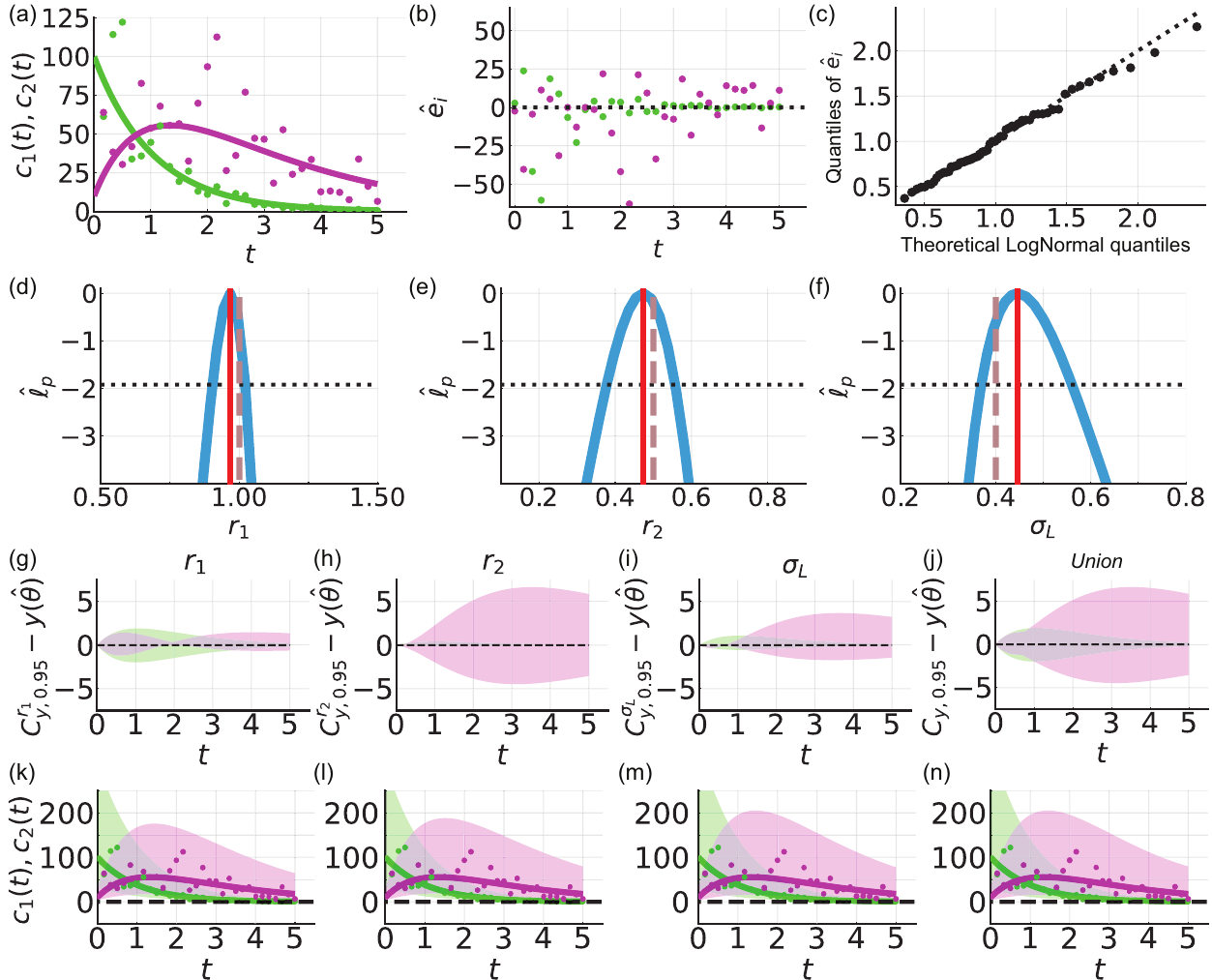}
	\caption{Caricature ODE model with linear reactions (Eq (\ref{eqn:linearodes})) and the correct model specification using the log-normal measurement error model. (a) Synthetic data (circles) at thirty-one equally--spaced time points from $t=0.0$ to $t=5.0$ are generated by simulating Eq (\ref{eqn:linearodes_sol}), the log-normal measurement error model, known model parameters $(r_{1}, r_{2}, \sigma_{\mathrm{L}}) = (1.0, 0.5, 0.4)$, and fixed initial conditions $(c_{1}(0), c_{2}(0)) = (100.0, 10.0)$. Equation (\ref{eqn:linearodes_sol}) simulated with the MLE $(r_{1}, r_{2},\sigma_{\mathrm{L}}) = (0.97, 0.47, 0.45)$ (solid). Throughout $c_{1}(t)$ (solid green) and $c_{2}(t)$ (solid magenta). (b) Difference between the observed data and the best-fit model solution, quantified through the ratios $\hat{e}_{i}=y_{i}^{\mathrm{o}}/y_{i}(\hat{\theta})$. (c) Log-normal quantile-quantile plot of ratios $\hat{e}_{i}=y_{i}^{\mathrm{o}}/y_{i}(\hat{\theta})$. (d)-(f) Profile log-likelihoods (blue) for (d) $r_{1}$, (e) $r_{2}$, and (f) $\sigma_{\mathrm{L}}$ with MLE (red-dashed), an approximate $95\%$ confidence interval threshold (horizontal black-dashed) and know model parameters (vertical brown dashed). (g)-(j) Difference between Bonferroni correction-based confidence set for model solution and the solution of the mathematical model evaluated at the MLE for (g) $r_{1}$, (h) $r_{2}$, (i) $\sigma_{\mathrm{L}}$, and (j) their union. (k-n) Profile-wise Bonferroni correction-based confidence sets for model realisations (shaded) and the solution of the mathematical model evaluated at the MLE (solid line).}
	\label{fig:fig5}
\end{figure}

	Re-analysing the data in Fig \ref{fig:fig3}a using the log-normal error model we avoid any non-physical results. The MLE, $\hat{\theta}=(r_{1}, r_{2}, \sigma_{L}) =(0.97, 0.47, 0.45)$, is close to the known values. The difference between the observed data and the best-fit model solution, quantified through the ratios $\hat{e}_{i}=y_{i}^{\mathrm{o}}/y_{i}(\hat{\theta})$, are reasonably described by the log-normal distribution (Fig \ref{fig:fig5}c). Profile log-likelihoods suggest model parameters are practically identifiable (Fig \ref{fig:fig5}d-f). Approximate $95\%$ confidence intervals, $r_{1} \in (0.90, 1.02)$, $r_{2} \in (0.38, 0.56)$ and $\sigma_{\mathrm{L}} \in (0.37, 0.56)$, capture known parameters and show that using the additive Gaussian error model overestimated uncertainty in $r_{1}$. Profile-wise confidence sets for data realisations and their union are non-negative and so physically realistic (Fig \ref{fig:fig5}k-n). Supplementary S6 presents additional quantile-quantile plots with and without misspecification of the measurement error model.

\clearpage

\subsection{Temporal nonlinear models}\label{sec:results_ode2}

It is straightforward to explore mathematical models of increasing complexity within the framework. A natural extension of Eq (\ref{eqn:linearodes}) assumes that chemical reactions are rate-limited and nonlinear, 
\begin{equation}\label{eqn:nonlinearodes}
	\begin{split}
		\frac{\mathrm{d}c_{1}(t)}{\mathrm{d}t} &= -\frac{V_{1}c_{1}(t)}{K_{1} + c_{1}(t)},\\
		\frac{\mathrm{d}c_{2}(t)}{\mathrm{d}t} &= \frac{V_{1}c_{1}(t)}{K_{1} + c_{1}(t)} -\frac{V_{2}c_{2}(t)}{K_{2} + c_{2}(t)}.
	\end{split}
\end{equation}
Here $V_{i}$ and $K_{i}$ represent maximum reaction rates and Michaelis-Menten constants for chemical species $C_{i}$, with concentrations $c_{i}(t)$, for $i=1,2$.  We solve Eq (\ref{eqn:nonlinearodes}) numerically. We treat the initial conditions $c_{1}(0)$ and $c_{2}(0)$ as known. Then Eq (\ref{eqn:nonlinearodes}) is characterised by four parameters $\theta=(V_{1}, K_{1}, V_{2}, K_{2})$ that we will estimate. These four parameters are structurally identifiable. Note that the previous example, Eq (\ref{eqn:linearodes}), only involved two mathematical model parameters and so our use of the profile log-likelihood in that case could have been avoided by working directly with the likelihood, however in this case we have four unknown parameters in the mathematical model and so visual interpretation of the full likelihood is not straightforward. While one could marginalise the full likelihood for each parameter this often involves sampling-based integration methods that are typically more computationally expensive than optimisation procedures that are required to obtain profile log-likelihoods for each parameter. Furthermore, working directly with the full likelihood to generate predictions can result in an order of magnitude increase in computational time in comparison to profile-wise predictions \cite{Simpson2023}.

\begin{figure}[p]
	\centering
	\includegraphics[width=\textwidth]{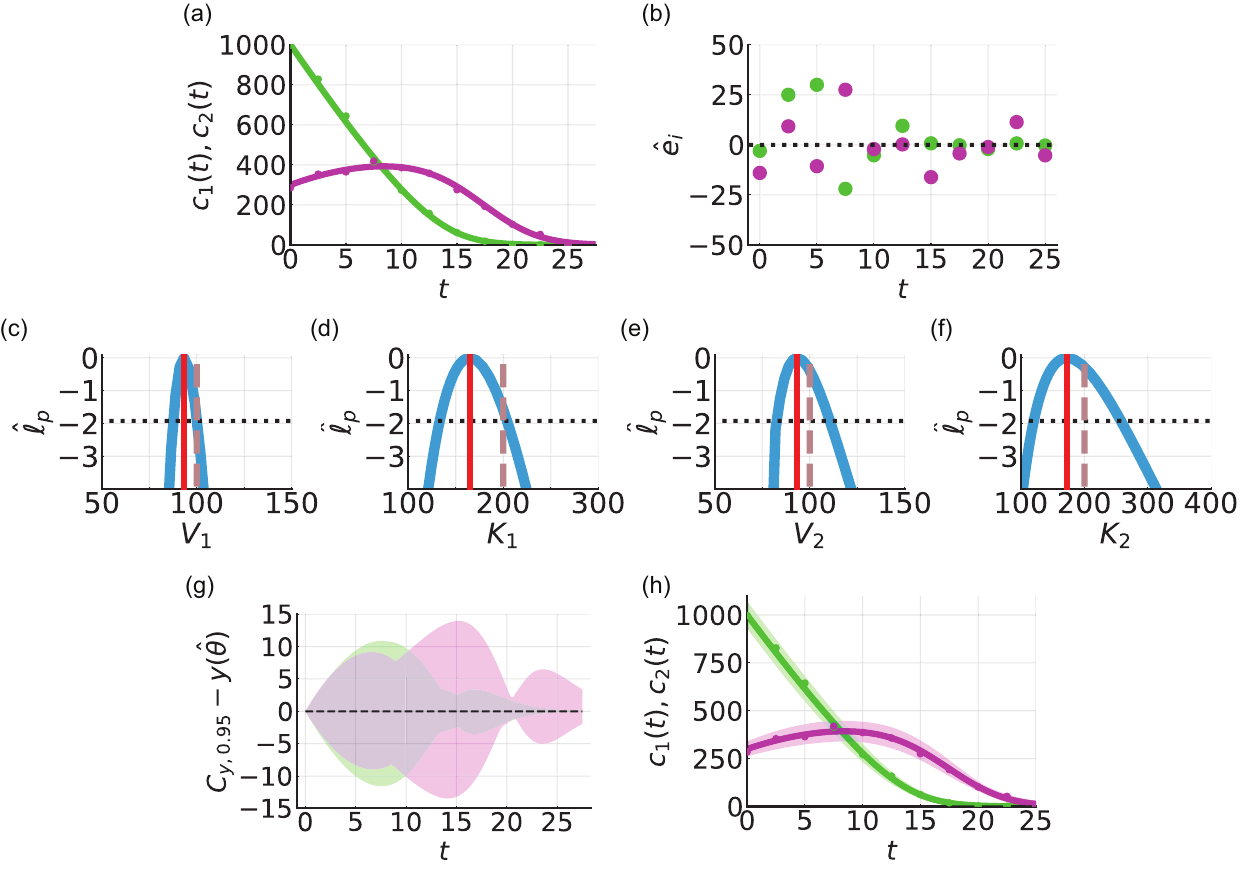}
	\caption{Caricature ODE model with nonlinear reaction terms (Eq (\ref{eqn:nonlinearodes})) and the Poisson measurement error model. (a) Synthetic data (circles) at eleven equally--spaced time points from $t=0.0$ to $t=25.0$ are generated by simulating Eq (\ref{eqn:nonlinearodes}), the Poisson measurement error model, known model parameters $(V_{1}, K_{1}, V_{2}, K_{2}) = (100, 200, 100, 200)$, and fixed initial conditions $(c_{1}(0), c_{2}(0))=(1000, 300)$. Solution of Eq (\ref{eqn:nonlinearodes}) evaluated at the MLE $(V_{1}, K_{1}, V_{2}, K_{2}) = (93.2, 164.8, 93.4, 172.7)$ (solid). Throughout $c_{1}(t)$ (solid green) and $c_{2}(t)$ (solid magenta). (b) Residuals $\hat{e}_{i}=y_{i}^{\mathrm{o}}-y_{i}(\hat{\theta})$ with time $t$. (c)-(f) Profile log-likelihoods (blue) for (c) $V_{1}$, (d) $K_{1}$, (e) $V_{2}$, and (f) $K_{2}$ with MLE (red-dashed), an approximate $95\%$ confidence interval threshold (horizontal black-dashed) and known model parameters (vertical brown dashed). (g)  Difference between union of Bonferroni correction-based confidence sets for the model solution and the solution of the mathematical model evaluated at the MLE. (h) Union of Bonferroni correction-based profile-wise confidence sets for data realisations. Approximate $95\%$ confidence intervals computed from profile log-likelihoods are $V_{1} \in (87.7, 100.0)$, $K_{1}\in (133.2, 204.0)$, $V_{2}\in (82.6, 110.6)$, and $K_{2}\in (120.1, 257.5)$. The MLE is $\hat{\theta}=(V_{1}, K_{1}, V_{2}, K_{2}) = (93.3, 164.8, 93.4, 172.7)$.}
	\label{fig:fig6}
\end{figure}

We generate synthetic data using Eq (\ref{eqn:nonlinearodes}), the Poisson measurement error model, model parameters, $\theta=(V_{1}, K_{1}, V_{2}, K_{2}) = (100, 200, 100, 200)$, and initial conditions $(c_{1}(0), c_{2}(0)) = (1000, 300)$ (Fig \ref{fig:fig6}a). Using Eq (\ref{eqn:nonlinearodes}) and the Poisson measurement error model, we seek estimates of $V_{1}$, $K_{1}$, $V_{2}$, and $K_{2}$ and generate predictions.  Simulating the mathematical model with the MLE, we observe excellent agreement with the data (Fig \ref{fig:fig6}a). Profile log-likelihoods for $V_{1}$, $K_{1}$, $V_{2}$ and $K_{2}$ capture known parameter values and show that these parameters are practically identifiable. Predictions, in the form of the union of profile-wise confidence sets for the means (Fig \ref{fig:fig6}(g)) and the union of profile-wise confidence sets for realisations (Fig \ref{fig:fig6}(h)), show greater uncertainty at higher concentrations. Re-analysing this data using the additive Gaussian measurement error model results in non-physical predictions as we predict negative concentrations at later times where $c_{1}(t)$ and $c_{2}(t)$ are close to zero. The framework is straightforward to apply to other ODEs with nonlinear reaction terms, for example the Lotka-Volterra predator-prey model (Supplementary S4.1).

\clearpage
\newpage
\subsection{Spatio-temporal models}\label{sec:results_pde1}

Throughout mathematical biology and ecology we are often interested in dynamics that occur in space and time \cite{Britton2005,EdelsteinKeshet2005,Murray2002a,Murray2002b,Kot2001}. This gives rise to spatio-temporal data that we analyse with spatio-temporal models such as reaction--diffusion models. Reaction-diffusion models have been used to interpret a range of applications including chemical and biological pattern formation, spread of epidemics, and animal dispersion, invasion, and interactions \cite{Britton2005,EdelsteinKeshet2005,Murray2002a,Murray2002b,Kot2001,Kondo2010,Okubo2001,Turing1952}. As a caricature example, consider a system of two diffusing chemical species in a spatial domain $-\infty < x < \infty$ subject to the reactions in Eq (\ref{eqn:linearodes}). The governing system of PDEs is,
\begin{equation}\label{eqn:linearpdes}
	\begin{split}
		\frac{\partial c_{1}(t,x)}{\partial t}    &=  D\frac{\partial^{2} c_{1}(t,x)}{\partial x^{2}} -r_{1}c_{1}(t,x)),\\
\frac{\partial c_{2}(t,x)}{\partial t}  &= D\frac{\partial^{2} c_{2}(t,x)}{\partial x^{2}}  + r_{1}c_{1}(t,x) - r_{2}c_{2}(t,x).
\end{split}
\end{equation}
Here, $D$ represents a constant diffusivity. We choose initial conditions to represent the release of chemical $C_{1}$ from a confined region, 
\begin{subequations}\label{eqn:linearpdes_ics}
		\renewcommand{\theequation}{\theparentequation.\arabic{equation}}
		\begin{alignat}{1}
		 c_{1}(0,x)    &= \begin{cases}
		 	C_{0}	\quad  \vert x \rvert < h,\\
		  	0 \quad   \vert x \rvert > h,
		 					\end{cases}\\ 
		c_{2}(0,x)   &= 0, \quad -\infty < x < \infty.
		\end{alignat}
\end{subequations}
Solving Eqs (\ref{eqn:linearpdes})-(\ref{eqn:linearpdes_ics}) analytically, for $r_{1}\neq r_{2}$, gives  (Supplementary S1) \cite{Clement2001,Crank1975}, 
\begin{subequations}\label{eqn:linearpdessolution}
		\renewcommand{\theequation}{\theparentequation.\arabic{equation}}
		\begin{alignat}{1}
			c_{1}(t,x) & = \frac{C_{0}}{2}  \left[ \mathrm{erf}\left( \frac{h-x}{2\sqrt{Dt}} \right) + \mathrm{erf}\left(\frac{h+x}{2\sqrt{Dt}}  \right) \right] \exp({-r_{1}t}), \\
			c_{2}(t,x) & = \left(\frac{r_{1}}{r_{2}-r_{1}}\right)\frac{C_{0}}{2}  \left[ \mathrm{erf}\left( \frac{h-x}{2\sqrt{Dt}} \right) + \mathrm{erf}\left(\frac{h+x}{2\sqrt{Dt}}  \right) \right] \bigg( \exp\left({-r_{1}t}\right) - \exp\left({-r_{2}t}\right) \bigg),
		\end{alignat}
	\end{subequations}
where $\mathrm{erf}(z)=2/\sqrt{\pi}\int_{0}^{z} \exp(\eta^{2}) \ \mathrm{d}\eta$ is the error function \cite{Crank1975}. An analytical solution for the special case $r_{1}=r_{2}$ can also be obtained and has a different format where again $c_2(t,x)$ is proportional to $c_1(t,x)$. Assuming that $C_{0}$ and $h$ are known, Eq (\ref{eqn:linearpdessolution}) is characterised by three unknown parameters ($D$, $r_{1}$, $r_{2}$). 

\begin{figure}[h!]
	\centering
	\includegraphics[width=\textwidth]{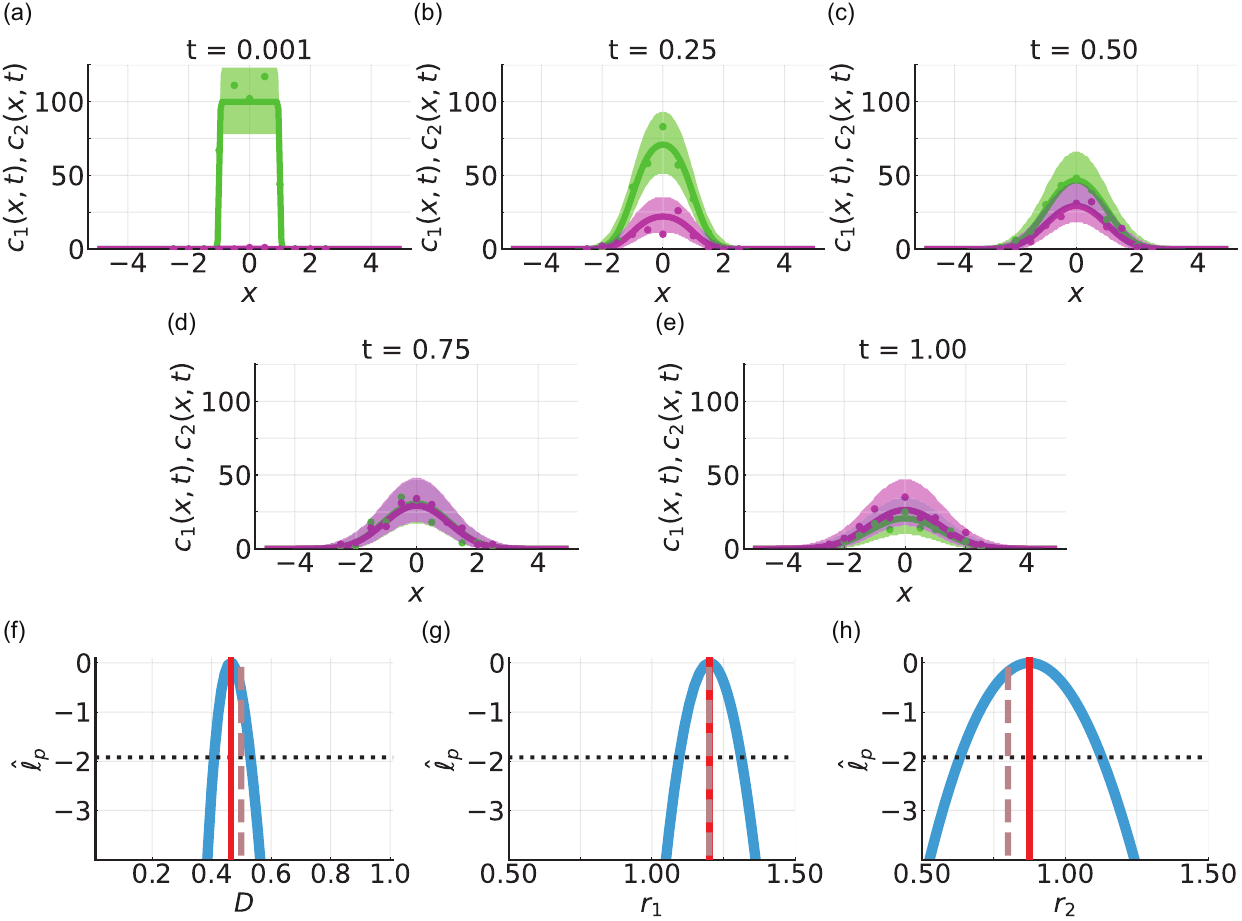}
	\caption{Caricature PDE model (Eq (\ref{eqn:linearpdessolution})) with Poisson measurement error model. Synthetic data (circles) generated by solving Eq (\ref{eqn:linearpdessolution}) and the Poisson measurement error model with known model parameters $\theta=(D, r_{1}, r_{2}) = (0.5, 1.2, 0.8)$. (a-d) Union of Bonferroni correction-based profile-wise confidence sets for data realisations of $c_{1}(t)$ (green shaded) and $c_{2}(t)$ (magenta shaded) for (a) $t=0.001$, (b) $t=0.25$, (c) $t=0.50$, (d) $t=0.75$, and (e) $t=1.0$. Data points are measured at eleven equally--spaced positions between $x = -2.5$ and $x=2.5$, inclusive. The solution of Eq (\ref{eqn:linearpdessolution}) evaluated at the MLE shown for $c_{1}(t)$ (green solid) and $c_{2}(t)$ (magenta solid). (f)-(h) Profile log-likelihoods (blue) for (f) $D$, (g) $r_{1}$, and (h) $r_{2}$ with MLE (red-dashed), an approximate $95\%$ confidence interval threshold (horizontal black-dashed) and known model parameters (vertical brown dashed).}
	\label{fig:fig7}
\end{figure}

We generate synthetic spatio-temporal data at eleven spatial points and  five different times (Fig \ref{fig:fig7}a-e). To generate the synthetic data we use Eq (\ref{eqn:linearpdessolution}), the Poisson measurement error model, and set $\theta=(D, r_{1}, r_{2})=(0.5, 1.2, 0.8)$ and fix $(C_{0}(0), h) = (100, 1)$. To obtain estimates of $D$, $r_{1}$, $r_{2}$ and generate predictions, we use Eq (\ref{eqn:linearpdessolution}) and the Poisson measurement error model. Simulating the mathematical model with the MLE, we observe excellent agreement with the data (Fig \ref{fig:fig7}a-f). Univariate profile log-likelihoods for $D$, $r_{1}$, and $r_{2}$ are well-formed, capture the known parameter values, and suggest that these parameters are practically identifiable. Predictions, in the form of the union of profile-wise confidence sets for realisations (Fig \ref{fig:fig6}h), show that there is greater uncertainty at higher chemical concentrations. This framework also applies to systems of PDEs that are solved numerically (Supplementary S2). Previous comments exploring measurement error model misspecification for systems of ODEs also hold for systems of PDEs.

\newpage
\clearpage

\clearpage

\clearpage
\newpage
\section{Coverage}\label{sec:coverage}

Frequentist methods for estimation, identifiability, and prediction are generally concerned with constructing estimation procedures with reliability guarantees, such as coverage of confidence intervals and sets. For completeness we explore coverage properties numerically. We present an illustrative example revisiting Eq (\ref{eqn:linearodes}) with the additive Gaussian noise model and now fix $\sigma_{N}=5$. This results in a model with two parameters, $\theta=(r_{1}, r_{2})=(1.0, 0.5)$, that we estimate. Initial conditions $(c_{1}(0), c_{2}(0))=(100, 25)$ are fixed. The same evaluation procedure can be used to assess coverage properties for non-Gaussian noise models, such as the log-normal error model (Supplementary S5).

We generate 5000 synthetic data sets using the same mathematical model, measurement error model, and model parameters, $\theta$. Each data set comprises measurements of $c_{1}(t)$ and $c_{2}(t)$ at sixteen equally--spaced time points from $t=0.0$ to $t=2.0$. For each data set we compute a univariate profile log-likelihood for $r_{1}$ and use this to form an approximate $95.0\%$ confidence interval for $r_{1}$. We then test whether this approximate $95.0\%$ confidence interval contains the true value of $r_{1}$. This holds for $95.2\%$ of the data sets, corresponding to an observed coverage probability of $0.952$. Similarly, the observed coverage probability for $r_{2}$ is $0.946$. Therefore, the observed coverage probabilities for both $r_{1}$ and $r_{2}$ are close to the target coverage probability of $0.950$. In contrast to our profile-wise coverage approach, a full likelihood-based approach recovers an observed coverage probability of $0.950$ for the confidence region for $r_{1}$ and $r_{2}$ (Supplementary S3).

For each data set, we propagate forward variability in $r_{1}$ to generate an approximate $95\%$ confidence set for the model solution, $C_{y,0.95}^{r_{1}}$. We consider coverage of this confidence set from two perspectives. First, we explore coverage from the perspective of testing whether or not the true model solution, $y(t;\theta)$, is entirely contained within the confidence set and refer to this as \textit{curvewise coverage}. Second, we discretise the model solution and for each point of the model solution, $y(t_{i};\theta)$ for $i=1,2,3,\ldots,N$, we test whether or not it is contained within the confidence set for the model solution and refer to this as \textit{pointwise coverage}. Note that the time points at which we discretise the model solution do not need to be the same time points where measurements are observed. Previous profile likelihood-based methods focus only on pointwise predictions \cite{Hass2016,Kreutz2012,Villaverde2022}. In our framework curvewise properties are natural for model trajectories since we are interested in the variability of model solutions obtained by propagating forward variability in model parameters using a continuous deterministic mathematical model. Curvewise coverage properties are more challenging to achieve in general and pointwise coverage properties can help to explain why. 

\begin{figure}[h!]
	\centering
	\includegraphics[width=\textwidth]{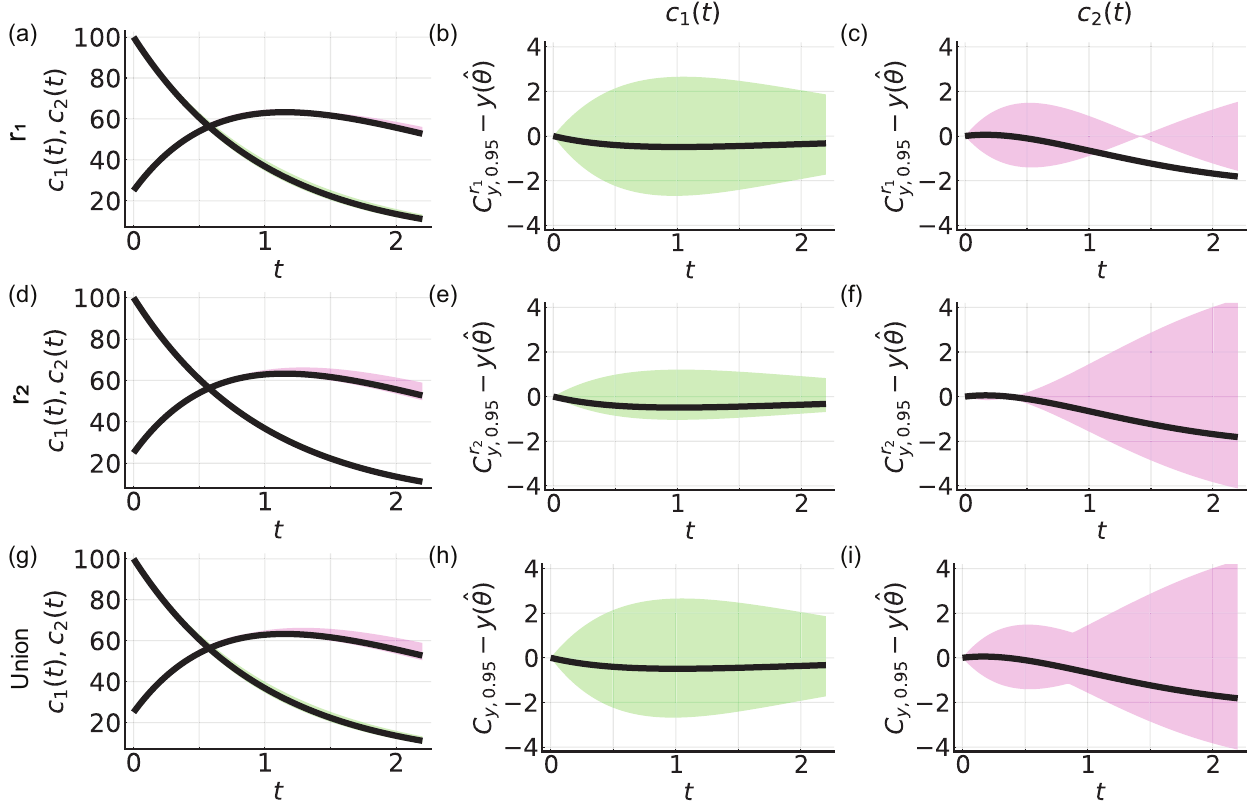}
	\caption{Curvewise confidence sets for the model solutions of a caricature ODE model with linear reaction terms (Eq \ref{eqn:linearodes}) and the additive Gaussian measurement error model with known $\sigma_{N}$. (a) Confidence sets for model solution generated from uncertainty in $r_{1}$, $C_{y,0.95}^{r_{1}}$ (shaded), and the true model solution, $y(\theta)$ (black). (b)-(c) Difference between curvewise confidence set and solution of the mathematical model evaluated at the MLE, $C_{y,0.95}^{r_{1}}-y(\hat{\theta})$ ($c_{1}(t)$ (shaded green) and $c_{2}(t)$ (shaded magenta) and the difference between the true model solution and the solution of the mathematical model evaluated at the MLE, $y(\theta) - y(\hat{\theta})$ (black). (d)-(f) Results based on uncertainty in $r_{2}$. (g)-(i) Results for the union of curvewise confidence sets. Throughout, to plot $y(\theta)$ the temporal domain is discretised into $100$ equally--spaced points ($0.022 \leq t \leq 2.200$) connected using a solid line. }	\label{fig:Fig8}
\end{figure}

\subsection{Curvewise coverage}

For the problems we consider the variation in the confidence set at each time point is narrow relative to the overall variation in $c_{1}(t)$ and $c_{2}(t)$ over time (Fig \ref{fig:Fig8}a). Therefore, we plot and examine the difference between the confidence set and the model solution at the MLE, $C_{y,0.95}^{r_{1}}-y(\hat{\theta})$, and the difference between the true model solution and the model solution at the MLE, $y(\theta)-y(\hat{\theta})$ (Fig \ref{fig:Fig8}b,c). The $c_{1}(t)$ component of the true model solution, $y(t;\theta)$, is contained within the confidence set (Fig \ref{fig:Fig8}b). However, the true model solution is only contained within the $c_{2}(t)$ component of the confidence set for $t \leq 1.056$ (Fig \ref{fig:Fig8}c). Hence, the true model solution is not contained within the confidence set $C_{y,0.95}^{r_{1}}$. We repeat this analysis for the confidence set $C_{y,0.95}^{r_{2}}$ (Figure \ref{fig:Fig8}d-f) and the union of the confidence sets $C_{y,0.95} =C_{y,0.95}^{r_{1}} \cup C_{y,0.95}^{r_{2}}$ (Figure \ref{fig:Fig8}g-i). By construction, the confidence set $C_{y,0.95}$ has coverage properties that are at least as good as $C_{y,0.95}^{r_{1}}$ and $C_{y,0.95}^{r_{2}}$. For example, in Fig \ref{fig:Fig8}h,i the true model solution is contained within $C_{y,0.95}$ whereas it is not contained within  $C_{y,0.95}^{r_{1}}$. Assessing whether the model solution is or is not entirely contained within the confidence sets $C_{y,0.95}^{r_{1}}$, $C_{y,0.95}^{r_{2}}$, and $C_{y,0.95}$ for each of the 5000 data sets, we obtain observed curvewise coverage probabilities of $0.007$, $0.018$, and $0.609$, respectively. These observed coverage probabilities are much lower than results for confidence intervals of model parameters. However, in contrast to our profile-wise coverage results, a full likelihood-based approach recovers an observed curvewise coverage probability of $0.956$ for the confidence set for model solutions (Supplementary S3).

Given the drastic differences in observed curvewise coverage probabilities between the profile likelihood-based method and full likelihood-based method one may expect that the confidence sets from the two methods are qualitatively very different. However, comparing the two confidence sets they appear to qualitatively very similar (Supplementary S3). This suggests that subtle differences in confidence sets may play an important role in observed curvewise coverage probabilities. Full likelihood-based approaches are computationally expensive relative to profile likelihood-based methods, especially for models with many parameters. Here we have only considered univariate profiles. However, an interesting approach is to use profile likelihood-based methods with higher-dimensional interest parameters. These have been shown to improve coverage properties relative to scalar valued interest parameters at a reduced computational expense relative to full likelihood-based methods \cite{Simpson2023}.

\subsection{Pointwise coverage} 

\begin{figure}[h]
	\centering
	\includegraphics[width=\textwidth]{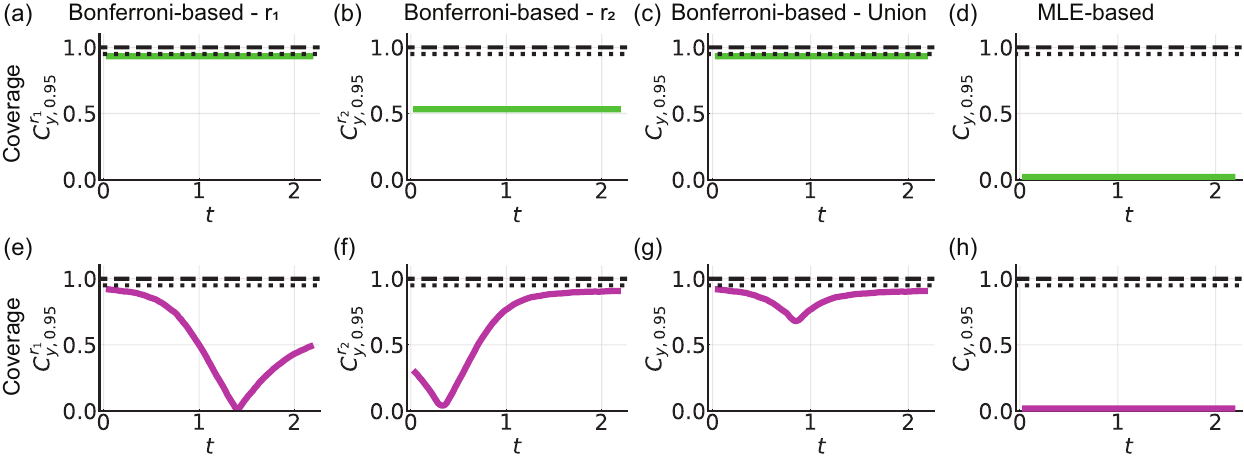}
	\caption{Pointwise coverage analysis for confidence sets for model solutions. Analysis performed using the caricature ODE model (Eq (\ref{eqn:linearodes})) as an illustrative example. (a)-(h) Pointwise coverage analysis of confidence sets for model solutions.  Results for Bonferroni correction-based confidence sets for $r_{1}$, $r_{2}$, and their union are shown in (a,e), (b,f), and (c,g), respectively. Results for MLE-based confidence sets are shown in (d,h). The temporal domain is discretised into $100$ equally--spaced points ($0.022 \leq t \leq 2.200$). Horizontal dotted and horizontal dashed lines correspond to observed probabilities of $0.95$ and $1.00$, respectively. }
	\label{fig:Fig9}
\end{figure}

Assessing pointwise coverage can help diagnose why we do not reach target curvewise coverage properties when propagating univariate profiles. This kind of diagnostic tool can be used to inform experimental design questions regarding when, and/or where, to collect additional data. In this context, the confidence sets can be interpreted as tools for sensitivity analysis. We discretise the temporal domain into $100$ equally--spaced points ($0.022 \leq t \leq 2.200$), and exclude $t=0$ because initial conditions are treated as fixed quantities in this instance. For each data set, time point, chemical concentration, and confidence set, we test whether the true model solution is contained within the confidence set. For the component of the confidence set $C_{y,0.95}^{r_{1}}-y(\hat{\theta})$ concerning $c_{1}(t)$, the observed pointwise coverage is constant throughout time and equal to $0.932$ which is relatively close to the desired value (Fig \ref{fig:Fig9}a). In contrast, for the component of the confidence set $C_{y,0.95}^{r_{1}}-y(\hat{\theta})$ concerning $c_{2}(t)$, the observed pointwise coverage is initially equal to $0.920$ at $t=0.022$, then decreases over time reaching a minimal value of $0.012$ at $t=1.408$ before increasing to $0.497$ at $t=2.200$ (Fig \ref{fig:Fig9}e). Similar behaviour is observed for the confidence set $C_{y,0.95}^{r_{2}}-y(\hat{\theta})$ (Fig \ref{fig:Fig9}b,f). Taking the union of the confidence sets we obtain more conservative confidence sets, with an observed pointwise coverage for $c_{1}(t)$ of $0.932$ throughout (Fig \ref{fig:Fig9}c) and an observed pointwise coverage for $c_{2}(t)$ of at least $0.681$ (Fig \ref{fig:Fig9}g). Note that the solution of the mathematical model evaluated at the MLE, $y(t; \hat{\theta})$, is not identical to the true model solution so, as expected, the observed pointwise coverage probability of this single trajectory is zero at all time points (Fig \ref{fig:Fig9}d,h).

\begin{figure}[h!]
	\centering
	\includegraphics[width=\textwidth]{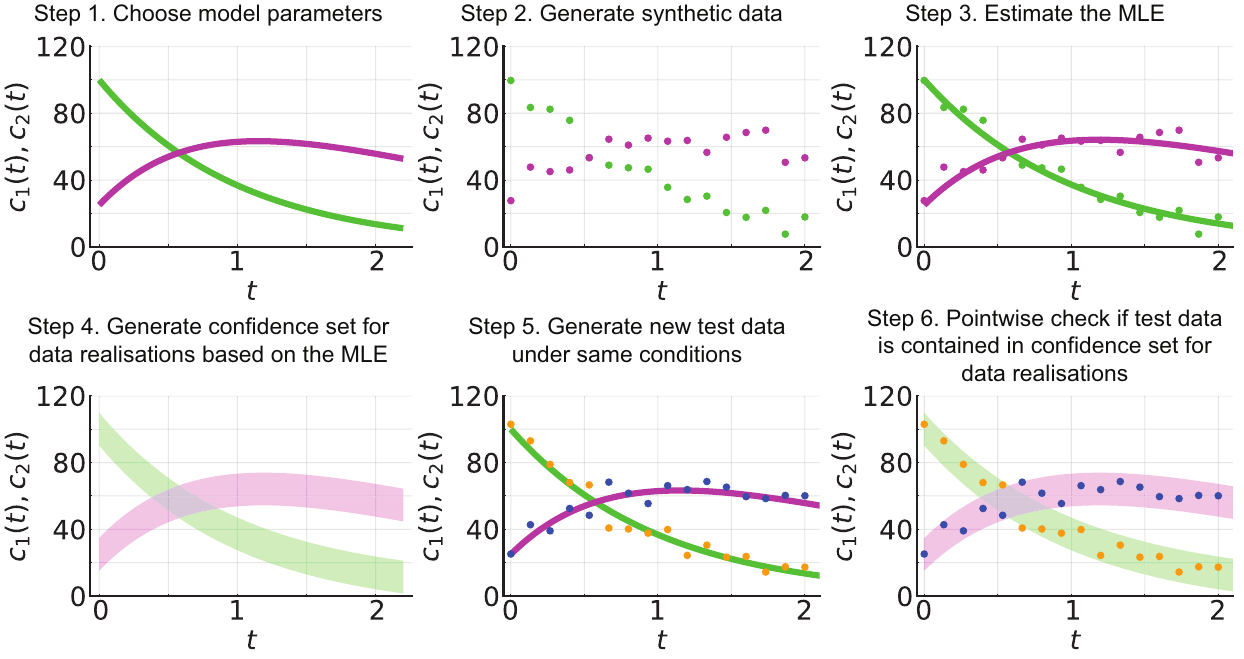}
	\caption{Schematic for evaluation procedure used to test coverage properties of confidence sets for model realisations. In this work we repeat these steps $5000$ times. Example presented using the MLE-based approach, and is readily adapted for Bonferroni correction-based confidence sets for model realisations by modifying step four. }
	\label{fig:Fig10}
\end{figure}

\begin{figure}[h!]
	\centering
	\includegraphics[width=\textwidth]{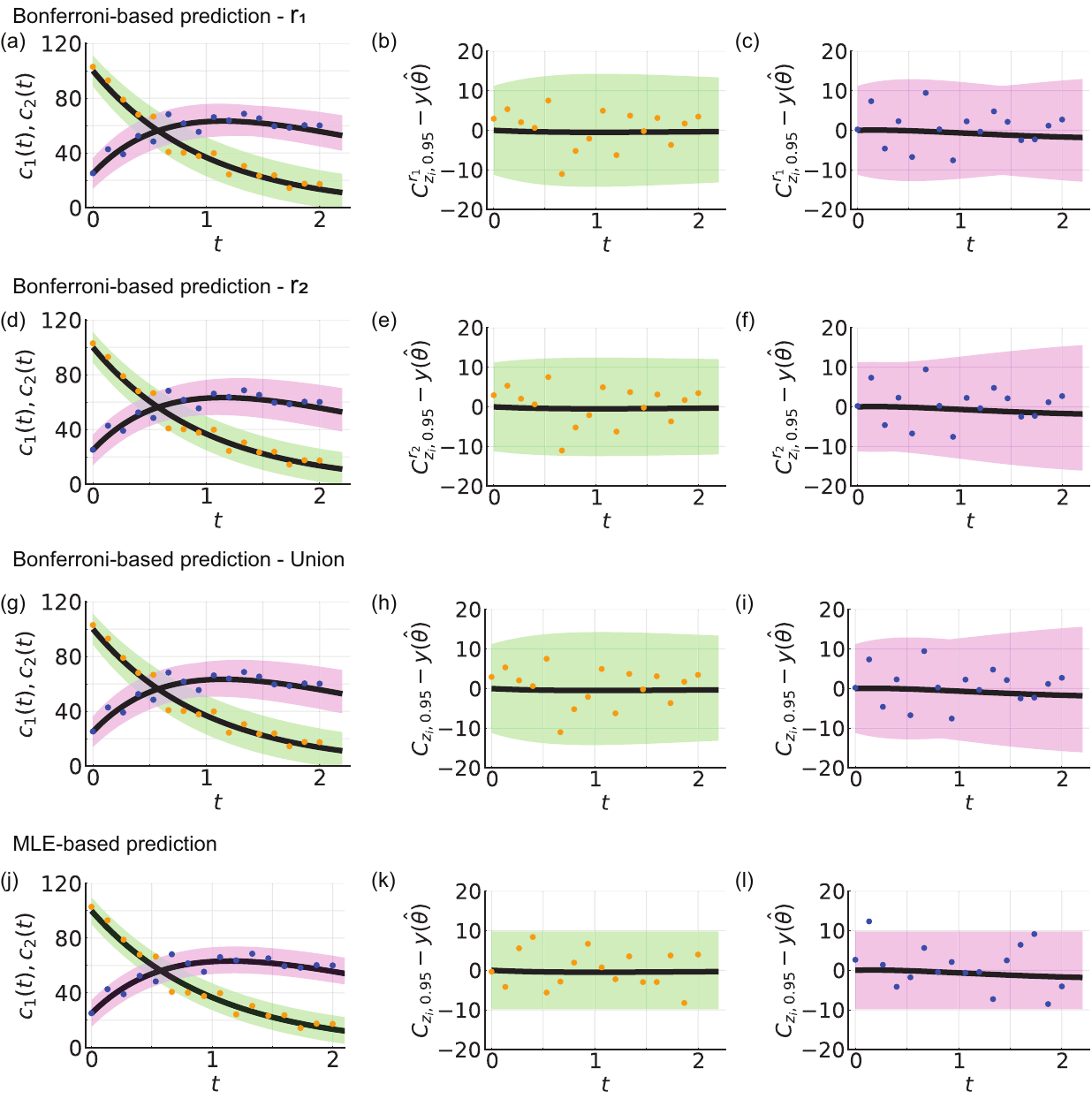}
	\caption{Confidence sets for model realisations of a caricature ODE model with linear reaction terms (Eq \ref{eqn:linearodes}) and the additive Gaussian measurement error model with known $\sigma_{N}$. (a) Bonferroni correction-based confidence set for $r_{1}$. (b)-(c) Difference between the Bonferroni correction-based confidence set for $r_{1}$ and solution of the mathematical model evaluated at the MLE, $C_{y,0.95}^{r_{1}}-y(\hat{\theta})$ ($c_{1}(t)$ (shaded green) and $c_{2}(t)$ (shaded magenta) and the difference between the true model solution and the solution of the mathematical model evaluated at the MLE, $y(\theta) - y(\hat{\theta})$ (black). (d-i) Results for Bonferroni correction-based confidence sets for (d-f) $r_{2}$ and (g-i) the union. (j-l) Results for MLE-based confidence set.}\label{fig:Fig11}
\end{figure}

\clearpage
\newpage
We now explore MLE-based and Bonferroni correction-based confidence sets for model realisations in a pointwise manner. For both methods we apply the same evaluation procedure (Fig \ref{fig:Fig10}). For each of the $5000$ synthetic data sets we generate the confidence set for the data realisations and then generate a new synthetic data set under the same conditions as the original synthetic data set. In particular, the new data set is generated at the same time points using the same mathematical model, measurement error model, and parameter values. This approach can be be thought of as a test of the predictions under replication of the experiment. For each new data point, which includes fifteen equally--spaced data points from $t=0.13$ to $t=2.00$, we test whether or not it is contained within the confidence set for the model realisation. Results for a single synthetic data set show that Bonferroni correction-based confidence sets for model realisations based on $r_{1}$ (Fig \ref{fig:Fig11}a-c), $r_{2}$ (Fig \ref{fig:Fig11}d-f), and their union
(Fig \ref{fig:Fig11}g-i) can overcover relative to the MLE-based approach (Figure \ref{fig:Fig11}j-l).

\clearpage

\begin{figure}[h]
	\centering
	\includegraphics[width=\textwidth]{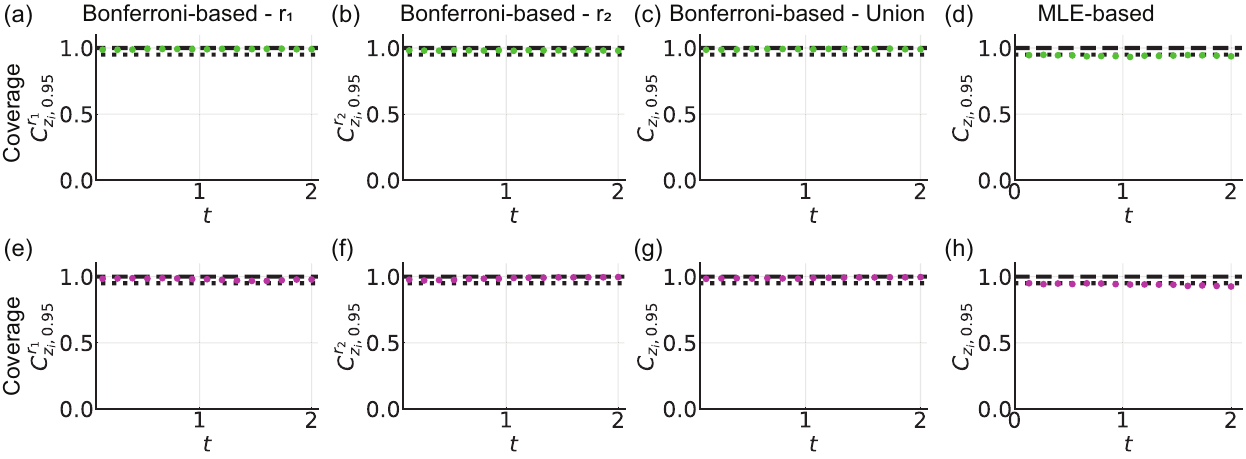}
	\caption{Pointwise coverage analysis for confidence sets for model realisations. Analysis performed using the caricature ODE model (Eq (\ref{eqn:linearodes})) as an illustrative example. (a)-(h) Pointwise coverage of confidence sets for model realisations. Results for Bonferroni correction-based confidence sets for $r_{1}$, $r_{2}$, and their union are shown in (a,e), (b,f), and (c,g), respectively. Results for MLE-based confidence sets are shown in (d,h). Horizontal dotted and horizontal dashed lines correspond to observed probabilities of $0.95$ and $1.00$, respectively.}
	\label{fig:Fig12}
\end{figure}

Analysing results for the $5000$ synthetic data sets we find that the average observed pointwise coverage probability for the MLE-based confidence set for model realisations across all time points and the two chemical species is $0.937$. Pointwise coverage properties per time point and chemical species for the MLE-based approach are shown in Fig \ref{fig:Fig12}d,h. In this example statistical noise is large relative to the difference in the true model solution and the solution of the mathematical model evaluated at the MLE, $y(t;\theta) - y(t;\hat{\theta})$, such that the coverage properties are relatively close to the target coverage probability of $0.950$. The average pointwise coverage for Bonferroni correction-based confidence set for model realisations is $0.985$ for $r_{1}$, $0.982$ for $r_{2}$, $0.990$ for their union. Pointwise coverage properties per time point and chemical species for the Bonferroni correction-based approaches are shown in Fig \ref{fig:Fig12}a-c,e-g. For this particular example the Bonferroni correction-based consistently exceeds the target coverage probability. Using a full likelihood-based method recovers an observed average pointwise coverage probability $0.994$ for the Bonferroni correction-based confidence set for model realisations (Supplementary S3). Note that since the MLE-based confidence set for model realisations depends only on the MLE, the confidence set independent of whether a profile likelihood-based or full likelihood-based approach is implemented.

 While the framework presented in this section is straightforward to apply to other mathematical models and measurement error models, coverage properties should be interpreted and assessed on a case-by-case basis. In Supplementary S5 we present such an example using the log-normal measurement error model and find similar results to those discussed here. Other frequentist evaluation procedures can also be used to explore coverage properties of confidence sets for model realisations. For example, for a data set with $I$ elements we could generate a confidence set for model realisations based on the first $k < I$ time points of data and then test if one, or more, of the remaining $I-k$ elements of the data set are contained in the confidence set.

\clearpage
\newpage

\section{Conclusion}\label{sec:conclusion}

In this review we demonstrate how to practically implement a variety of measurement error models in a general profile likelihood-based framework for parameter estimation, identifiability analysis, and prediction. Illustrative case studies explore additive, multiplicative, discrete, and continuous measurement error models and deal with the commonly-encountered situation of noisy and incomplete data. Mathematical models in the case studies are motivated by the types of models commonly found in the systems biology literature and the mathematical biology literature. Within the framework, assessing uncertainties in parameter estimates and propagating forward these uncertainties to form predictions allows us to assess the appropriateness of measurement error models and make direct comparisons to data. Furthermore, techniques to assess pointwise and curvewise coverage properties provide useful tools for experimental design and sensitivity analysis. The profile likelihood-based methods, based on numerical optimisation procedures, are computationally efficient and a useful approximation to full likelihood-based methods (Supplementary S3) \cite{Simpson2023}. Open source Julia code to reproduce results is freely available on \href{https://github.com/ryanmurphy42/Murphy2023ErrorModels}{GitHub}. These implementations can be adapted to deal with other forms of mathematical models or they could be adapted for implementation within other software frameworks, however we prefer Julia because it is freely available and computationally efficient.

We illustrate the framework using simple caricature models to emphasise the practical implementation of the methods and how to interpret results, rather than the details of each mathematical model. This includes systems of ODEs that are often used in the systems biology literature (\cref{sec:results_ode1}, \cref{sec:results_ode2}, Supplementary S4) and systems of PDEs routinely used in the mathematical biology literature (\cref{sec:results_pde1}). ODE--based models are also routinely used to described biological population dynamics \cite{Toni2009} and disease transmission \cite{He2010}. As parameter estimation, identifiability analysis, and prediction within the profile likelihood-based framework depend only on the solution of the mathematical model, the solution can be obtained analytically or numerically. Analytical solutions are preferred over numerical solutions for computational efficiency, however closed-form exact solutions cannot always be found. For this reason we implement a number of case studies that involve working with simple exact solutions, as well as working with numerical solutions obtained using standard discretisations of the governing differential equations. One can also consider other mathematical models with the framework, such as difference equations are often used in applications about ecology (Supplementary S4) \cite{Ricker1954,Hefley2013a,AugerMethe2021,Valpine2002,Murray2002a}. More broadly the framework can apply to stochastic differential equation-based models \cite{Browning2020} and stochastic simulation-based models \cite{Simpson2021,Simpson2022b,Breto2018,Breto2019,Ionides2017}. Extensions to models that incorporate process noise are of interest \cite{Valpine2002,Breto2009,Dennis2006,King2016,MartinezBakker2015,Valpine2002a}.

The framework is well-suited to consider a variety of measurement error models. Illustrative case studies explore the additive Gaussian error model, the multiplicative log-normal model, and the discrete Poisson model. All example calculations presented in this review take an approach where synthetic data are generated using a mathematical model rather than working with experimental measurements. This is a deliberate choice that allows us to explicitly explore questions of model misspecification and model choice unambiguously since we have complete control of the underlying data generating process. By definition, samples from the log-normal distribution are positive so we deliberately avoid situations where the observed data is zero when using the log-normal measurement error model. A different error model should be considered in such a case, for example, based on the zero-modified log-normal distribution \cite{Johnson2005,Aitchison1955}. For both the log-normal and Poisson error models we also avoid situations where the observed data is positive and the model solution is identically zero. For example, our solutions of ODE-based models approach zero at late time but remain positive for all time considered in this work. Exploring error models for reaction-diffusion PDEs with nonlinear diffusion is of interest, for example those that give rise to travelling wave solutions describing biological invasion with sharp boundaries \cite{Crank1987,ElHachem2019,Murphy2021}. In such an example we may expect to evaluate the error model, and so the likelihood function, at points in space where the data is positive but the model solution is zero. How to handle such a situation and which measurement error model to incorporate is an interesting question that could be explored by extending the tools developed in this review.

Within the framework one could also consider other forms of multiplicative error models, for example based on the gamma distribution \cite{Agamennoni2012,Firth1987}, of which the exponential and Erlang distributions are special cases, or based on the beta distribution \cite{Valpine2002}. A different form of the log-normal distribution with mean equal to $y_{i}(\theta)$ could also be considered within the framework and is given by $y_{i} \mid \theta \sim \mathrm{LogNormal}(\log(y_{i}(\theta))-\sigma_{L}^{2}/2, \sigma_{L})$. Multiplicative noise can be also be implemented in other forms. We have considered multiplicative noise of the form $y_{i}^{\mathrm{o}}=y_{i}(\theta)\eta_{i}$ with $\eta_{i} \sim \mathrm{LogNormal}(0,\sigma_{\mathrm{L}}^2)$ (Eq \ref{eqn:noise_lognormal1}), which for a straight line model, $y(\theta)=c+mx$, would be $y_{i}^{\mathrm{o}}=(c+mx_{i})\eta_{i}$. However, multiplicative noise could also be associated with a component of the model solution. As a specific example from a protein quantification study \cite{Kreutz2007} consider the straight line model where multiplicative noise is incorporated into the slope of the equation but not the $y$-intercept, i.e $y_{i}^{\mathrm{o}} = c + mx_{i}\eta_{i}$ with $\eta_{i} \sim \mathrm{LogNormal}(0,\sigma_{\mathrm{L}}^2)$. One could also relax assumptions in the Poisson distribution that the variance is equal to the mean, in which case the negative binomial distribution may be useful \cite{King2016}. The framework also applies to other discrete distributions such as the binomial model \cite{Maclaren2017,Simpson2023a}. Different measurement error models could also be studied for example the proportional, exponential, and combined additive and proportional error models that are used in pharmacokinetic modelling \cite{Mould2013}. Throughout we assume that errors are independent and identically distributed. Extending the noise model to consider correlated errors is also of interest \cite{Lambert2023,Lei2020}. Assessing coverage properties using different evaluation procedures and assessing predictive capability through the lens of tolerance intervals is also of interest \cite{Miller1981,Lieberman1963}. Overall, the choice of which mathematical model and measurement error model to use should be considered on a case-by-case basis and can be explored within this framework.

\newpage
\appendix

\section{Code}

 Julia implementations of all computations are available on \href{https://github.com/ryanmurphy42/Murphy2023ErrorModels}{GitHub}. Here we highlight key packages and code used in our implementation.

Throughout we assess structural identifiability using the \texttt{StructuralIdentifiability} package \cite{Dong2022}). To estimate parameters and explore practical identifiability using profile log-likelihoods we find that it is straightforward to compute the log-likelihood for a range of error models using the \texttt{loglikelihood} function in the \texttt{Distributions} package \cite{JuliaDistributionsZenodo}. For example, we evaluate the log-likelihood for the additive Gaussian, multiplicative log-normal, and discrete Poisson measurement error models using \texttt{loglikelihood}$(\mathrm{Normal}(y_{i}(\theta), \sigma^{2}),y_{i}^{\mathrm{o}})$, \texttt{loglikelihood}$(\mathrm{LogNormal}(\log(y_{i}(\theta)), \sigma_{\mathrm{L}}^{2}),y_{i}^{\mathrm{o}})$, and \texttt{loglikelihood}$(\mathrm{Poisson}(y_{i}(\theta)),y_{i}^{\mathrm{o}})$, respectively. Approximate confidence interval thresholds are obtained computationally by \texttt{c=quantile(Chisq($\nu$), $1-\alpha$)/2}, using the Distributions package \cite{JuliaDistributionsZenodo}. For example, $90\%$, $95\%$, $99\%$ and $99.9\%$ ($\alpha = 0.100, 0.050, 0.010, 0.001 $ respectively) correspond to threshold values of  $\ell_{c}=-1.35$, $-1.92$, $-3.32$, and $-5.51$, respectively \cite{Pawitan2001,Royston2007}.

All systems of differential equations are solved numerically using the default \texttt{ODEproblem} solver in the \texttt{DifferentialEquations} package \cite{JuliaDifferentialequations}. To perform numerical maximisation, we find that the Nelder-Mead local optimisation routine, with default stopping criteria, within the \texttt{NLopt} optimisation package performs well for the problems in this study \cite{JuliaNLopt}.

\newpage

\noindent 

\subsubsection*{Code Availability}
Julia implementations of all computations are available on GitHub,

\href{https://github.com/ryanmurphy42/Murphy2023ErrorModels}{https://github.com/ryanmurphy42/Murphy2023ErrorModels} 

\subsubsection*{Author's contributions}

All authors conceived and designed the study. RJM performed the research and drafted the article. All authors provided comments and approved the final version of the manuscript. 

\subsubsection*{Competing interests} We declare we have no competing interest. 

\subsubsection*{Funding} MJS is supported by the Australian Research Council (DP200100177). The funders had no role in study design, data collection and analysis, decision to publish, or preparation of the manuscript.

\end{document}


\title{Supplementary Material: \\Implementing measurement error models with mechanistic mathematical models in a likelihood-based framework for estimation, identifiability analysis, and prediction in the life sciences\date{}}

\author{Ryan J. Murphy$^{1,*}$, Oliver J. Maclaren$^{2}$, Matthew J. Simpson$^{1}$}

\maketitle

		\vspace{-0.25in}
\begin{center} 
	\textit{$^{1}$ Mathematical Sciences, Queensland University of Technology, Brisbane, Australia}\\
	\textit{$^{2}$ The Department of Engineering Science and Biomedical Engineering, University of Auckland, Auckland, New Zealand}
\end{center}

\footnotetext[1]{Corresponding author: r23.murphy@qut.edu.au}

\clearpage
\newpage

\renewcommand{\contentsname}{Supplementary Material}

\begin{small}
\tableofcontents
\end{small}

\newpage
\section{Analytical solutions}\label{sec:method_analyticalsolutions}

Here we present a transformation method to obtain analytical solutions to systems of linear ordinary differential equations and systems of linear partial differential equations with coupling in the source terms of the partial differential equation models. These methods, based on diagonalisation \cite{Clement2001}, can be applied to systems with $n$ chemical species; to systems of partial differential equations in higher spatial dimensions, and to systems of partial differential equations describing additional mechanisms such as advection \cite{Clement2001}; and to a range of initial conditions for which there are exact solutions for the analogous uncoupled problems \cite{Crank1975}.

\newpage
\subsection{Systems of ordinary differential equations with linear reaction terms}\label{sec:supp_analytical_ODE}

The general method follows \cite{Clement2001}. Consider a system of ordinary differential equations $\mathrm{d}c(t)/\mathrm{d}t = Kc(t)$, where $c(t)$ is an $n$-dimensional vector valued function and $K$ is an $n\times n$ constant diagonalisable matrix. We first determine an $n\times n$ constant matrix $S$ whose columns are the eigenvectors of $K$. Next, we define a new $n$-dimensional vector valued function, $b(t)$, via the relationship $c(t)=Sb(t)$. Assuming $S$ is invertible, $b(t)=S^{-1}c(t)$ and the system of equations can be written as an uncoupled system of equations $\mathrm{d}b(t)/\mathrm{d}t = \tilde{K}b(t)$ that we solve for $b(t)$. Here $\tilde{K}= S^{-1}KS$ is a $n\times n$ constant diagonal matrix. We obtain the solution $c(t)$ using $c(t)=Sb(t)$.

As an explicit example, consider Eq (16) with $c(t) = (c_{1}(t), c_{2}(t))$ and $r_{1}\neq r_{2}$, 
\begin{equation}\label{eqn:supp_linearodes1}
	\begin{bmatrix} \dfrac{\mathrm{d}c_{1}(t)}{\mathrm{d}t} \\[5pt]  \dfrac{\mathrm{d}c_{2}(t)}{\mathrm{d}t} \end{bmatrix} = \begin{bmatrix} -r_{1} & 0 \\ r_{1} &  - r_{2}  \end{bmatrix}	\begin{bmatrix} c_{1}(t) \\ c_{2}(t) \end{bmatrix}.
\end{equation}
Here,
\begin{equation}\label{eqn:supp_linearodes2}
	S = \begin{bmatrix} 1 & 0 \\ \dfrac{r_{1}}{r_{2}-r_{1}} &  1  \end{bmatrix}	\\ \quad \mathrm{ and } \quad 
	S^{-1} = \begin{bmatrix} 1 & 0 \\ \dfrac{-r_{1}}{r_{2}-r_{1}} &  1  \end{bmatrix}	.
\end{equation}
So we can write Eq (\ref{eqn:supp_linearodes1}) in terms of $b(t)=(b_{1}(t), b_{2}(t))$ as,
\begin{equation}\label{eqn:supp_linearodes4}
	\begin{bmatrix} \dfrac{\mathrm{d}b_{1}(t)}{\mathrm{d}t} \\[5pt]  \dfrac{\mathrm{d}b_{2}(t)}{\mathrm{d}t} \end{bmatrix} = \begin{bmatrix} -r_{1} & 0 \\ 0 &  - r_{2}  \end{bmatrix}	\begin{bmatrix} b_{1}(t) \\ b_{2}(t) \end{bmatrix}.
\end{equation}
Equation (\ref{eqn:supp_linearodes4}) represents two uncoupled equations that we solve analytically for $b_{1}(t)$ and $b_{2}(t)$,
\begin{equation}\label{eqn:supp_linearodes5}
	\begin{split}
		b_{1}(t) &= b_{1}(0)\exp(-r_{1}t),\\		b_{2}(t) &= b_{2}(0)\exp(-r_{2}t).
	\end{split}
\end{equation}
Using $c(t)=Sb(t)$, we transform $b(t)$ to obtain,
\begin{equation}\label{eqn:supp_linearodes6}
	\begin{bmatrix} c_{1}(t) \\ c_{2}(t) \end{bmatrix} = \begin{bmatrix} 1 & 0 \\ \dfrac{r_{1}}{r_{2}-r_{1}} &  1  \end{bmatrix}		\begin{bmatrix} b_{1}(t) \\ b_{2}(t) \end{bmatrix}.
\end{equation}
The unknowns $b_{1}(0)$ and $b_{2}(0)$ are determined using the initial conditions for $c_{1}(0)$ and $c_{2}(0)$, giving the solution for $c(t)$,
\begin{equation}\label{eqn:supp_linearodes7}
	\begin{split}
	c_{1}(t) &= c_{1}(0)\exp(-r_{1}t),\\		
	c_{2}(t) &= c_{1}(0)\exp(-r_{1}t)\left( \frac{r_{1}}{r_{2}-r_{1}} \right) + \left[c_{2}(0) - \frac{c_{1}(0)r_{1}}{r_{2}-r_{1}} \right] \exp(-r_{2}t).
\end{split}
\end{equation}
Solutions in Eq (\ref{eqn:supp_linearodes7}) are restricted to $r_{1} \neq r_{2}$. When $r_{1}=r_{2}$ the solutions can be evaluated and in this case $c_{2}(t)$ is a multiple of $c_{1}(t)$ \cite{Boyce2021}.

\newpage
\subsection{Systems of partial differential equations with linear reaction terms}

The general method follows \cite{Clement2001} and extends Supplementary \cref{sec:supp_analytical_ODE}.Consider a system of partial differential equations $\partial c(t,x)/\partial t - D\partial^{2}c(t,x)/\partial x^{2} = Kc(t,x)$, where $c(t,x)$ is an $n$-dimensional vector valued function, $K$ is an $n\times n$ constant diagonalisable matrix, and $D$ is a constant parameter. We first determine an $n\times n$ constant matrix $S$ whose columns are the eigenvectors of $K$. Next, we define a new $n$-dimensional vector valued function, $b(t,x)$, via the relationship $c(t,x)=Sb(t,x)$. Assuming $S$ is invertible, $b(t,x)=S^{-1}c(t,x)$ and the system of equations can be written as as an uncoupled system of equations $\partial b(t,x)/\partial t - D\partial^{2}b(t,x)/\partial x^{2} = \tilde{K}b(t,x)$ that we solve for $b(t,x)$ using standard methods (e.g. similarity solutions, integral transforms). Here $\tilde{K}= S^{-1}KS$ is a $n\times n$ constant diagonal matrix.  We obtain the solution $c(t,x)$ using $c(t,x)=Sb(t,x)$.

As an explicit example, consider Eq (19) with $c(t,x) = (c_{1}(t,x), c_{2}(t,x))$ on the spatial domain $\infty < x< \infty$, 
\begin{equation}\label{eqn:supp_linearpdes1}
	\begin{bmatrix} \dfrac{\partial c_{1}(t,x)}{\partial t} \\[5pt]  \dfrac{\partial c_{2}(t,x)}{\partial t} \end{bmatrix}  - D\begin{bmatrix}  \dfrac{\partial^{2} c_{1}(t,x)}{\partial x^{2}} \\[1em] \dfrac{\partial^{2} c_{2}(t,x)}{\partial x^{2}} \end{bmatrix}= \begin{bmatrix} -r_{1} & 0 \\ r_{1} &  - r_{2}  \end{bmatrix}	\begin{bmatrix} c_{1}(t,x) \\ c_{2}(t,x) \end{bmatrix}.
\end{equation}
Here, for $r_{1}\neq r_{2}$,
\begin{equation}\label{eqn:supp_linearpdes2}
	S = \begin{bmatrix} 1 & 0 \\ \dfrac{r_{1}}{r_{2}-r_{1}} &  1  \end{bmatrix}	\\ \quad \mathrm{ and } \quad 
S^{-1} = \begin{bmatrix} 1 & 0 \\ \dfrac{-r_{1}}{r_{2}-r_{1}} &  1  \end{bmatrix}	.
\end{equation}
So we can write Eq (19) in terms of $b(t,x)=(b_{1}(t,x), b_{2}(t,x))$ as,
\begin{equation}\label{eqn:supp_linearpdes4}
	\begin{bmatrix} \dfrac{\partial b_{1}(t,x)}{\partial t} \\[5pt]  \dfrac{\partial b_{2}(t,x)}{\partial t} \end{bmatrix}  - D\begin{bmatrix}  \dfrac{\partial^{2} b_{1}(t,x)}{\partial x^{2}} \\[1em] \dfrac{\partial^{2} b_{2}(t,x)}{\partial x^{2}} \end{bmatrix} = \begin{bmatrix} -r_{1} & 0 \\ 0 &  - r_{2}  \end{bmatrix}	\begin{bmatrix} b_{1}(t,x) \\ b_{2}(t,x) \end{bmatrix}.
\end{equation}
We also transform the initial conditions (Eq (20)), using $b(0,x)= S^{-1} c(0,x)$, to obtain,
\begin{subequations}\label{eqn:supp_linearpdes6}
		\renewcommand{\theequation}{\theparentequation.\arabic{equation}}
		\begin{alignat}{1}
			b_{1}(0,x)    &= \begin{cases}
				C_{0}	\quad  \vert x \rvert < h,\\
				0 \quad \vert x \rvert > h,
			\end{cases}\\ 
			b_{2}(0,x)   &=  \begin{cases}
				\dfrac{-C_{0}r_{1}}{r_{2}-r_{1}}	\quad  \vert x \rvert < h,\\
				0 \quad \vert x \rvert > h.
			\end{cases}
		\end{alignat}
	\end{subequations}
Equation (\ref{eqn:supp_linearpdes4}) represents two uncoupled equations that we can solve analytically \cite{Crank1975}, to obtain,
\begin{subequations}\label{eqn:supp_liearpdes8}
		\renewcommand{\theequation}{\theparentequation.\arabic{equation}}
	\begin{alignat}{1}
		b_{1}(t,x) & = \frac{C_{0}}{2} \left[ \mathrm{erf}\left( \frac{h-x}{2\sqrt{Dt}} \right) + \mathrm{erf}\left(\frac{h+x}{2\sqrt{Dt}}  \right) \right] e^{-r_{1}t}, \\
		b_{2}(t,x) & = \left( \frac{-r_{1}}{r_{2}-r_{1}}\right)\frac{C_{0}}{2}\left[ \mathrm{erf}\left( \frac{h-x}{2\sqrt{Dt}} \right) + \mathrm{erf}\left(\frac{h+x}{2\sqrt{Dt}}  \right) \right] e^{-r_{2}t}. 
	\end{alignat}
\end{subequations}
Using $c(t,x)=Sb(t,x)$, we transform $b(t,x)$ to obtain the solution for $c(t,x)$,
\begin{subequations}\label{eqn:supp_liearpdes10}
		\renewcommand{\theequation}{\theparentequation.\arabic{equation}}
		\begin{alignat}{1}
			c_{1}(t,x) & = \frac{C_{0}}{2}  \left[ \mathrm{erf}\left( \frac{h-x}{2\sqrt{Dt}} \right) + \mathrm{erf}\left(\frac{h+x}{2\sqrt{Dt}}  \right) \right] \exp({-r_{1}t}), \\
			c_{2}(t,x) & = \left(\frac{r_{1}}{r_{2}-r_{1}}\right)\frac{C_{0}}{2}  \left[ \mathrm{erf}\left( \frac{h-x}{2\sqrt{Dt}} \right) + \mathrm{erf}\left(\frac{h+x}{2\sqrt{Dt}}  \right) \right] \bigg( \exp({-r_{1}t}) - \exp({-r_{2}t}) \bigg).
		\end{alignat}
	\end{subequations}
Solutions in Eq (\ref{eqn:supp_liearpdes10}) are restricted to $r_{1} \neq r_{2}$. An analytical solution can also be found for $r_{1}=r_{2}$.

\newpage
\section{Numerical solution of system of partial differential equations} \label{sec:supp_numericalpde}

The framework for parameter estimation, identifiability analysis, and prediction in the main manuscript can be applied to mathematical models that are solved analytically and/or numerically. Here we present an explicit example using the system of partial differential equations in Eq (19) that form the mathematical biology case study.

\begin{figure}[h!]
	\centering
	\includegraphics[width=\textwidth]{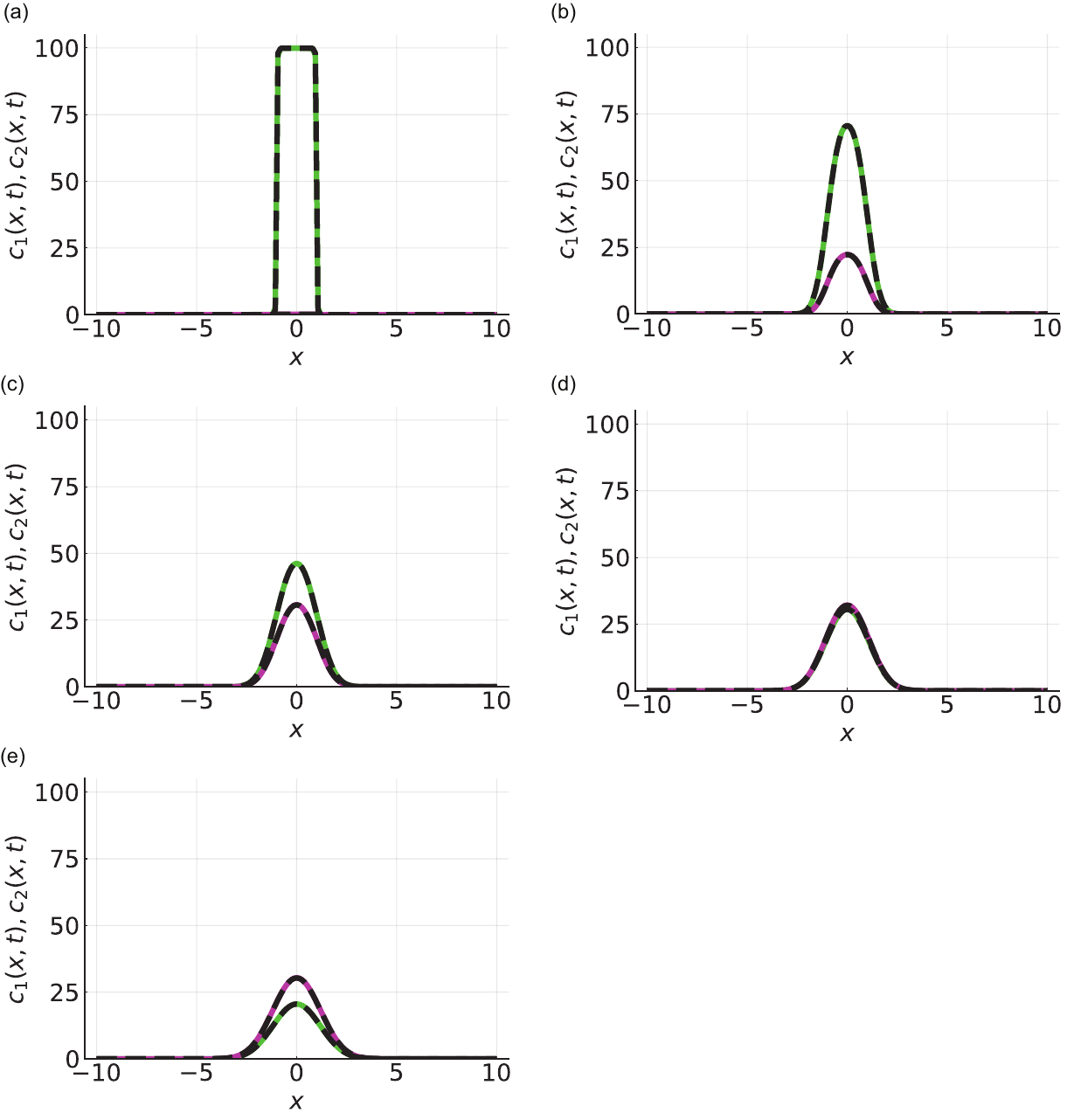}
	\caption{Analytical and numerical solutions of the system of partial differential equations in Eq (19) show excellent agreement. Results shown at (a) $t=0.001$, (b) $t=0.250$, (c) $t=0.500$, (d) $t=0.750$, and (e) $t=1.000$, for parameter values $(D, r_{1}, r_{2}, C_{0}(0), h, L)=(0.5, 1.2, 0.8, 100.0, 1.0, 10.0)$. Results show numerical solution of $c_{1}(t)$ (green), numerical solution of $c_{2}(t)$ (magenta), and analytical solutions (black-dashed).}
	\label{fig:figPDEs_numericalandanalytical}
\end{figure}

In the main manuscript we solve Eq (19) analytically. We obtain the same profile log-likelihoods, confidence sets for model solutions and confidence sets for model realisations when solving Eq (19) numerically. This is because the output of the mathematical model is independent of the solution method. In particular, comparing the analytical and numerical solutions of Eq (19) we observe excellent agreement (Fig \ref{fig:figPDEs_numericalandanalytical}). This is a useful result. We often require numerical methods to solve systems of partial differential equations. 

In the following we present numerical methods to solve Eq (19). Rewriting Eq (19) gives,
	\begin{equation}\label{eqn:supp2_linearpdes}
		\begin{split}
			\frac{\partial c_{1}(t,x)}{\partial t}    &=  D\frac{\partial^{2} c_{1}(t,x)}{\partial x^{2}} -r_{1}c_{1}(t,x),\\
			\frac{\partial c_{2}(t,x)}{\partial t}  &= D\frac{\partial^{2} c_{2}(t,x)}{\partial x^{2}}  + r_{1}c_{1}(t,x) - r_{2}c_{2}(t,x).
		\end{split}
	\end{equation}
To numerically solve Eq (\ref{eqn:supp2_linearpdes}) we consider a truncated spatial domain $- L < x < L$, where $L$ is chosen sufficiently large such that the time evolution of $c_{1}(t,x)$ and $c_{2}(t,x)$ is not impacted by the boundary condition applied on the truncated domain. In agreement with the main manuscript, $D$ represents a constant diffusivity and initial conditions are chosen to represent the release of chemical $C_{1}$ from a confined region, 
\begin{subequations}\label{eqn:supp2_linearpdes_ics}
		\renewcommand{\theequation}{\theparentequation.\arabic{equation}}
		\begin{alignat}{1}
			c_{1}(0,x)    &= \begin{cases}
				C_{0},	\quad  \vert x \rvert < h,\\
				0,  \quad    \vert x \rvert > h,
			\end{cases}\\ 
			c_{2}(0,x)   &= 0, \quad -\infty < x < \infty.
		\end{alignat}
	\end{subequations}
We solve Eqs (\ref{eqn:supp2_linearpdes})-(\ref{eqn:supp2_linearpdes_ics}) numerically using the default solver in the \texttt{DifferentialEquations.jl} \cite{JuliaDifferentialequations} together with the \texttt{ModelingToolkit} \cite{JuliaModelinToolkit}, \texttt{MethodOfLines} \cite{JuliaMethodOfLines}, \texttt{DomainSets} \cite{JuliaDomainSets} packages in Julia. We set $L=10$, discretise the domain with $151$ equally--spaced nodes, and impose Neumann boundary conditions at $x=\pm L$.

\newpage
\clearpage
\section{Comparison to a full likelihood-based approach}\label{sec:supp_fulllikelihood}

Here we present an example comparing results from the profile log-likelihood-based method described in Section 2 of the main manuscript and a gold-standard full likelihood-based method that we describe here \cite{Simpson2023}. Consider the two-parameter two-species chemical reaction model from Section 4 of the main manuscript. This is given by Eq (16) with the additive Gaussian error model. We fix the measurement error model parameter ($\sigma_{N}=5$) and initial conditions $(c_{1}(0), c_{2}(0))=(100, 25)$. This results in a model with two unknown parameters, $\theta=(r_{1}, r_{2})$, that we will estimate. We choose to fix $\sigma_{N}$ and the initial conditions in this illustrative example as it is simpler to interpret and visualise results in two dimensions.

\begin{figure}[h!]
	\centering
	\includegraphics[width=0.75\textwidth]{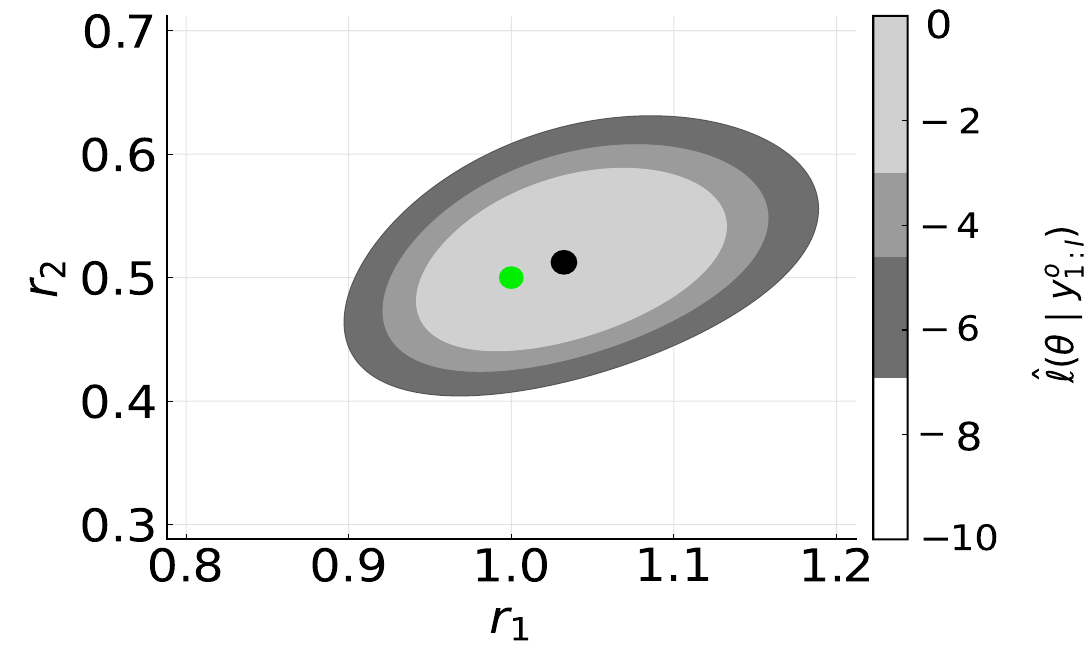}
	\caption{Normalised log-likelihood in $(r_{1}, r_{2})$ parameter space using a full likelihood-based approach.  Shaded regions represent $95\%$ (light grey), $99\%$ (medium grey), and $99.9\%$ (dark grey) confidence regions, given by the threshold values $-3.00$, $-4.61$, and $-6.91$ for the two-dimensional full parameter $\theta$, respectively. The true parameter values used to generate the data, $\theta=(1.00,0.50)$ (green circle), and MLE $\hat{\theta}=(r_{1}, r_{2})=(1.03, 0.51)$ (black circle). }
	\label{fig:FigSupp_FullLikelhood_ODELinear_NormalNoise_NormalFit_df2}
\end{figure}

We generate synthetic data for $\theta=(r_{1}, r_{2})=(1.0, 0.5)$ (this is the same data as shown in Fig 1a). Computing the MLE, we obtain $\hat{\theta}=(r_{1}, r_{2})=(1.03, 0.51)$. We now take a full likelihood-based approach and work with the normalised log-likelihood (Eq (8). We discretise $(r_{1}, r_{2})$-parameter space uniformly into a two-dimensional grid with $201 \times 201$ points. Parameter bounds, $0.8 \leq r_{1} \leq 1.2$ and $0.3 \leq r_{2} \leq 0.7$, are chosen to ensure that we capture the $95\%$ confidence region for $\theta$ and so that the true value used to generate the data is at the centre of the domain. At each of the $40,401$ points we evaluate the normalised log-likelihood (Fig \ref{fig:FigSupp_FullLikelhood_ODELinear_NormalNoise_NormalFit_df2}).  Contour lines in Fig \ref{fig:FigSupp_FullLikelhood_ODELinear_NormalNoise_NormalFit_df2} represent threshold values $-3.00$, $-4.61$, and $-6.91$ corresponding to $95.0\%$, $99.0\%$, and $99.9\%$ approximate confidence intervals for $\theta$. Each threshold, $\ell_{c}$, is calibrated using the chi-squared distribution with two degrees of freedoms since $\theta$ is two-dimensional (i.e. $\ell_{c}=-\Delta_{\nu,1-\alpha}/2$, where $\Delta_{\nu,1-\alpha}$ refers to the $(1-\alpha)$ quantile of a chi-squared distribution with $\nu$ degrees of freedom) \cite{Pawitan2001,Royston2007}.  The MLE, $\hat{\theta}=(r_{1}, r_{2})=(1.03, 0.51)$ (black circle) and true parameter value, $\theta=(r_{1}, r_{2})=(1.00, 0.50)$ (green circle), are both contained within the $95\%$ confidence region for $\theta$.

\newpage

\begin{figure}[p]
	\centering
	\includegraphics[width=\textwidth]{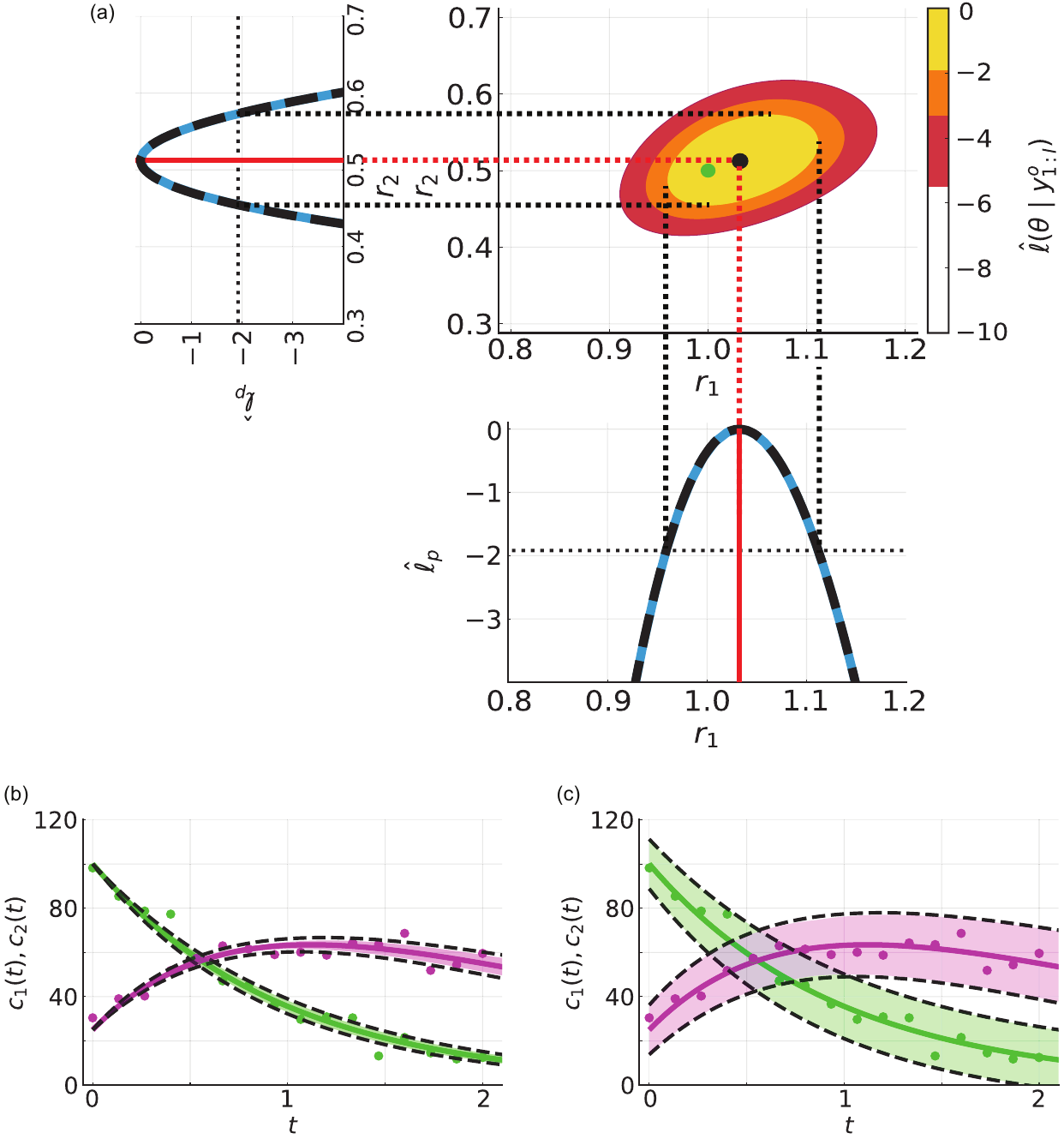}
	\caption{Comparing the profile likelihood-based approach with a full likelihood-based approach we observe excellent agreement. (a) Normalised log-likelihood in $(r_{1}, r_{2})$ parameter space. Shaded regions represent $95\%$ (yellow), $99\%$ (orange), and $99.9\%$ (red) confidence regions, given by the threshold values $-1.92$, $-3.32$, and $-5.51$ for univariate parameters, respectively. The true parameter values used to generate the data, $\theta=(1.00,0.50)$ (green circle), and MLE $\hat{\theta}=(r_{1}, r_{2})=(1.03, 0.51)$ (black circle). The profile log-likelihood for $r_{1}$ shows results obtained from optimisation procedures in the profile likelihood method (blue) and from maximising over the grid of the normalised log-likelihood values (black-dashed). Vertical red dashed corresponds to the MLE and the horizontal black-dashed line represents the $95\%$ confidence interval threshold for a univariate parameter. Similarly, for the profile likelihood for $r_{2}$. (b) Profile likelihood-based union of profile-wise confidence sets for the model solution (shaded) and full likelihood-based confidence set for model solution (black-dashed). (c) Profile log-likelihood-based union of Bonferroni correction-based profile-wise confidence sets for data realisations (shaded) and full likelihood-based Bonferroni correction-based confidence set for data realisations (black-dashed). Throughout results from the profile log-likelihood-based approach agree closely with the results from the full likelihood approach.}
	\label{fig:FigSupp_FullLikelhood_ODELinear_NormalNoise_NormalFit}
\end{figure}

For comparison with profile log-likelihood-based methods we plot the normalised log-likelihood with contour lines representing the threshold values $-1.92$, $-3.32$, and $-5.51$ corresponding to $95.0\%$, $99.0\%$, and $99.9\%$ approximate confidence intervals for the univariate interest parameters (Fig \ref{fig:FigSupp_FullLikelhood_ODELinear_NormalNoise_NormalFit}a). Each threshold is calibrated using the chi-squared distribution with one degree of freedom. The MLE, $\hat{\theta}=(r_{1}, r_{2})=(1.03, 0.51)$ (black circle) and true parameter value, $\theta=(r_{1}, r_{2})=(1.00, 0.50)$ (green circle), are both contained within the region defined by the approximate $95\%$ approximate confidence interval threshold for a univariate parameter. Furthermore, due to the fine mesh, profile log-likelihoods obtained by maximising over the normalised log-likelihood values in the grid and by optimisation procedures (Section 2 of the main manuscript) show excellent agreement (Fig \ref{fig:FigSupp_FullLikelhood_ODELinear_NormalNoise_NormalFit}a).

Proceeding with the full likelihood-based approach, we simulate the model solution at all points in $(r_{1}, r_{2})$ parameter space where $\hat{\ell}(\theta \mid y_{1:I}^{\mathrm{o}}) \geq -3.00$. In Fig \ref{fig:FigSupp_FullLikelhood_ODELinear_NormalNoise_NormalFit}b, we present the minimum and maximum of these projections using black-dashed lines and observe qualitatively excellent agreement with the union of the profile log-likelihood-based confidence sets for the model solutions (shaded). Incorporating measurement noise, we observe qualitatively excellent agreement between Bonferroni correction-based confidence sets for data realisations obtained from the full likelihood-based approach and the union of Bonferroni correction-based profile-wise confidence sets for data realisations from the profile log-likelihood-based method (Fig \ref{fig:FigSupp_FullLikelhood_ODELinear_NormalNoise_NormalFit}c).

\clearpage

\begin{figure}[h!]
	\centering
	\includegraphics[width=0.85\textwidth]{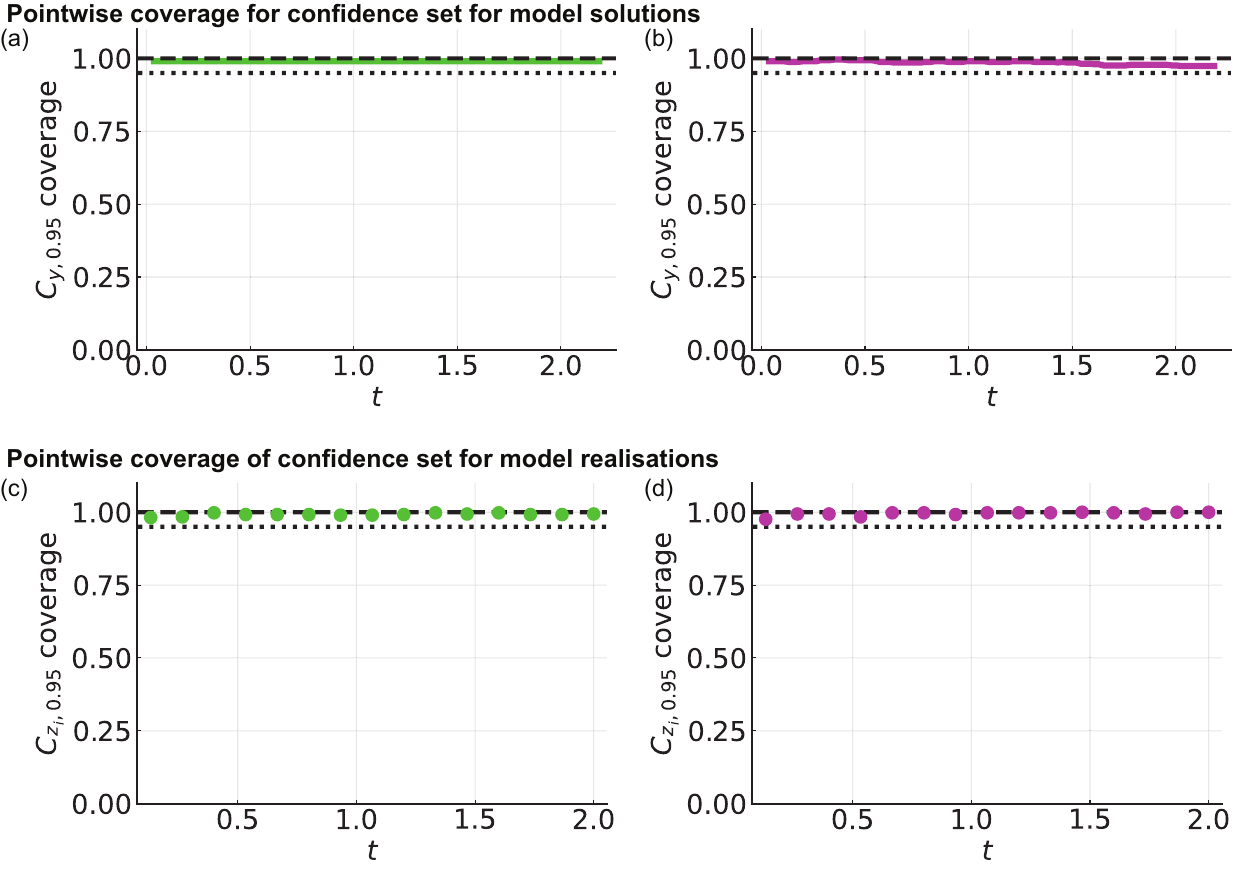}
	\caption{Pointwise coverage analysis for full likelihood-based confidence sets for model solutions and Bonferroni correction-based confidence sets for model realisations. Analysis performed using the two-parameter two-chemical reaction model as an illustrative example. Pointwise coverage of confidence sets for model solutions for (a) $c_{1}(t)$ and (b) $c_{2}(t)$. The temporal domain is discretised into $100$ equally--spaced points ($0.022 \leq t \leq 2.200$). Pointwise coverage of confidence sets for model realisations for (c) $c_{1}(t)$ and (d) $c_{2}(t)$. Results are obtained by analysing $500$ data sets generated by simulating Eq (16), the additive Gaussian measurement error model with $\sigma_{\mathrm{N}}=5.0$, known model parameters $(r_{1}, r_{2}) = (1.0, 0.5)$, and fixed initial conditions $(c_{1}(0), c_{2}(0))=(100.0, 25.0)$.  Horizontal dotted and horizontal dashed lines correspond to coverage probabilities of $0.95$ and $1.00$, respectively.}
	\label{fig:FigSupp_Coverage_FullLikelihood}
\end{figure}

Statistical coverage properties can be evaluated for the confidence sets numerically. We generate $5000$ synthetic data sets. For each data set we compute the $95\%$ confidence region for $r_{1}$ and $r_{2}$ and test whether the true model parameter is contained within the region. This gives an observed coverage probability of $0.950$ which is very close to the target coverage probability of $0.950$. For each data set we also construct a $95\%$ confidence set for the model solution and test whether the true model solution is entirely contained within the confidence set. This gives an observed curvewise coverage probability of $0.954$, which is much greater than results obtained using the profile log-likelihood-based methods (Section 4 in the main manuscript). To compare with results from the profile log-likelihood-based approach we also compute the pointwise coverage of the model solutions for $c_{1}(t)$ and $c_{2}(t)$ (Fig \ref{fig:FigSupp_Coverage_FullLikelihood}). Note that confidence sets for the model solutions from the profile log-likelihood-based method and the full likelihood-based method appear to agree very well qualitatively (Fig \ref{fig:FigSupp_FullLikelhood_ODELinear_NormalNoise_NormalFit}b) but observed differences in observed curvewise and pointwise coverage do not agree (Fig \ref{fig:FigSupp_Coverage_FullLikelihood}a,b, 8c,f). This suggests that subtle differences in confidence sets can result in drastic changes to coverage properties. Using the full likelihood-based method, the average observed pointwise coverage of the $95\%$ Bonferroni correction-based confidence set for model realisations is found to be conservative and equal to $99.4\%$ (Fig \ref{fig:FigSupp_Coverage_FullLikelihood}c,d).

\clearpage
\newpage
\section{Additional results}\label{sec:appendix_additional_results}

In the main manuscript we demonstrate the framework for systems of ordinary differential equations and partial differential equations. Here we demonstrate the framework with different models including the Lotka-Volterra model (Supplementary \cref{sec:supp_predatorprey}) and a discrete-time population growth model (Supplementary \cref{sec:appendix_discretetime}).

\subsection{System of ordinary differential equations: Predator-prey}\label{sec:supp_predatorprey}

Here, we demonstrate that the framework works well for systems of ordinary differential equations that give rise to oscillatory solutions. We consider the Lotka-Volterra predator-prey model for two chemical species $C_{1}$ (the `prey') and $C_{2}$ (the `predator') \cite{Britton2005,Murray2002},
\begin{equation}\label{eqn:lotkavolterra}
	\begin{split}
		\frac{\mathrm{d}c_{1}(t)}{\mathrm{d}t} &= V_{1}c_{1}(t) - K_{1}c_{1}(t)c_{2}(t),\\
		\frac{\mathrm{d}c_{2}(t)}{\mathrm{d}t} &= V_{2}c_{1}(t)c_{2}(t) - K_{2}c_{2}(t).
	\end{split}
\end{equation}
where $c_{1}(t)$ and $c_{2}(t)$ represent the concentrations of $C_{1}$ and $C_{2}$; and $V_{1}, K_{1}, V_{2}, K_{2}$ are positive constants. We treat initial conditions $c_{1}(0)$ and $c_{2}(0)$ as known. Then Eq (\ref{eqn:lotkavolterra}) is characterised by four parameters $\theta=(V_{1}, K_{1}, V_{2}, K_{2})$. For parameter estimation we solve Eq (\ref{eqn:lotkavolterra}) numerically using the default \texttt{ODEproblem} solver in Julia (\texttt{DifferentialEquations} package) \cite{JuliaDifferentialequations}.

\begin{figure}[h!]
	\centering
	\includegraphics[width=\textwidth]{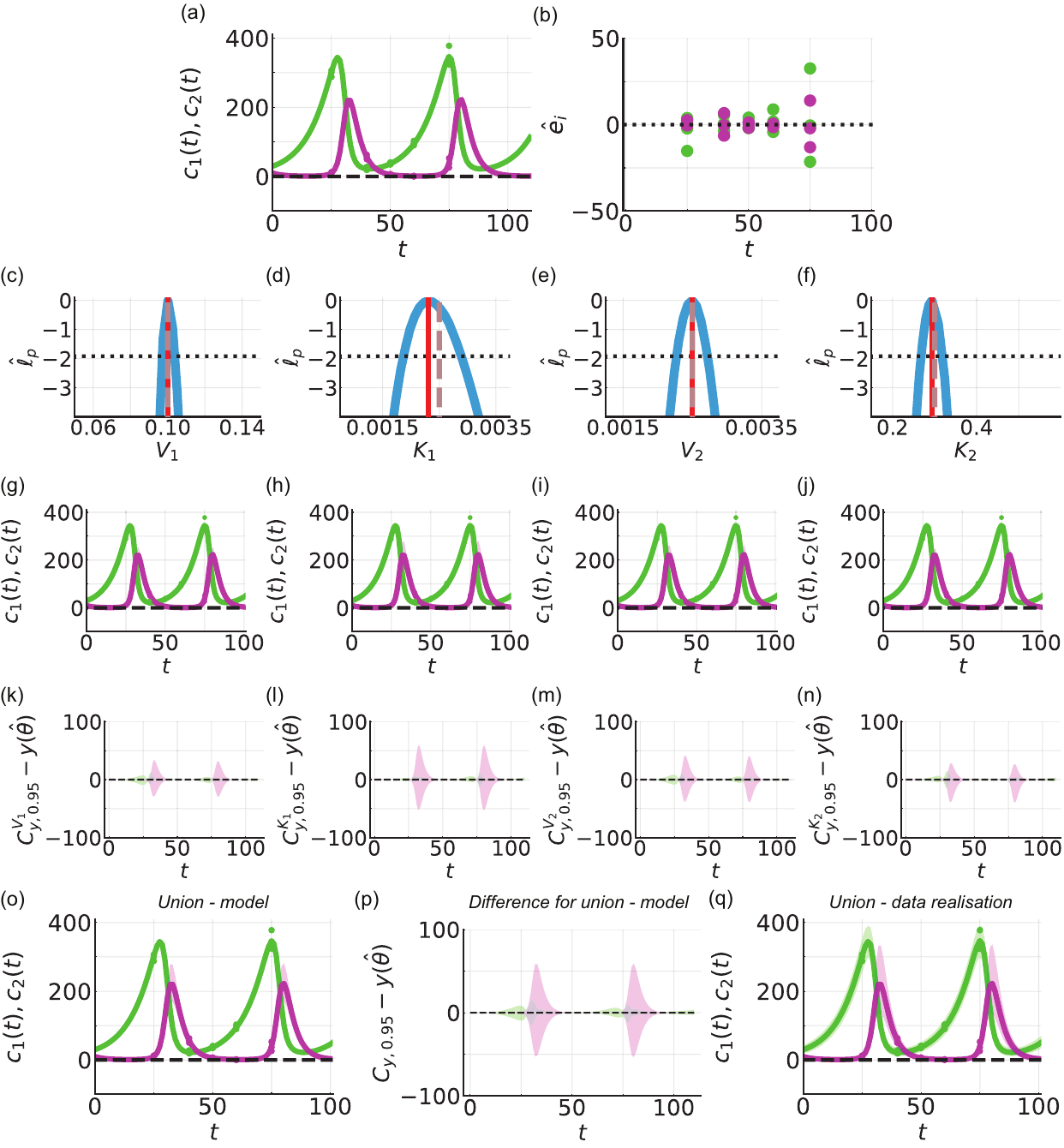}
	\caption{Lotka-Volterra predator-prey case study. (a) Synthetic data (circles) generated by simulating Eq (\ref{eqn:lotkavolterra}) and the Poisson measurement error model with known model parameters $\theta=(V_{1}, K_{1}, V_{2}, K_{2}) = (0.1000, 0.0025, 0.0025, 0.3000)$, and initial conditions $(c_{1}(0), c_{2}(0))=(30.0, 10.0)$. (b) Residuals $\hat{e}_{i}=y_{i}^{\mathrm{o}}-y_{i}(\hat{\theta})$ with time, $t$. (c-f) Profile log-likelihoods for (c) $V_{1}$, (d) $K_{1}$, (e) $V_{2}$, and (f) $K_{2}$, with MLE (red-dashed), an approximate $95\%$ confidence interval threshold (horizontal black-dashed) and known model parameters (vertical brown dashed). (g-j,o) Confidence sets for model solutions generated by propagating forward uncertainty in (g) $V_{1}$, (h) $K_{1}$, (i) $V_{2}$, (j) $K_{2}$, and (o) their union. (k-n,p) Difference between confidence set for model solution and model solution at MLE. (q) Union of confidence sets for model realisations.}
	\label{fig:figA2}
\end{figure}

We generate synthetic data using Eq (\ref{eqn:lotkavolterra}), the Poisson measurement error model, model parameters, $\theta=(V_{1}, K_{1}, V_{2}, K_{2}) = (0.1000, 0.0025, 0.0025, 0.3000)$, and initial conditions $(c_{1}(0), c_{2}(0))=(30.0, 10.0)$ (Fig \ref{fig:figA2}a). We consider three measurements of $c_{1}(t)$ and three measurements of $c_{2}(t)$ at $t=25$, $t=40$, $t=50$, $t=60$, and $t=75$. These parameters and time points are chosen deliberately so that residuals are not normally distributed with zero mean and constant variance (Fig \ref{fig:figA2}b). Using Eq (\ref{eqn:lotkavolterra}) and the Poisson measurement error model, we seek estimates of $\theta=(V_{1}, K_{1}, V_{2}, K_{2})$ and generate predictions.  Simulating the mathematical model with the MLE, we observe excellent agreement with the data (Fig \ref{fig:figA2}a). Profile log-likelihoods for $V_{1}$, $K_{1}$, $V_{2}$, and $K_{2}$ capture known parameter values and show that these parameters are practically identifiable (Fig \ref{fig:figA2}c-f). Predictions, in the form of the confidence sets for model solutions (Fig \ref{fig:figA2}k-p) and confidence sets for data realisations (Fig \ref{fig:figA2}q) show greater uncertainty at the peaks of the oscillations. 

\begin{figure}[h!]
	\centering
	\includegraphics[width=\textwidth]{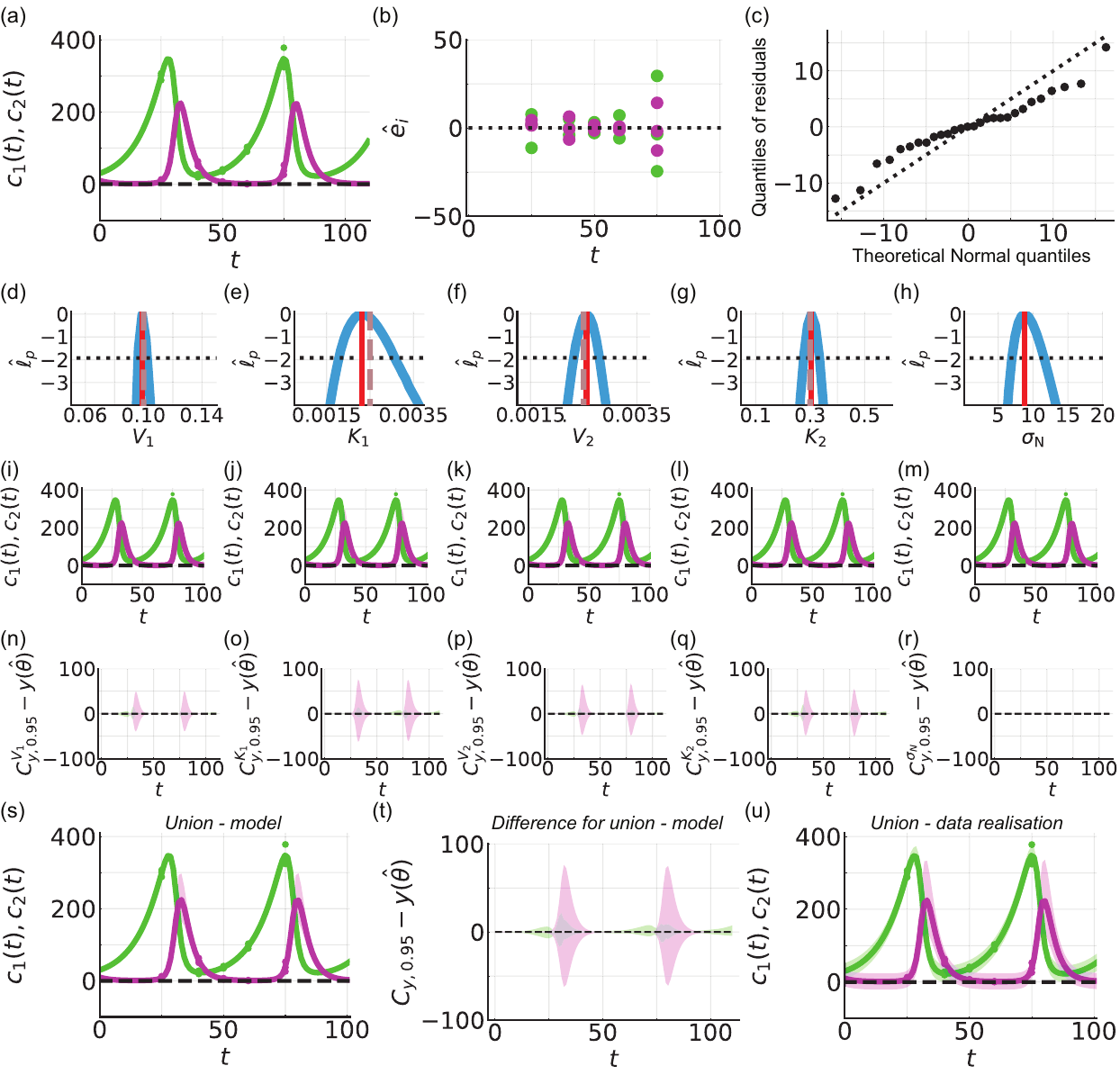}
	\caption{Lotka-Volterra predator-prey case study with deliberate measurement error model misspecification. (a) Synthetic data (circles) generated by simulating Eq (\ref{eqn:lotkavolterra}) and the Poisson measurement error model with known model parameters $\theta=(V_{1}, K_{1}, V_{2}, K_{2}) = (0.1000, 0.0025, 0.0025, 0.3000)$, and initial conditions $(c_{1}(0), c_{2}(0))=(30.0, 10.0)$. Analysis performed using the additive Gaussian measurement error model. (b) Residuals  $\hat{e}_{i}=y_{i}^{\mathrm{o}}-y_{i}(\hat{\theta})$ with time, $t$. (c) normal quantile-quantile plot of residuals. (d-h) Profile log-likelihoods for (d) $V_{1}$, (e) $K_{1}$, (f) $V_{2}$, (g) $K_{2}$, and (h) $\sigma_{N}$ with MLE (red-dashed), an approximate $95\%$ confidence interval threshold (horizontal black-dashed) and known model parameters (vertical brown dashed). (i-m,s) Confidence sets for model solutions generated by propagating forward uncertainty in (i) $V_{1}$, (j) $K_{1}$, (k) $V_{2}$, (l) $K_{2}$, (m) $\sigma_{N}$ and (r) their union. (n-r,t) Difference between confidence set for model solution and model solution at MLE. (u) Union of confidence sets for model realisations.}
	\label{fig:figA3}
\end{figure}

We now repeat this analysis and deliberately misspecify the measurement error model. We use the additive Gaussian measurement error model and find that this leads to non-physical predictions. Simulating the mathematical model with the MLE, we observe good agreement with the data (Fig \ref{fig:figA3}a). Profile log-likelihoods for $V_{1}$, $K_{1}$, $V_{2}$, $K_{2}$, $\sigma_{N}$ capture known parameter values and show that these parameters are practically identifiable (Fig \ref{fig:figA3}c,d). Furthermore, profile log-likelihoods using the additive Gaussian error model are qualitatively similarly to profile log-likelihoods obtained using the true Poisson error model. Predictions, in the form of the confidence sets for model solutions (Fig \ref{fig:figA3}i-m,s) and realisations (Fig \ref{fig:figA3}u), show greater uncertainty than results obtained using the Poisson error model. Confidence sets for data realisations give non-physical results with negative concentrations.

\clearpage

\clearpage
\newpage
\subsection{Difference equations}\label{sec:appendix_discretetime}

In the main manuscript we present case studies using ordinary differential equations and partial differential equations. Here, we present an example demonstrating that the framework also naturally handles difference equations that frequently appear in ecological applications \cite{Ricker1954,Hefley2013a,AugerMethe2021,Valpine2002,Murray2002} 

As a simple caricature example consider the discrete-time Ricker logistic model for population growth \cite{Ricker1954}. In this model $N_{t}$ represents the population at time $t$ and the population at the next time point, $t+1$, is
\begin{equation}\label{eqn:Rickermodel}
	N_{t+1} = N_{t} \exp\left[ r\left(1 - \frac{N_{t}}{K} \right)\right],
\end{equation}
where $r$ is the maximum intrinsic growth rate and $K$ is the carrying capacity. We treat the initial condition $N_{0}$ as known. Then Eq (\ref{eqn:Rickermodel}) is characterised by two parameters $\theta=(r, K)$ that we will estimate.

\begin{figure}[h!]
	\centering
	\includegraphics[width=0.95\textwidth]{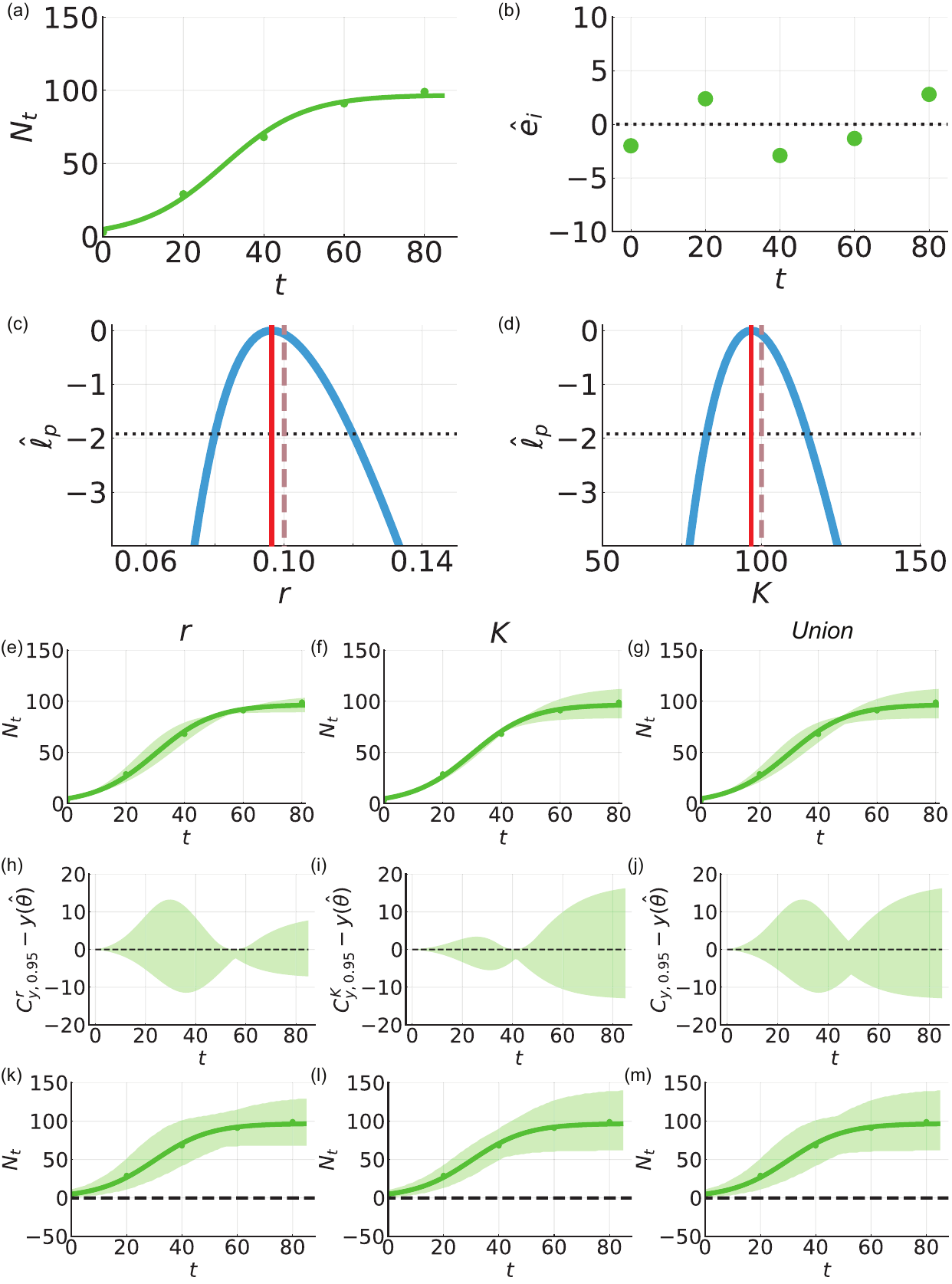}
	\caption{Discrete-time Ricker logistic model describing population growth. (a) Synthetic data (circles) at five equally--spaced time points from $t=0.0$ to $t=80.0$ are generated by simulating Eq (16), the Poisson measurement error model, known model parameters $(r, K) = (0.1, 100.0)$, and initial condition $N_{0} = 5.0$. The MLE is computed assuming an Poisson measurement error model to be $(r, K) = (0.096, 95.830)$. Eq (\ref{eqn:Rickermodel}) simulated with the MLE (solid). (b) Residuals $\hat{e}_{i}=y_{i}^{\mathrm{o}}-y_{i}(\hat{\theta})$ with time $t$. (c-d) Profile log-likelihoods (blue) for (c) $r$ and (d) $K$ with MLE (red-dashed), an approximate $95\%$ confidence interval threshold (horizontal black-dashed) and know model parameters (vertical brown dashed). (e-g) Profile-wise confidence sets for the model solution (e) $r$, (f) $K$, and (g) their union. (h-j) Difference between confidence set and mathematical model simulated at the MLE. (k-m) Confidence sets for the model realisations (k) $r$, (l) $K$, and (m) their union.}
	\label{fig:figB1}
\end{figure}

We generate synthetic data using Eq (\ref{eqn:Rickermodel}), the Poisson measurement error model, model parameters, $\theta=(r, K) = (0.1, 100.0)$, and initial condition $N_{0}=5.0$ (Fig \ref{fig:figB1}a). Using Eq (\ref{eqn:Rickermodel}) and the Poisson measurement error model, we seek estimates of $\theta=(r, K)$ and generate predictions.  Simulating the mathematical model with the MLE, we observe excellent agreement with the data (Fig \ref{fig:figB1}a). Profile log-likelihoods for $r$ and $K$ capture known parameter values and show that these parameters are practically identifiable (Fig \ref{fig:figB1}c,d). Predictions, in the form of the confidence sets for model solutions (Fig \ref{fig:figB1}e-j)  and confidence sets for realisations (Fig \ref{fig:figB1}k-m), demonstrate how uncertainty in parameter estimates results propagates forward into uncertainty in $N_{t}$. The framework can also be applied to systems of difference equations.

\clearpage

\newpage
\clearpage
\section{Coverage: Log-normal measurement error model}\label{sec:supp_coverage}

In Section 4 of the main manuscript we explore coverage properties using an example that considers the additive Gaussian measurement error model. Here we show that the same evaluation procedure can be used to assess coverage properties for different measurement error models. As an illustrative example consider Eq (16) with the log-normal error model. After fixing $\sigma_{L}=0.4$, this results in a model with two parameters, $\theta=(r_{1}, r_{2})=(1.0, 0.5)$, that we estimate. Initial conditions $(c_{1}(0), c_{2}(0))=(100, 10)$ are fixed. Analysing the coverage properties of this model gives similar results to those described in Section 4 of the main manuscript.

We generate $5000$ synthetic data sets using the same mathematical model, measurement error model, and model parameters, $\theta$. Each data set comprises measurements of $c_{1}(t)$ and $c_{2}(t)$ at thirty--one equally--spaced time points from $t=0.0$ to $t=5.0$. For each data set we compute a univariate profile log-likelihood for $r_{1}$ and use this to form an approximate $95.0\%$ confidence interval for $r_{1}$. We then test whether this approximate $95.0\%$ confidence interval contains the true value of $r_{1}$. This holds for $95.0\%$ of the data sets, corresponding to an observed coverage probability of $0.950$. Similarly, the observed coverage probability for $r_{2}$ is $0.948$. Therefore, the observed coverage probabilities for both $r_{1}$ and $r_{2}$ are close to the target coverage probability of $0.950$. In contrast to our profile-wise coverage approach, a full likelihood-based approach, also using $5000$ synthetic data sets, recovers an observed coverage probability of $0.951$ for the confidence region for $r_{1}$ and $r_{2}$.

For Bonferroni correction-based confidence sets for model solutions for $r_{1}$, $r_{2}$, and their union we observe curvewise coverage probabilities of $0.0004$, $0.0800$, and $0.7234$, respectively. These are much lower than the observed coverage probabilities of model parameters. Pointwise coverage properties per time point and chemical species for these confidence sets are shown in Fig \ref{fig:FigSupp_Coverage3_LogNormal}a-h. A full likelihood-based approach recovers an observed curvewise coverage probability of $0.961$ for the confidence set for model solutions (Supplementary \cref{sec:supp_fulllikelihood}). Pointwise coverage properties per time point and chemical species for the full likelihood-based confidence set for model solutions is shown in Fig \ref{fig:FigSupp_Coverage_FullLikelihood_LogNormal}a-b.

For Bonferroni correction-based confidence sets for model realisations for $r_{1}$, $r_{2}$, and their union the average pointwise coverage probabilities are $0.981$ for $r_{1}$, $0.980$ for $r_{2}$, $0.987$ for their union. For the MLE-based confidence set for model realisations we observe an average pointwise coverage probability of $0.946$. Pointwise coverage properties per time point and chemical species for these confidence sets for model realisations are shown in Fig \ref{fig:FigSupp_Coverage3_LogNormal}i-p. A full likelihood-based approach recovers an average pointwise coverage probability of $0.990$. Pointwise coverage properties per time point and chemical species for the full likelihood-based confidence sets for model realisations is shown in Fig \ref{fig:FigSupp_Coverage_FullLikelihood_LogNormal}c-d.

\begin{figure}[h!]
	\centering
	\includegraphics[width=\textwidth]{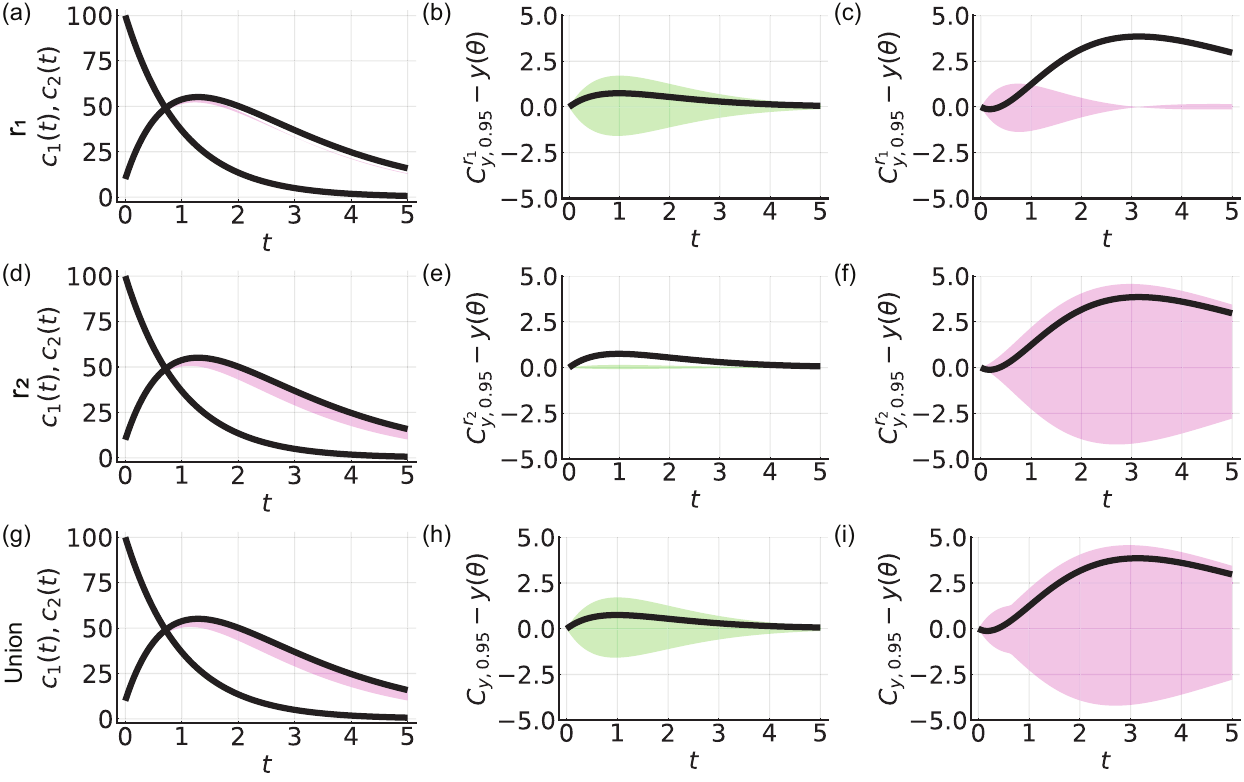}
	\caption{Curvewise confidence sets for the model solutions of a caricature ODE model with linear reaction terms (Eq 16) and the log-normal measurement error model with known $\sigma_{L}$. (a) Confidence sets for model solution generated from uncertainty in $r_{1}$, $C_{y,0.95}^{r_{1}}$ (shaded), and the true model solution, $y(\theta)$ (black). (b)-(c) Difference between curvewise confidence set and solution of the mathematical model evaluated at the MLE, $C_{y,0.95}^{r_{1}}-y(\hat{\theta})$ ($c_{1}(t)$ (shaded green) and $c_{2}(t)$ (shaded magenta) and the difference between the true model solution and the solution of the mathematical model evaluated at the MLE, $y(\theta) - y(\hat{\theta})$ (black). (f)-(h) Results based on uncertainty in $r_{2}$. (i)-(k) Results for the union of curvewise confidence sets. Throughout, to plot $y(\theta)$ the temporal domain is discretised into $101$ equally--spaced points ($0.00 \leq t \leq 5.00$) connected using a solid line. Results are obtained by analysing a single data set generated by simulating Eq (16), the log-normal measurement error model with $\sigma_{{L}}=0.4$, known model parameters $(r_{1}, r_{2}) = (1.0, 0.5)$, and fixed initial conditions $(c_{1}(0), c_{2}(0))=(100.0, 10.0)$.}	\label{fig:FigSupp_Coverage1_LogNormal}
\end{figure}

\begin{figure}[h!]
	\centering
	\includegraphics[width=\textwidth]{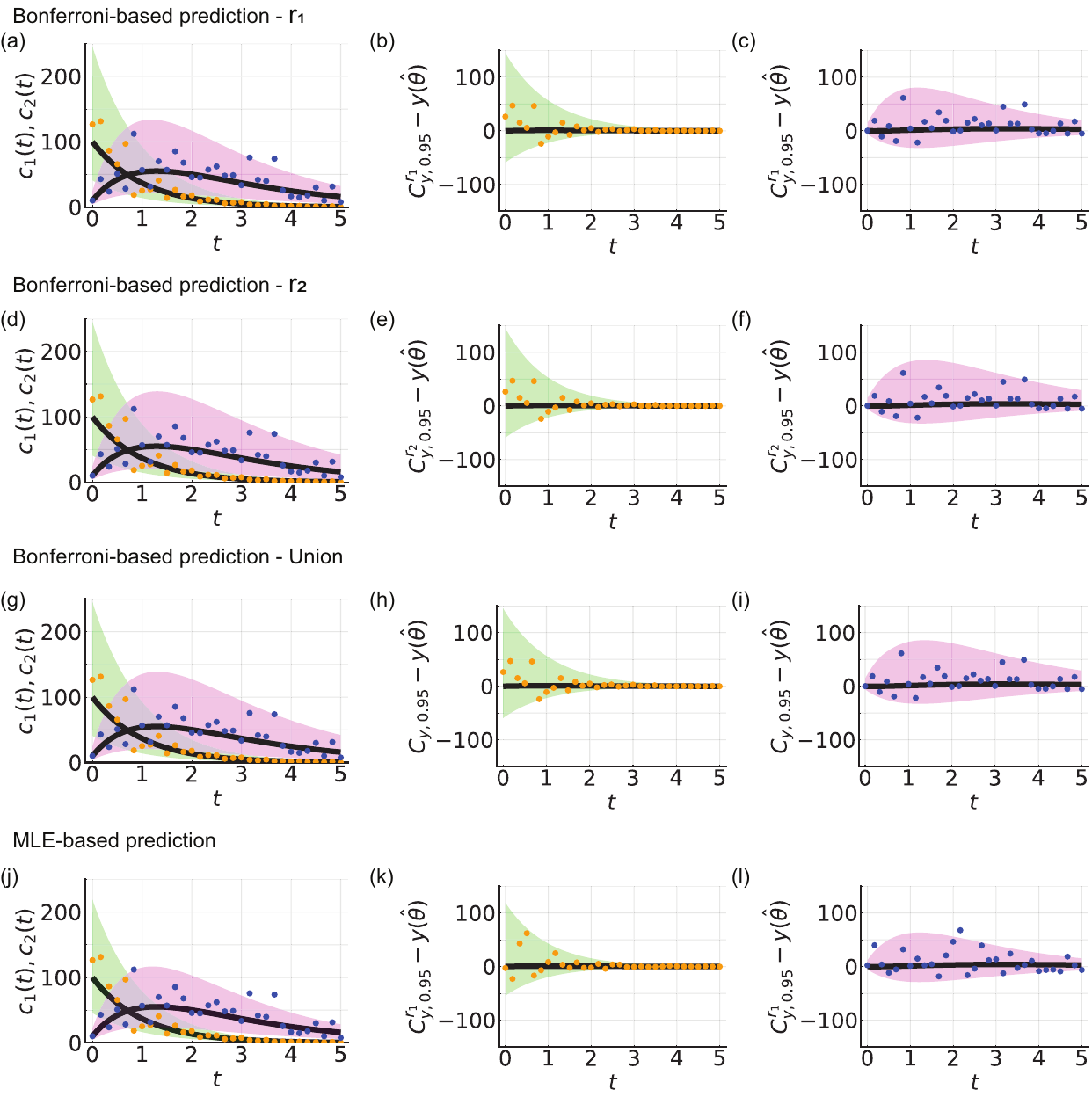}
	\caption{Confidence sets for model realisations of a caricature ODE model with linear reaction terms (Eq 16) and the log-normal measurement error model with known $\sigma_{L}$. (a) Bonferroni correction-based confidence set for $r_{1}$. (b)-(c) Difference between the Bonferroni correction-based confidence set for $r_{1}$ and solution of the mathematical model evaluated at the MLE, $C_{y,0.95}^{r_{1}}-y(\hat{\theta})$ ($c_{1}(t)$ (shaded green) and $c_{2}(t)$ (shaded magenta) and the difference between the true model solution and the solution of the mathematical model evaluated at the MLE, $y(\theta) - y(\hat{\theta})$ (black). (d-i) Results for Bonferroni correction-based confidence sets for (d-f) $r_{2}$ and (g-i) the union. (j-l) Results for MLE-based confidence set.}	\label{fig:FigSupp_Coverage1_MLE_LogNormal}
\end{figure}

 \begin{figure}[h]
 	\centering
 	\includegraphics[width=\textwidth]{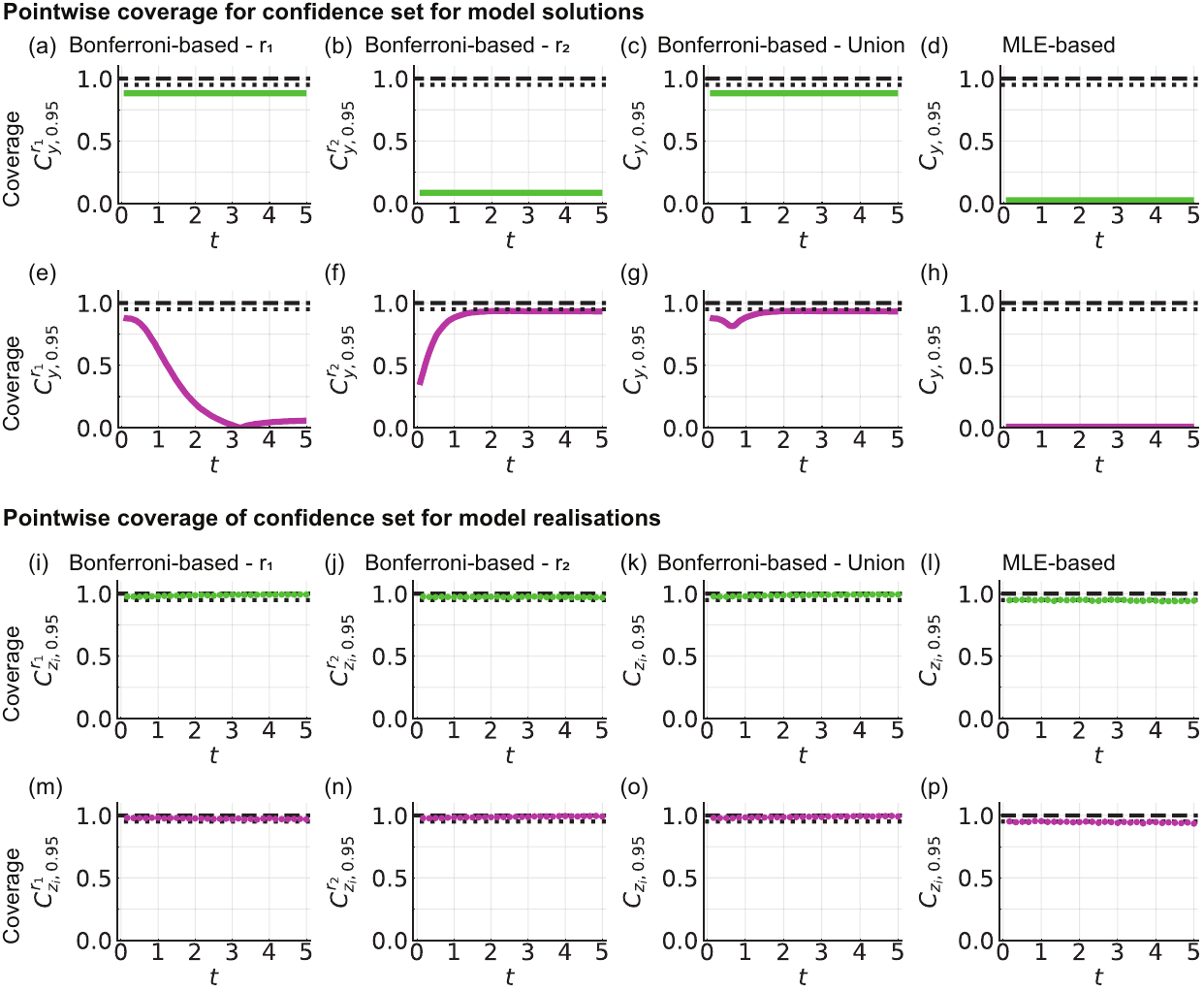}
 	\caption{Pointwise coverage analysis for confidence sets for model solutions and pointwise coverage analysis for confidence sets for model realisations. Analysis performed using the caricature ODE model (Eq (16)) and the log-normal error model as an illustrative example. (a)-(h) Pointwise coverage analysis of confidence sets for model solutions. (i)-(p) Pointwise coverage of confidence sets for model realisations. Results for MLE-based confidence sets are shown in (a,e,i,m). Results for Bonferroni correction-based confidence sets for $r_{1}$, $r_{2}$, and their union are shown in (b,f,j,n), (c,g,j,n), and (d,h,l,p), respectively. Results are obtained by analysing $5000$ data sets generated by simulating Eq (16), the log-normal measurement error model with $\sigma_{{L}}=0.4$, known model parameters $(r_{1}, r_{2}) = (1.0, 0.5)$, and fixed initial conditions $(c_{1}(0), c_{2}(0))=(100.0, 10.0)$. The temporal domain is discretised into $101$ equally--spaced points ($0.00 \leq t \leq 5.00$). Horizontal dotted and horizontal dashed lines correspond to observed probabilities of $0.95$ and $1.00$, respectively.}
 	\label{fig:FigSupp_Coverage3_LogNormal}
 \end{figure}

 \begin{figure}[h!]
 	\centering
 	\includegraphics[width=\textwidth]{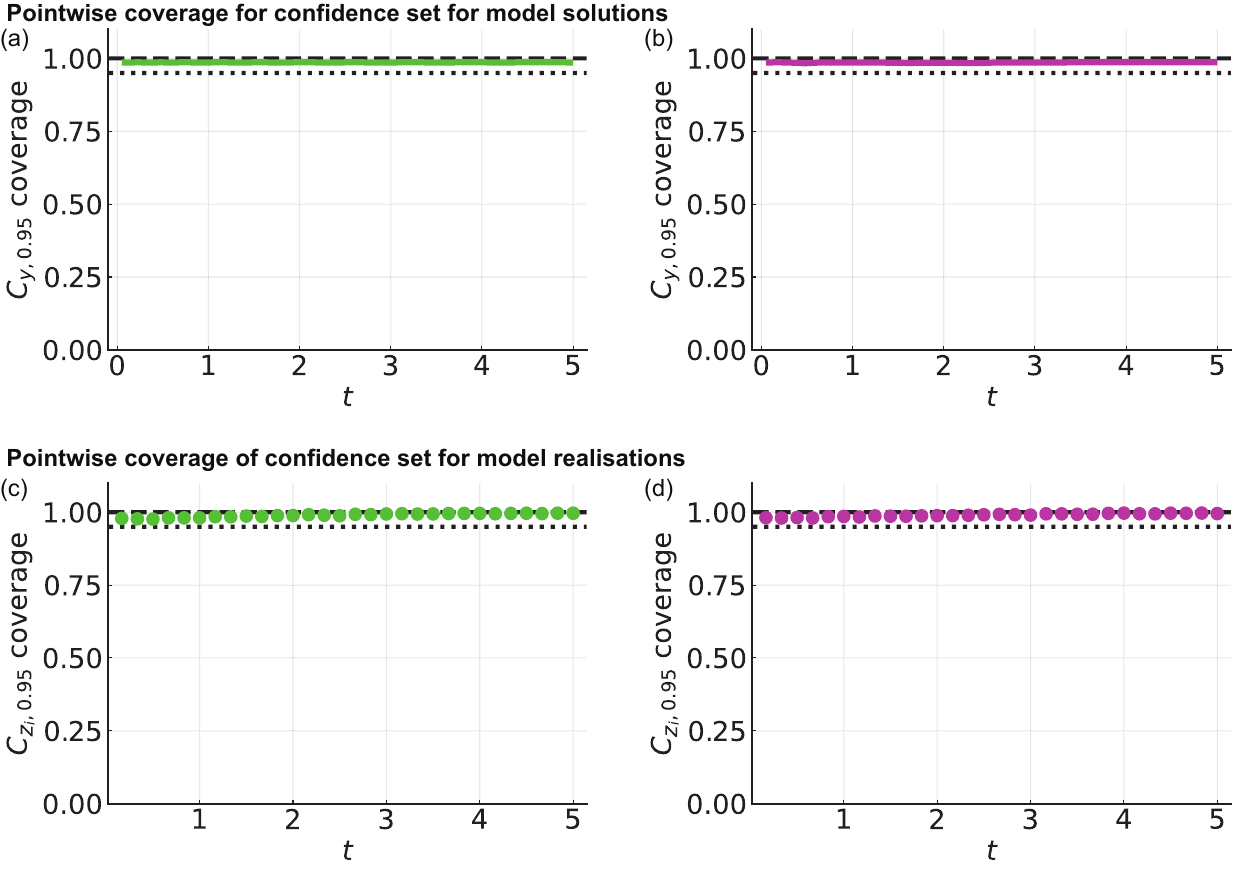}
 	\caption{Pointwise coverage analysis for full likelihood-based confidence sets for model solutions and Bonferroni correction-based confidence sets for model realisations. Analysis performed using the two-parameter two-chemical reaction model as an illustrative example. Pointwise coverage of confidence sets for model solutions for (a) $c_{1}(t)$ and (b) $c_{2}(t)$. The temporal domain is discretised into $100$ equally--spaced points ($0.00 \leq t \leq 5.00$). Pointwise coverage of confidence sets for model realisations for (c) $c_{1}(t)$ and (d) $c_{2}(t)$. Results are obtained by analysing $5000$ data sets generated by simulating Eq (16), the log-normal measurement error model with $\sigma_{{L}}=0.4$, known model parameters $(r_{1}, r_{2}) = (1.0, 0.5)$, and fixed initial conditions $(c_{1}(0), c_{2}(0))=(100.0, 10.0)$.  Horizontal dotted and horizontal dashed lines correspond to coverage probabilities of $0.95$ and $1.00$, respectively.}
 	\label{fig:FigSupp_Coverage_FullLikelihood_LogNormal}
 \end{figure}

 \newpage
 \clearpage
 \section{Residual analysis for error models}\label{sec:qqplots}
 
 In the main manuscript we take a simple and common graphical approach to analyse residuals using quantile-quantile plots. Here we generate a range of synthetic data and explore how quantile-quantile plots can look with and without measurement error model misspecification. This extends analysis performed in the main manuscript for data generated using the log-normal error model (Figs 3-5). Throughout, when we refer to residuals analysed using the additive Gaussian error model we intend the standard additive form, $\hat{e}_{i}=y_{i}^{\mathrm{o}}-y_{i}(\hat{\theta})$, and for residuals analysed using the log-normal error model we intend the ratio $\hat{e}_{i} = y_{i}^{\mathrm{o}}/y_{i}(\hat{\theta})$. We choose not to define residuals for the Poisson error model and discuss this in further detail in the following. Note that there are many approaches to analyse the appropriateness of error models and which approach to use should be considered on a case-by-case basis.
 
\begin{figure}[h!]
 	\centering
 	\includegraphics[width=\textwidth]{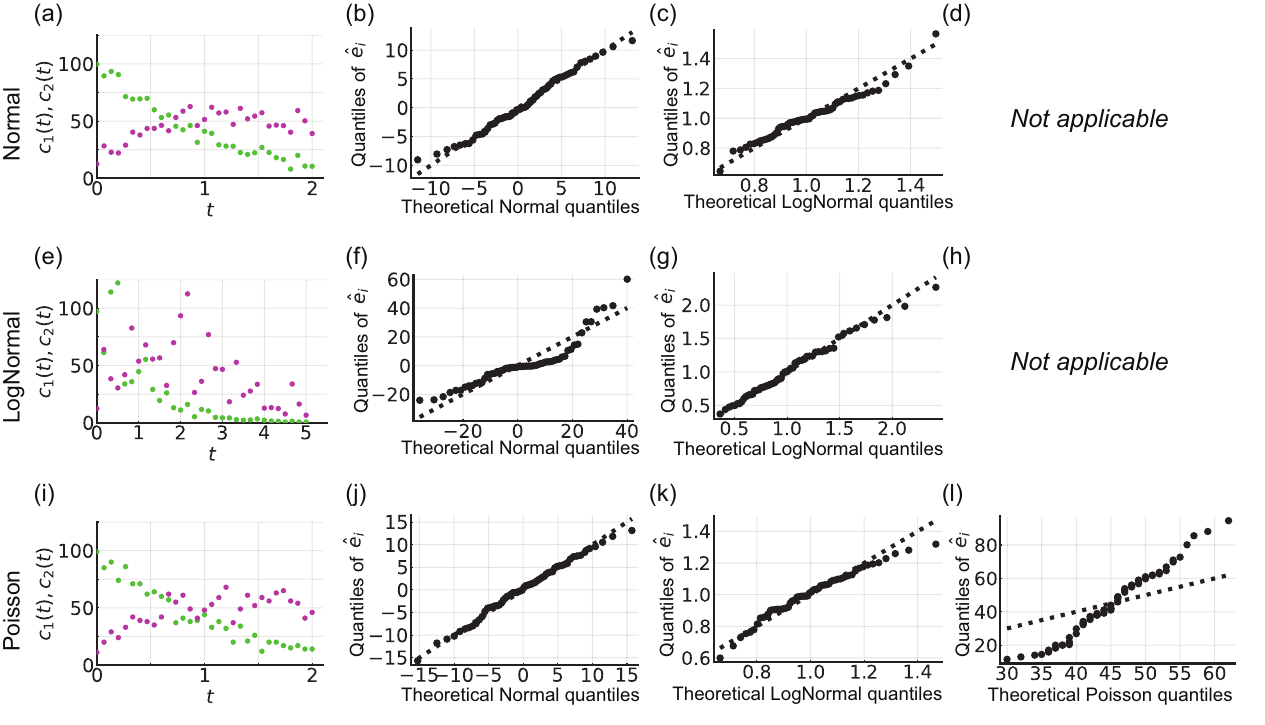}
 	\caption{Example quantile-quantile plots with and without measurement error model misspecification. Synthetic generated using (a) additive Gaussian error model, (e) log-normal error model, (i) Poisson error model. Equations and parameters are described in Section \ref{sec:qqplots}. Quantile-quantile plots for data generated using (b-d) the additive Gaussian error model, (f-h) the log-normal error model, and (j-l) the Poisson error model. The data is analysed using the (b,f,j) the additive Gaussian error model, (c,g,k) the log-normal error model, and (d,h,l) a Poisson error model and a Poisson quantile-quantile model that assumes data is generated from a single rate parameter.}
 	\label{fig:FigSuppqqplots}
 \end{figure}

 In the main manuscript we generate synthetic data at thirty-one equally--spaced time points from $t~=~0.0$ to $t=5.0$ by simulating Eq (17), the log-normal error model, known model parameters $(r_{1}, r_{2}, \sigma_{\mathrm{L}}) = (1.0, 0.5, 0.4)$, and fixed initial conditions $(c_{1}(0), c_{2}(0)) = (100.0, 10.0)$. We then estimate the MLE of model parameters, assuming a log-normal error model, and simulate the solution of the mathematical model evaluated at the MLE. Analysing the ratios $\hat{e}_{i} = y_{i}^{\mathrm{o}}/y_{i}(\hat{\theta})$  using a log-normal quantile-quantile plot we observe that the data are close to the reference line suggesting that the log-normal error model is reasonable (Fig 3c, Fig \ref{fig:FigSuppqqplots}g). In contrast, assuming an additive Gaussian error model for parameter estimation, we observe that residuals on a normal quantile-quantile data deviate from the reference line (Fig 5, Fig \ref{fig:FigSuppqqplots}f). This deviation from the reference line suggests that an additive Gaussian error model is not appropriate in this situation. Recall that the additive Gaussian and log-normal error models are both based on continuous probability distributions whereas the Poisson error model is based on a discrete probability distribution with non-negative integer support. Therefore, directly inspecting the data we observe that it cannot be generated from a Poisson error model as many data values are not non-negative integers. Hence, we do not present a Poisson quantile-quantile plot for data generated using a log-normal error model (Fig \ref{fig:FigSuppqqplots}h).
 
 We now repeat this analysis generating data using the additive Gaussian error model. We generate synthetic data at thirty-one equally--spaced time points from $t=0.0$ to $t=2.0$ using the same initial conditions, and with known model parameters $(r_{1}, r_{2}, \sigma_{\mathrm{N}}) = (1.0, 0.5, 5.0)$. We consider time points over a shorter duration to ensure that all data are non-negative. Assuming an additive Gaussian error model for parameter estimation, we observe that residuals on a normal quantile-quantile plot are close to the reference line suggesting that the additive Gaussian error model is reasonable (Fig \ref{fig:FigSuppqqplots}b). In practice, further analysis should be performed to assess the appropriateness of the additive Gaussian error model for these data, for example by comparing predictions to data. Assuming a log-normal error model for parameter estimation, we observe that the ratios $\hat{e}_{i} = y_{i}^{\mathrm{o}}/y_{i}(\hat{\theta})$ on a log-normal quantile-quantile plot deviate from the reference line suggesting the log-normal error model is not appropriate in this situation (Fig \ref{fig:FigSuppqqplots}c). Again we do not present a Poisson quantile-quantile plot since many data values are not non-negative integers (Fig \ref{fig:FigSuppqqplots}d).
 
 \newpage 
 
 We now repeat this analysis generating data using the Poisson error model. We generate synthetic data at the same time points and initial conditions, and with known model parameters $(r_{1}, r_{2}) = (1.0, 0.5)$. Assuming an additive Gaussian error model for parameter estimation, we observe that residuals on a normal quantile-quantile plot are close to the reference suggesting that the additive Gaussian error model is reasonable (Fig \ref{fig:FigSuppqqplots}j). Assuming a log-normal error model for parameter estimation, we observe that the ratios $\hat{e}_{i} = y_{i}^{\mathrm{o}}/y_{i}(\hat{\theta})$ on a log-normal quantile-quantile plot deviate from the reference line at a tail of the data (Fig \ref{fig:FigSuppqqplots}k). In practice, further analysis should be performed to assess the appropriateness of the additive Gaussian error model and multiplicative log-normal error model for these data, for example by examining the variance of the data through time and comparing predictions to data. In this example, all data are non-negative integers so a Poisson error model could be appropriate. Due to the form of Eq (5) we analyse the data directly rather than considering residuals. It is not as straightforward to interpret the data using a Poisson quantile-quantile plot. A Poisson quantile-quantile plot assumes that all data are generated from a single rate parameter. However, due to the form of Eq (5) data at each time point and for each variable correspond to different rate parameters of the Poisson error model. Therefore, it is not surprising to observe that data deviate from the reference line on a Poisson quantile-quantile plot (Fig \ref{fig:FigSuppqqplots}l). One approach to overcome this challenge is to analyse the assumptions of the Poisson error model per time point and variable, provided there are sufficient measurements.